\documentclass[11pt]{article}
\usepackage{a4}
\usepackage{amstext}
\usepackage{amsfonts}
\usepackage{amssymb}
\usepackage{color}
\usepackage{cite}
\usepackage{epsfig}
\usepackage{mathtools}
\usepackage{ulem}
\usepackage{dsfont}
\usepackage{caption}         
\usepackage{subcaption}   
\usepackage{framed}
\usepackage{multirow}
\usepackage{booktabs}
\usepackage{braket,slashed}
\usepackage{dsfont}

\usepackage{placeins}
\usepackage[english]{babel}
\usepackage[babel]{csquotes}
\usepackage{xspace}

\newcommand{\gtapprox}{\raisebox{-0.5ex}{$\,\stackrel{>}{\scriptstyle\sim}\,$}}
\newcommand{\ltapprox}{\raisebox{-0.5ex}{$\,\stackrel{<}{\scriptstyle\sim}\,$}}

\DeclareMathOperator{\tr}{tr}
\DeclareMathOperator{\Li}{Li}
\DeclareMathOperator{\arcosh}{arcosh}
\newcommand{\fref}[1]{Fig.~\ref{#1}}
\newcommand{\tref}[1]{Tab.~\ref{#1}}
\newcommand{\sref}[1]{Sec.~\ref{#1}}
\newcommand{\srefs}[1]{Secs.~\ref{#1}}
\renewcommand{\eqref}[1]{Eq.\ (\ref{#1})}

\newcommand{\Ns}{\ensuremath{N_{s}}}
\newcommand{\Nt}{\ensuremath{N_{0}}}
\newcommand{\Nf}{\ensuremath{N_{f}}}

\newcommand{\e}{e}
\newcommand{\D}{d}
\newcommand{\I}{\ensuremath{\mathds{1}}}
\newcommand{\curlD}{\ensuremath{\mathcal{D}}}
\newcommand{\ii}{\ensuremath{i}}
\newcommand{\R}{\ensuremath{\mathbb{R}}}
\newcommand{\Z}{\ensuremath{\mathbb{Z}}}

\newcommand{\Q}{Q}
\newcommand{\Seff}{S_{\text{eff}}}
\newcommand{\Tr}{\mathrm{Tr}}
\renewcommand{\det}{\mathrm{Det}}

\definecolor{mcol}{rgb}{1,0,1}
\definecolor{wincol}{rgb}{0,0.4,0.15}
\definecolor{lkcol}{rgb}{0.4,0.4,0}

\usepackage{ulem} 

\begin{document}


\begin{center}

{\huge \bf Regulator dependence of }

{\huge \bf$\phantom{g}$ inhomogeneous phases in the$\phantom{g}$}

{\huge \bf 2+1-dimensional Gross-Neveu model}

\vspace{0.5cm}

\textbf{ Michael Buballa$^{a,c}$, Lennart Kurth$^a$, Marc Wagner$^{b,c}$, Marc Winstel$^b$}

$^a$Technische Universit\"at Darmstadt, Department of Physics, Institut f\"ur Kernphysik, Theoriezentrum,
Schlossgartenstr.\ 2, D-64289 Darmstadt, Germany

$^b$Goethe-Universit\"at Frankfurt, Institut f\"ur Theoretische Physik, Max-von-Laue-Stra{\ss}e 1, D-60438 Frankfurt am Main, Germany

$^c$Helmholtz Research Academy Hesse for FAIR, Campus Riedberg, Max-von-Laue-Stra{\ss}e 12, D-60438 Frankfurt am Main, Germany

\vspace{0.5cm}

December 18, 2020

\end{center}

\begin{tabular*}{16cm}{l@{\extracolsep{\fill}}r} \hline \end{tabular*}

\vspace{-0.4cm}
\begin{center} \textbf{Abstract} \end{center}
\vspace{-0.4cm}

The phase diagram of the Gross-Neveu model in $2+1$ space-time dimensions at non-zero temperature and chemical potential
is studied in the limit of infinitely many flavors,
focusing on the possible existence of inhomogeneous phases, where the order parameter $\sigma$ is non-uniform in space.  
To this end, we analyze the stability of the energetically favored homogeneous configuration 
$\sigma(\textbf{x}) = \bar\sigma = \textrm{const}$ with respect to small inhomogeneous fluctuations, 
employing lattice field theory with two different lattice discretizations as well as a continuum approach with Pauli-Villars 
regularization. Within lattice field theory, we also perform a full minimization of the effective action, allowing for arbitrary
1-dimensional modulations of the order parameter.
For all methods special attention is paid to the role of cutoff effects.
For one of the two lattice discretizations, no inhomogeneous phase was found. 
For the other lattice discretization and within the continuum approach with a finite Pauli-Villars cutoff parameter $\Lambda$,
we find a region in the phase diagram where an inhomogeneous order parameter is favored. 
This inhomogeneous region shrinks, however, when the lattice spacing is decreased or $\Lambda$ is increased,
and finally disappears for all non-zero temperatures when the cutoff is removed completely.
For vanishing temperature, we find hints for a degeneracy of homogeneous and inhomogeneous solutions,
in agreement with earlier findings.

\begin{tabular*}{16cm}{l@{\extracolsep{\fill}}r} \hline \end{tabular*}

\thispagestyle{empty}


\newpage

\setcounter{page}{1}

\section{Introduction}

Mapping the phase diagram of Quantum Chromodynamics (QCD) at non-zero temperature $T$ and quark chemical potential $\mu$ is one of the major challenges in the field of strong-interaction physics \cite{Kumar:2013cqa,Friman:2011zz}. From the theoretical side, the situation is complicated by the fact that perturbative techniques are not applicable in the regime of interest, so that non-perturbative methods must be applied. 
At $\mu = 0$, precise and reliable results from lattice gauge theory with realistic quark masses are available, 
which have revealed that chiral symmetry, which is spontaneously broken in the vacuum, gets approximately restored in a crossover transition around $T \approx 156 \, \textrm{MeV}$ \cite{Borsanyi:2010bp,Bazavov:2011nk,Bellwied:2015rza,Bazavov:2018mes}. While the range of validity of present lattice QCD simulations is restricted to chemical potentials $\mu \lesssim T$ (corresponding to $\mu_B \lesssim 3T$ with the baryon chemical potential $\mu_B = 3 \mu$), continuum approaches to QCD, i.e., Dyson-Schwinger equations \cite{Fischer:2018sdj} or the Functional Renormalization Group \cite{Fu:2019hdw}, predict that at higher chemical potentials there is a first-order phase transition, ending at a chiral critical endpoint (CEP) at $\mu_\textrm{CEP} \approx (1.4 \ldots 2.0)$ $T_\textrm{CEP}$. Qualitatively similar results were also obtained quite some time ago within QCD inspired models, like the Nambu-Jona-Lasinio (NJL) model \cite{Asakawa:1989bq} or the Quark-Meson (QM) model \cite{Scavenius:2000qd,Schaefer:2006ds}. In the chiral limit, i.e., at vanishing bare quark mass, the crossover gets replaced by a second-order phase transition, which is joined to the first-order one at a tricritical point (TCP).

Most of these investigations have been performed assuming that the order parameter (the quark condensate 
$\langle \bar\psi \psi \rangle$) is homogeneous. Several model studies have revealed, however, that, at least in the mean-field approximation, there are certain regions in the phase diagram where spatially varying condensates are favored over homogeneous ones (see Ref.\ \cite{Buballa:2014tba} for a review). Such inhomogeneous phases have been analyzed in detail in models with $1+1$ spacetime dimensions, like the $1+1$-dimensional Gross-Neveu (GN) model \cite{Thies:2003kk,Schnetz:2004vr,Thies:2006ti}, but are also found in $3+1$ dimensions in the NJL and QM models \cite{Nakano:2004cd,Nickel:2009wj,Carignano:2014jla,Heinz:2015lua} as well as in a Dyson-Schwinger approach to QCD \cite{Muller:2013tya}. Typically, the inhomogeneous regions cover the first-order boundary between the homogeneous phases and reach up to the CEP. In the chiral limit the TCP is then replaced by a Lifshitz point (LP), where three phases, the homogeneous symmetry-broken, the restored and the inhomogeneous phase, meet \cite{Nickel:2009ke}.

When considering inhomogeneous phases, the difficulty arises that the determination of the ground state corresponds to the functional minimization of the effective action with respect to the condensate $\langle \bar\psi \psi \rangle(\mathbf{x})$ with arbitrary spatial shape. Obviously this is a very hard problem, which, until now, has only been solved in $1+1$-dimensional models \cite{Thies:2003kk,Schnetz:2004vr,Thies:2006ti,Basar:2009fg} but not in higher dimensions. Instead of a full minimization, most authors, therefore, use certain ansatz functions for the condensate (e.g., single plane waves \cite{Nakano:2004cd} or embedding the known solutions from the $1+1$-dimensional into $3+1$ spacetime dimensions \cite{Nickel:2009wj}) or perform Ginzburg-Landau or stability analyses \cite{Nickel:2009ke,Abuki:2011pf,Buballa:2018hux,Carignano:2019ivp,Buballa:2020xaa}. An interesting alternative is to investigate inhomogeneous phases with lattice field theory or related numerical methods, where, at least in principle, the effective action can be minimized without restricting the condensate to a specific ansatz. This has been demonstrated in Refs.\ \cite{deForcrand:2006zz,Wagner:2007he}, where the $1+1$-dimensional GN model with an infinite number of fermion flavors $N_f$
was investigated numerically, reproducing the known analytical results for the phase diagram and the condensate functions within numerical precision. Recently lattice field theory was also used to simulate the $1+1$-dimensional GN model at finite $N_f$, showing for the first time that an inhomogeneous phase also exists in this case \cite{Lenz:2020bxk,Lenz:2020cuv}.

In the present article we extend these studies to investigate the GN model in $2+1$ dimensions. 
In doing so, we restrict our investigations to the limit $N_f \rightarrow \infty$, corresponding to a mean-field approximation.
For homogeneous phases this model has been analyzed more than three decades ago \cite{Klimenko:1987gi,Rosenstein:1988dj}. It was shown that it is renormalizable in the $N_f$ expansion and it was found that the phase transition is of second order at any non-zero temperature, while at $T=0$ the order parameter changes discontinuously at the phase boundary. Two of us presented a first lattice field theory study of inhomogeneous phases within that model at a recent conference \cite{Winstel:2019zfn}, showing indications for the existence of such phases. Shortly afterwards, however, it was found that the inhomogeneous phase disappears in the continuum limit \cite{Narayanan:2020uqt}, thus, suggesting that the result of Ref.\ \cite{Winstel:2019zfn} was an artifact of the lattice discretization. In the present paper we investigate this more systematically, comparing the dependence of the results on the lattice spacing for two different discretization prescriptions. We also complement the lattice field theory calculations by a stability analysis in a continuum approach. In particular, guided by the idea that a non-zero lattice spacing might be similar to a finite momentum cutoff, we investigate the dependence of the inhomogeneous phase on the cutoff.

The paper is organized as follows. In \sref{SEC597} we introduce the GN model in $2+1$ spacetime dimensions,
discuss the symmetries of the model and address the issue of fermion representations (2-component versus 4-component spinors). In \sref{SEC499} we explain our continuum approach, in \sref{SEC500} the lattice field theory techniques. Our results are presented and discussed in \sref{sec:results}. We draw our conclusions in \sref{SEC608}.



\section{\label{SEC597}Theoretical basics}


\subsection{\label{SEC456}The Gross-Neveu model in $2+1$ dimensions in the limit of infinitely many flavors}

The Gross-Neveu (GN) model \cite{Gross:1974jv} is a relativistic quantum field theory describing $\Nf$ fermion flavors with a four-fermion interaction. We consider the model in $2+1$-dimensional Euclidean spacetime, where action and partition function are
\begin{align}
S[\bar{\psi},\psi] = \int \D^3x \, \bigg(\bar\psi \Big(\gamma_\nu \partial_\nu + \gamma_0 \mu \Big) \psi - \cfrac{\lambda}{2\Nf} \Big(\bar\psi \psi\Big)^2\bigg) \quad , \quad Z = \int \curlD\bar{\psi} \, \curlD\psi \, \e^{-S[\bar{\psi},\psi]} \, . \label{eq:fermi_action}
\end{align}
$\psi = \left( \psi_1 , \dots, \psi_{\Nf} \right)$ represents $\Nf$ massless fermion fields, $\lambda$ is the coupling constant and $\mu$ is the chemical potential. Representations of the $\gamma$ matrices are discussed in \sref{sec:fermion_repr_disc_symmetr}. At non-vanishing temperature the spacetime integral is over $[0,\beta] \times V$, where $\beta = 1/T$ is the inverse temperature and $V$ denotes the 2-dimensional spatial volume.

To get rid of the four-fermion interaction, we introduce an auxiliary scalar field $\sigma$ and perform a Hubbard-Stratonovich transformation \cite{Hubbard:1959ub},
\begin{align}
\label{eq:partboson_action} S_\sigma[\bar{\psi}, \psi, \sigma] = \int \D^3x \, \bigg(\bar\psi \Q \psi + \frac{\Nf}{2 \lambda} \sigma^2\bigg) \quad , \quad Z = \int \curlD\bar{\psi} \, \curlD\psi \, \curlD\sigma \, \e^{-S_\sigma[\bar{\psi},\psi, \sigma]} ,
\end{align}
where 
\begin{equation}
\label{eq:Dirac_operator} \Q = \gamma_\nu \partial_\nu + \gamma_0 \mu + \sigma
\end{equation}
is the Dirac operator. One can show that the expectation value of the scalar field $\sigma$ is proportional to the condensate $\langle \bar\psi \psi \rangle$, i.e.,
\begin{align}
\label{eq:sigm_eqofmotion} \langle\sigma\rangle = -\frac{\lambda}{\Nf} \langle \bar\psi \psi \rangle .
\end{align}   
Thus, $\langle \sigma \rangle$ can be used as an order parameter for chiral symmetry breaking (for more details see \sref{sec:fermion_repr_disc_symmetr}).

After integrating over the fermion fields, one obtains an effective action, which only depends on the scalar field $\sigma$,
\begin{align}
\label{eq:bos_action} S_{\text{eff}}[\sigma] = \Nf \bigg(\frac{1}{2 \lambda} \int \D^3x \, \sigma^2 - \ln(\textrm{Det}(\Q))\bigg) \quad , \quad Z = \int \curlD \sigma \, \e^{-S_{\text{eff}}[\sigma]} .
\end{align}
Here $\textrm{Det}$ denotes a functional determinant in spacetime and in spinor space, while the degenerate flavor degrees of freedom have been factored out. Strictly speaking, the dimensionful operator $\Q$ should be measured with respect to some energy scale, e.g., the temperature in thermal field theory or the inverse lattice spacing in a lattice field theory. While this scale is needed to have a dimensionless argument of the logarithm, it only leads to a constant (i.e., $\sigma$-independent) shift of $S_{\text{eff}}$ and does not affect our results.

In this work we restrict the dependence of $\sigma$ to the spatial coordinates, i.e., $\sigma = \sigma(x_1,x_2)$. With this restriction $S_{\text{eff}}$ is real, which is shown in App.\ \ref{APP002}. 

As one can see from \eqref{eq:bos_action}, the action is proportional to the number of fermion flavors $\Nf$. Since we exclusively consider the limit $\Nf \rightarrow \infty$, only field configurations $\sigma$ corresponding to global minima of $S_{\text{eff}}$ contribute to the partition function $Z$. Thus, instead of integrating over the scalar field $\sigma$ in \eqref{eq:bos_action} it is sufficient to find a global minimum of $S_{\text{eff}}$. Observables $O(\sigma)$ are then evaluated on the minimizing field $\sigma=\langle\sigma\rangle$, i.e., $\langle O(\sigma) \rangle =  O(\langle\sigma\rangle)$.


\subsection{\label{sec:fermion_repr_disc_symmetr}Fermion representations and the discrete symmetry $\sigma \rightarrow -\sigma$}

In App.\ \ref{APP001} we show that the effective action (\ref{eq:bos_action}) has a discrete symmetry
\begin{equation}
\label{eq:sig_to_minus_sig} \sigma \rightarrow -\sigma ,
\end{equation}
i.e., $S_{\text{eff}}[\sigma] = S_{\text{eff}}[-\sigma]$. A non-vanishing $\langle\sigma\rangle$ indicates spontaneous breaking of this symmetry. Moreover, as discussed in the context of \eqref{eq:sigm_eqofmotion}, $\langle\sigma\rangle$ is proportional to the condensate $\langle \bar{\psi}\psi \rangle$. 
 
We also note that a suitable set of $\gamma$ matrices has to fulfill the Dirac algebra in Euclidean spacetime, %
\begin{align}
\label{eq:clifford} \{\gamma_\mu , \gamma_\nu\} = \gamma_\mu \gamma_\nu + \gamma_\nu \gamma_\mu = 2 \delta_{\mu \nu} \I , 
\end{align}
where $\I$ is the identity matrix in spinor space. 


\subsubsection{\label{SEC544}The GN model in $1+1$ dimensions}

We start with a brief discussion of the fermion representation typically used for the GN model in $1+1$ spacetime dimensions, where the situation is less complicated than in $2+1$ spacetime dimensions. A possible irreducible $2 \times 2$ representation of the Dirac algebra (\ref{eq:clifford}) is
\begin{align}
\label{eq:2d_gamma} \gamma_0 = \tau_1 \quad , \quad \gamma_1 = \tau_2 ,
\end{align}
where $\tau_j$ denote Pauli matrices. A suitable $\gamma_5$ matrix, which anti-commutes with both $\gamma_0$ and $\gamma_1$, i.e., fulfilling $\{\gamma_5 , \gamma_\mu\} = 0$, can be defined according to
\begin{align}
\gamma_5 = \tau_3 .
\end{align}
The free fermion action is then invariant under continuous chiral transformations generated by $\gamma_5$,
\begin{equation}
\label{eq:contin_sym} \psi \rightarrow e^{\ii \theta^a \gamma_5 \lambda^a} \psi \quad , \quad \bar{\psi} \rightarrow \bar{\psi} e^{\ii \theta^a \gamma_5 \lambda^a} ,
\end{equation}
where $\lambda^a$ are the generators of the $\mathrm{U}(\Nf)$ flavor symmetry, e.g., the generalized Gell-Mann matrices and the identity, and $\theta^a$ are the parameters of the transformation. The four-fermion interaction term is, however, not invariant under this chiral transformation. For example for $\psi \rightarrow e^{\ii \theta \gamma_5} \psi$,
\begin{align}
\Big(\bar{\psi} \psi\Big)^2 \rightarrow \Big(\bar{\psi} e^{2 \ii \theta \gamma_5} \psi\Big)^2 = \Big(\bar{\psi} \Big(\cos(2 \theta) \I + \ii \sin(2 \theta) \gamma_5\Big) \psi\Big)^2 ,
\end{align} 
i.e., $(\bar{\psi} \psi)^2$ is invariant only for $\theta = n \pi / 2$, $n \in \Z$. One can show that only a discrete chiral symmetry remains,
\begin{align}
\label{eq:disc_sym} \psi \rightarrow \gamma_5 \psi \quad , \quad \bar{\psi} \rightarrow -\bar{\psi} \gamma_5 .
\end{align} 
Due to \eqref{eq:sigm_eqofmotion}, a non-vanishing $\langle\sigma\rangle$ implies a non-vanishing fermion condensate $\langle\bar{\psi}\psi\rangle$. This in turn indicates spontaneous breaking of the discrete chiral symmetry, because this symmetry implies
\begin{equation}
\langle \bar{\psi} \psi \rangle \rightarrow -\langle \bar{\psi} \psi \rangle .
\end{equation}
Consequently, in $1+1$ dimensions with fermion representation (\ref{eq:2d_gamma}) $\langle\sigma\rangle$ is an order parameter for spontaneous breaking of the discrete chiral symmetry \eqref{eq:disc_sym}.


\subsubsection{The GN model in $2+1$ dimensions}

In $2+1$ dimensions there are two inequivalent irreducible $2 \times 2$ representations of the Dirac algebra (\ref{eq:clifford}), which can be written as
\begin{eqnarray}
\label{eq:2comp_gamm1} & & \hspace{-0.7cm} \gamma_0 = +\tau_2 \quad , \quad	\gamma_1 = +\tau_3 \quad , \quad \gamma_2 = +\tau_1 , \\
\label{eq:2:comp_gamm2} & & \hspace{-0.7cm} \tilde{\gamma_0} = -\tau_2 \quad , \quad \tilde{\gamma_1} = -\tau_3 \quad , \quad \tilde{\gamma_2} = -\tau_1 .
\end{eqnarray} 
Neither for the representation (\ref{eq:2comp_gamm1}) nor the representation (\ref{eq:2:comp_gamm2}) there is an appropriate $\gamma_5$ matrix, which anticommutes with all three $\gamma_\mu$. Consequently, there is no discrete chiral symmetry (\ref{eq:disc_sym}) and a non-vanishing $\langle\sigma\rangle$ cannot be interpreted as indication for chiral symmetry breaking. There is, however, another discrete symmetry,
\begin{align}
\label{eq:Parity_2+1} (x_0,x_1,x_2)^T \stackrel{P}{\rightarrow} (x_0,x_1,-x_2)^T \quad , \quad \psi \stackrel{P}{\rightarrow} -\ii \gamma_2 \psi \quad , \quad \bar{\psi} \stackrel{P}{\rightarrow} -\bar{\psi} \ii \gamma_2 ,
\end{align}
changing the sign of $\langle \bar{\psi} \psi \rangle$, i.e., \eqref{eq:Parity_2+1} implies
\begin{equation}
\langle \bar{\psi} \psi \rangle \stackrel{P}{\rightarrow} - \langle \bar{\psi} \psi \rangle .
\end{equation}
This symmetry, which is the reflection of the $x_2$ coordinate, is usually referred to as parity $P$ in $2+1$ dimensions, since the reflection of both spatial coordinates amounts to a rotation by the angle $\pi$.
Thus, a non-vanishing $\langle\sigma\rangle$ indicates spontaneous breaking of parity.

Since the GN model is often used as a toy model for chiral symmetry breaking in QCD, also reducible fermion representations with a corresponding $\gamma_5$ matrix play an important role. One possibility is to combine the representations (\ref{eq:2comp_gamm1}) and (\ref{eq:2:comp_gamm2}) to a $4 \times 4$ representation,
\begin{eqnarray}
\nonumber & & \hspace{-0.7cm} \gamma_0 = \tau_3 \otimes \tau_2 =  \bigg(\begin{array}{cc}
	 +\tau_2 & 0 \\
	 0 & -\tau_2
	 \end{array}\bigg) \quad , \quad \gamma_1 = \tau_3 \otimes \tau_3 =  \bigg(\begin{array}{cc}
	 +\tau_3 & 0 \\
	 0 & -\tau_3
	 \end{array}\bigg) , \\
\label{eq:eq:4comp_gamma} & & \hspace{-0.7cm} \gamma_2 = \tau_3 \otimes \tau_1 =  \bigg(\begin{array}{cc}
	 +\tau_1 & 0 \\
	 0 & -\tau_1
	 \end{array}\bigg)
\end{eqnarray}
(see e.g., Refs.\ \cite{Appelquist:1986fd,Gies:2010st,Scherer:2012nn}). The three matrices are block-diagonal with the upper block corresponding to representation (\ref{eq:2comp_gamm1}) and the lower block to representation (\ref{eq:2:comp_gamm2}). There are two linearly independent matrices that anti-commute with the three $\gamma_0$, $\gamma_1$ and $\gamma_2$,
\begin{align}
\gamma_4 = \tau_1 \otimes \I_2 = \bigg(\begin{array}{cc}
0 & +\I_2 \\
+\I_2 & 0
\end{array}\bigg) \quad , \quad \gamma_5 = -\tau_2 \otimes \I_2 = \bigg(\begin{array}{cc}
0 & +i \I_2 \\
-i \I_2 & 0
\end{array}\bigg) ,
\end{align}
i.e., both fulfill the necessary properties for a suitable $\gamma_5$ matrix.\footnote{We follow the notation of Refs.\ \cite{Gies:2010st, Scherer:2012nn} and denote the two ``$\gamma_5$ candidates'' by $\gamma_4$ and $\gamma_5$, respectively.} Chiral transformations are defined by taking both $\gamma_4$ and $\gamma_5$ into account
\begin{align}
\psi \rightarrow e^{\ii (\phi^a \gamma_4 + \theta^a \gamma_5) \lambda^a} \psi \quad , \quad \bar{\psi} \rightarrow \bar{\psi} e^{\ii (\phi^a \gamma_4 + \theta^a \gamma_5) \lambda^a} , \label{eq:axial_trafo}
\end{align}
where $\phi^a$ and $\theta^a$ are the parameters of the transformation. While free massless fermions are chirally symmetric according to \eqref{eq:axial_trafo}, the four-fermion interaction of the GN model reduces this continuous symmetry to
\begin{equation}
\label{eq:disc_sym2_g4} \psi \rightarrow \gamma_4 \psi \quad , \quad \bar{\psi} \rightarrow -\bar{\psi} \gamma_4 \phantom{.}
\end{equation} 
or, equivalently,
\begin{equation}
\label{eq:disc_sym2} \psi \rightarrow \gamma_5 \psi \quad , \quad \bar{\psi} \rightarrow -\bar{\psi} \gamma_5 .
\end{equation}
Note, however, that \eqref{eq:disc_sym2_g4} and \eqref{eq:disc_sym2} are not independent, but related by a vector transformation in flavor space (see Ref.\ \cite{Schmidt:2017} for details). Thus, in the following,
it is sufficient to consider the discrete chiral symmetry (\ref{eq:disc_sym2}). $\langle\sigma\rangle$ is an order parameter for spontaneous breaking of this symmetry, as in the $1+1$-dimensional case discussed in \sref{SEC544}. 
 

\subsection{Equivalence of $2$- and $4$-component fermion representations  \label{sec:fermion_det}}

In the following we will show a simple relation between expectation values $\langle O(\sigma) \rangle$ obtained with either of the two irreducible 2-component fermion representations (\ref{eq:2comp_gamm1}) and (\ref{eq:2:comp_gamm2}) and the 4-component fermion representation (\ref{eq:eq:4comp_gamma}).

We denote the Dirac operators for fermion representations (\ref{eq:2comp_gamm1}), (\ref{eq:2:comp_gamm2}) and (\ref{eq:eq:4comp_gamma}) with $\Q^{(2)}$, $\tilde{\Q}{}^{(2)}$ and $\Q^{(4)}$, respectively. $\Q^{(4)}$ has block-diagonal structure in spinor space,
\begin{equation}
\Q^{(4)}[\sigma] = \bigg(\begin{array}{cc}
\Q^{(2)}[\sigma] & 0 \\
0 & \tilde{\Q}{}^{(2)}[\sigma]
\end{array}\bigg) .
\end{equation}
Thus,
\begin{equation}
\label{eq:detQ4_equiv} \det(\Q^{(4)}[\sigma]) = \det(\Q^{(2)}[\sigma]) \det(\tilde{\Q}{}^{(2)}[\sigma]) .
\end{equation}
Using
\begin{equation}
\det(\tilde{\Q}{}^{(2)}[+\sigma]) = \det(-\Q^{(2)}[-\sigma]) = \det(\Q^{(2)}[-\sigma]) = \det(\Q^{(2)}[+\sigma]) ,
\end{equation}
where the last step is shown in App.\ \ref{APP001}, \eqref{eq:detQ4_equiv} simplifies to
\begin{equation}
\label{eq:detQ4_equiv_detQ2_sq} \det(\Q^{(4)}[\sigma]) = \Big(\det(\Q^{(2)}[\sigma])\Big)^2 = \Big(\det(\tilde{\Q}{}^{(2)}[\sigma])\Big)^2 .
\end{equation}
From \eqref{eq:bos_action} and \eqref{eq:detQ4_equiv_detQ2_sq} one can conclude
\begin{align}
\label{eq:eff_action_repr} S_{\text{eff}}^{(4)}[\sigma,\lambda] = 2 S_{\text{eff}}^{(2)}[\sigma,2 \lambda] = 2 \tilde{S}_{\text{eff}}^{(2)}[\sigma,2 \lambda] ,
\end{align}
where $S_{\text{eff}}^{(2)}[\sigma,\lambda]$, $\tilde{S}_{\text{eff}}^{(2)}[\sigma,\lambda]$ and $S_{\text{eff}}^{(4)}[\sigma,\lambda]$ denote the effective actions for fermion representations (\ref{eq:2comp_gamm1}), (\ref{eq:2:comp_gamm2}) and (\ref{eq:eq:4comp_gamma}), respectively, and coupling constant $\lambda$. Consequently, expectation values $\langle O(\sigma) \rangle$ are related according to
\begin{align}
\langle O(\sigma) \rangle\Big|_{4 \times 4 \text{ rep.\ } (\ref{eq:eq:4comp_gamma})}^\lambda = \langle O(\sigma) \rangle\Big|_{2 \times 2 \text{ rep.\ } (\ref{eq:2comp_gamm1})}^{2 \lambda} = \langle O(\sigma) \rangle\Big|_{2 \times 2 \text{ rep.\ } (\ref{eq:2:comp_gamm2})}^{2 \lambda} .
\end{align}
Note in particular that the phase diagram with respect to the order parameter $\langle\sigma\rangle$ is the same for all three representations. In practice this is useful, because all numerical computations can be performed with the computationally cheaper $2 \times 2$ fermion representation (\ref{eq:2comp_gamm1}) (or (\ref{eq:2:comp_gamm2})), while the corresponding results are also valid for the $4 \times 4$ fermion representation (\ref{eq:eq:4comp_gamma}), where an interpretation in terms of chiral symmetry and its spontaneous breaking is possible. In the following we denote the dimension of the fermion representation by $N_d$.



\section{\label{SEC499}Continuum approach}

In this section, before introducing our lattice field theory techniques in \sref{SEC500}, we describe, how we study the phase diagram of the $2+1$-dimensional GN model using continuum methods. As explained in \sref{SEC597}, the ground state of the system in the limit $\Nf \rightarrow \infty$ corresponds to the field configuration $\sigma(x)$, $x = (x_0, x_1, x_2)$, which minimizes the effective action $\Seff$, given in \eqref{eq:bos_action}. Despite being a considerable simplification compared to the situation at finite $\Nf$, 
finding this configuration in $2+1$ spacetime dimensions is still extremely difficult.
Since the ground state is static, we can safely ignore the (imaginary) time coordinate $x_0$, but we must retain the possible dependence of the $\sigma$ field on the spatial coordinates $\mathbf{x} = (x_1,x_2)$, if we want to find the true
minimum. The minimization of $\Seff$ is thus a functional minimization with respect to $\sigma(\mathbf{x})$.
This is a very hard problem, both analytically and numerically, which has not been solved so far.

While in \sref{sec:fullmini} we will present first steps towards such a full minimization using lattice field theory,
here we restrict ourselves to a simpler problem and perform a stability analysis of the lowest homogeneous state 
with respect to small inhomogeneous fluctuations. 
This method, which has already been applied in $3+1$ dimensions to investigate inhomogeneous phases in the NJL model \cite{Nakano:2004cd,Buballa:2018hux,Carignano:2019ivp}
and the QM model \cite{Buballa:2020xaa}, 
corresponds to searching for a sufficient, although not necessary, condition for an inhomogeneous phase:
If the lowest homogeneous state turns out to be unstable against small inhomogeneous fluctuations, it is clear that the ground state
must be inhomogeneous. On the other hand, the lowest homogeneous state can be stable against small inhomogeneous fluctuations but still be unstable against large ones. 
The inhomogeneous phases found by a full minimization of the effective action could thus be larger than the unstable regions of  the stability analysis, but the latter are always a part of the former. In the following we explain this method in more detail.


\subsection{\label{sec:stabana}Stability analysis}

In the first step we minimize the effective action at given chemical potential and temperature with respect to spatially constant 
fields,  $\sigma = \bar{\sigma}$. 
We will give a few technical details in \sref{sec:conthomo}, but basically this is a standard procedure, which is obviously much simpler than the functional minimization with respect to arbitrary space-dependent fields $\sigma = \sigma(\mathbf{x})$. 
In the second step, we consider small fluctuations around these homogeneous solutions and inspect their effect on the effective action. To this end we write
\begin{equation}
	\sigma(\mathbf{x}) = \bar\sigma + \delta\sigma(\mathbf{x}),
\end{equation}
where $\bar\sigma$ is a homogeneous field and $\delta\sigma(\mathbf{x})$ denotes a fluctuation, which can have an arbitrary spatial shape but is assumed to have an infinitesimally small amplitude. 

Decomposing the Dirac operator in the same way, $\Q = \bar{\Q} + \delta\sigma$ with $\bar{\Q} = \Q(\bar\sigma)$,
and noting that $\ln(\mathrm{Det}(\Q))  = \Tr(\ln(\Q))$, 
this term can straightforwardly be expanded in powers of $\delta\sigma$,
\begin{equation}
	\ln(\mathrm{Det}(\Q))  
	= \mathrm{Tr}(\ln(\bar{\Q})) +  \mathrm{Tr}\Big(\ln\Big(1+ \bar{\Q}^{-1}\delta\sigma\Big)\Big)
	= \mathrm{Tr}(\ln(\bar{\Q})) -  \sum_{n=1}^\infty  \frac{1}{n}  \Tr\Big(-\bar{\Q}^{-1}\delta\sigma\Big)^n .
\end{equation}
Accordingly, the effective action can be expanded as
\begin{align}
	\Seff =\sum_{n=0}^\infty \Seff^{(n)},
	\label{eq:Seffexpansion}
\end{align}
where $\Seff^{(n)}$ corresponds to the contribution of the $n$-th order in the fluctuations. 
Specifically we find for the three lowest-order terms
\begin{align}
	\Seff^{(0)} &= N_f \bigg( \frac{\beta V}{2\lambda} \bar\sigma^2 - \Tr(\ln(\bar \Q))\bigg)
	\\
	\Seff^{(1)} &= N_f \bigg( \frac{\beta}{\lambda} \bar\sigma \int d^2 x\, \delta\sigma(\mathbf{x}) - \Tr \Big(\bar \Q^{-1} \delta\sigma\Big)
	\bigg)
	\label{eq:Seff1}
	\\
	\Seff^{(2)} &= N_f \bigg(  \frac{\beta}{2\lambda}  \int d^2 x\, \big(\delta\sigma(\mathbf{x})\big)^2 +
	\frac{1}{2}  \Tr \Big(\bar \Q^{-1} \delta\sigma \bar \Q^{-1} \delta\sigma\Big) \bigg),
	\label{eq:Seff2}
\end{align}
where the integrals are over the spatial coordinates $x_1$ and $x_2$, while the factors of $\beta$ originate from the 
$x_0$ integrations.  
Demanding that $\bar\sigma$ minimizes $\Seff^{(0)}$, the corresponding stationary condition
$\partial \Seff^{(0)} / \partial\bar\sigma = 0$ yields the gap equation
\begin{equation}
	\bar\sigma = \frac{\lambda}{\beta V} \Tr\Big(\bar Q^{-1}\Big) .
	\label{eq:gap1}
\end{equation}
We note that $\bar Q^{-1} = (\gamma_\nu\partial_\nu+\gamma_0\mu + \bar\sigma)^{-1}$ is just the propagator of a
non-interacting fermion with mass $\bar\sigma$ at chemical potential $\mu$. In coordinate space it thus depends on
(the difference of) two space-time variables. The functional traces above are defined as 
\begin{equation} 
	\Tr \bigg(\Big(\delta\sigma \bar{\Q}^{-1} \Big)^n\bigg)
	= \int \prod\limits_{j=1}^n d^3x^{(j)}\, \tr\Big(\delta\sigma(\mathbf{x}^{(1)}) \bar Q^{-1}(x^{(1)},x^{(2)}) \ldots  \delta\sigma(\mathbf{x}^{(n)}) \bar Q^{-1}(x^{(n)},x^{(1)})\Big) ,
\end{equation} 
where $\tr$ denotes a trace in spinor space.
These expressions are most easily evaluated using the Fourier representation, 
\begin{equation}
	\bar Q^{-1}(x,x') 
	= \frac{1}{\beta V} \sum\limits_p e^{ip \cdot (x-x')} \bar Q^{-1}(p),
\end{equation}  
or, in the infinite-volume limit,
\begin{equation}
	\bar Q^{-1}(x,x')
	= 
	\int \frac{d^2 p}{(2\pi)^2}
	\frac{1}{\beta} \sum\limits_n e^{i(\omega_n (x_0-x'_0) + \mathbf{p} (\mathbf{x}-\mathbf{x}'))} \bar Q^{-1}(\omega_n,\mathbf{p}),
\end{equation}  
where $\omega_n = 2\pi(n-1/2) /\beta$ are fermionic Matsubara frequencies, and 
\begin{equation}
	\bar Q^{-1}(\omega_n,\mathbf{p}) = \frac{-i\gamma_\lambda \tilde p_\lambda + \bar\sigma}{\tilde p^2 + \bar\sigma^2} \quad , \quad
	\tilde p = \left(\begin{array}{c}  \omega_n -i\mu \\ \mathbf{p} \end{array}  \right)
\end{equation} 
is the Euclidean propagator in momentum space. 

Inserting this into Eq.~(\ref{eq:gap1}), the gap equation becomes
\begin{equation}
	\bar\sigma = \lambda\ell_1\bar\sigma ,
	\label{eq:gap}
\end{equation}
where we have defined
\begin{equation}
	\ell_1= 
	\frac{N_d}{\beta V}  \sum\limits_{p} \frac{1}{{\tilde p}^2 + \bar\sigma^2}
	\ \underset{V\rightarrow\infty}{\longrightarrow} \
	N_d \int \frac{\mathrm{d}^2p}{(2\pi)^2}  \frac{1}{\beta} \sum\limits_n \frac{1}{(\omega_n-i\mu)^2 + \mathbf{p}^2 + \bar\sigma^2} .
	\label{eq:l1def}
\end{equation}
After performing the Matsubara sum, this takes the form
\begin{equation}
	\ell_1= N_d\int \frac{\mathrm{d}^2p}{(2\pi)^2} \frac{1}{2E_\mathbf{p}}\Big(1-n(E_\mathbf{p})-\bar{n}(E_\mathbf{p})\Big)
	\label{eq:l1}
\end{equation}
with 
$E_\mathbf{p}=\sqrt{\mathbf{p}^2+ \bar\sigma^2}$
and the Fermi functions
\begin{align}
	n(E)=\frac{1}{1+\e^{\beta(E-\mu)}}
	\quad , \quad
	\bar{n}(E)=\frac{1}{1+\e^{\beta(E+\mu)}} .
\end{align}

The fluctuating field can be Fourier transformed as 
\begin{equation}
	\delta\sigma(\mathbf{x}) = \int\frac{d^2q}{(2\pi)^2}  e^{i\mathbf{q} \mathbf{x}} \delta\tilde\sigma(\mathbf{q}) ,
\end{equation}
where the field in momentum space must obey the relation $\delta\tilde\sigma(\mathbf{-q}) = \delta\tilde\sigma^\ast(\mathbf{q})$
to ensure that $\delta\sigma(\mathbf{x})$ is a real function.
For the first-order contribution to the effective action we then obtain from Eq.~(\ref{eq:Seff1})
\begin{equation}
	\Seff^{(1)} = \delta\tilde\sigma(\mathbf{0}) \frac{N_f \beta}{\lambda} \Big( \bar\sigma -  \lambda \ell_1 \bar\sigma\Big) = 0 ,
\end{equation}
where the second equality follows from the gap equation (\ref{eq:gap}).
In fact, since according to the first equality only the  homogeneous ($\mathbf{q} = \mathbf{0}$) fluctuations contribute to $\Seff^{(1)}$, 
it has to vanish if we expand about the ``homogeneous ground state'', i.e., the lowest homogeneous solution.

To find instabilities we therefore have to investigate the second-order contribution to the effective action. Evaluating \eqref{eq:Seff2} in momentum space we obtain
\begin{align}
	\Seff^{(2)}=\frac{1}{2} \beta  \int \frac{d^2q}{(2\pi)^2} |\delta\tilde\sigma(\mathbf{q})|^2 \Gamma^{-1}(\mathbf{q}^2),
	\label{eq:Seff2q}
\end{align}
where
\begin{align}
	\Gamma^{-1}(\mathbf{q}^2) =
	\Nf \bigg(\frac{1}{\lambda}-\ell_1-\frac{1}{2}(\mathbf{q}^2+4\bar\sigma^2)\ell_2(\mathbf{q}^2)\bigg)
	\label{eq:Gamma}
\end{align}
with
\begin{align}
	\ell_2(\mathbf{q}^2)
	=
	-N_d \int \frac{\mathrm{d}^2p}{(2\pi)^2}  \frac{1}{\beta} \sum\limits_n 
	\frac{1}{((\omega_n-i\mu)^2 + \mathbf{p}^2 + \bar\sigma^2)((\omega_n-i\mu)^2 + (\mathbf{p}+\mathbf{q})^2 + \bar\sigma^2)} .
	\label{eq:l2def}
\end{align}
Carrying out the Matsubara sum, one obtains
\begin{eqnarray}
\nonumber & & \hspace{-0.7cm}	\ell_2(\mathbf{q}^2) = \\
\nonumber & & = \frac{N_d}{2}\int\frac{\mathrm{d}^2p}{(2\pi)^2}
	\frac{1}{E_{\mathbf{p}+\mathbf{q}}^2 - E_{\mathbf{p}}^2 }
	\bigg(
	\frac{1}{E_{\mathbf{p}+\mathbf{q}}}\Big(1-n(E_{\mathbf{p}+\mathbf{q}})-\bar{n}(E_{\mathbf{p}+\mathbf{q}})\Big)
	-\frac{1}{E_{\mathbf{p}}}\Big(1-n(E_{\mathbf{p}})-\bar{n}(E_{\mathbf{p}})\Big)\bigg) . \\
\label{eq:ell2} & &
\end{eqnarray}

From \eqref{eq:Seff2q} we can see that, unlike in the first-order contribution,  also inhomogeneous ($\mathbf{q} \neq \mathbf{0}$) fluctuations contribute to $\Seff^{(2)}$. 
In particular, if $\Gamma^{-1}(\mathbf{q}^2) <0$ in some momentum region, small fluctuations in that region
will lower the effective action with respect to the homogeneous ground state.
A sufficient condition for an instability of the homogeneous ground state 
with respect to developing inhomogeneities is
therefore to find $\Gamma^{-1}(\mathbf{q}^2) <0$ in some interval around any momentum $\mathbf{q} \neq \mathbf{0}$.
A second-order phase boundary between a homogeneous and an inhomogeneous phase is thus given by the values of $T$ and
$\mu$ for which $\Gamma^{-1}(\mathbf{q})$ just touches the zero axis, i.e.,
both $\Gamma^{-1} = 0$ and $d\Gamma^{-1} / d\mathbf{q}^2 = 0$ at some non-vanishing momentum.


\subsection{Homogeneous phase diagram, tricritical points and Lifshitz points}
\label{sec:conthomo}

We stress again that, in order to arrive at the above conclusions, we must expand about the 
homogeneous ground state, i.e., we first have to determine the constant field $\bar\sigma$ which minimizes $\Seff^{(0)}$.
To this end we first solve the gap equation (\ref{eq:gap}) numerically, which yields all stationary points of $\Seff^{(0)}$. These can be minima, maxima or saddle points, but since there is only a (small) finite number of solutions, it is easy to identify the absolute minima. 
Strictly speaking, since the effective action is symmetric under $\sigma \rightarrow -\sigma$ (see \eqref{eq:sig_to_minus_sig}), 
there is a degeneracy between $\Seff^{(0)}(\bar\sigma)$ and $\Seff^{(0)}(-\bar\sigma)$.
However, this is of no further relevance and we may pick either of these solutions, e.g., the positive one.\footnote{Formally this can be achieved by adding a small fermion mass term to the original action and taking the zero-mass limit at the end.}
Another consequence of this symmetry is the fact that $\bar\sigma = 0$ is always a solution of the gap equation. 

Minimizing $\Seff^{(0)}$ at a large number of points in the $\mu$-$T$ plane we get the homogeneous phase diagram as a by-product.
The phase boundaries between the homogeneous symmetry-broken phase and the restored phase are then simply given by the lines where the minimum changes from a non-vanishing to a vanishing value of $\bar\sigma$ and can in general be obtained by a bisection procedure to the desired accuracy. For first-order phase transitions, i.e., when $\bar\sigma$ discontinuously drops to zero this is indeed what we do in our numerical calculations 
The method is less efficient for second-order phase transitions, i.e., when $\bar\sigma$ continuously goes to zero. 
In this case we determine the phase boundary by a stability analysis of the restored phase against homogeneous fluctuations of the sigma field. The idea is essentially the same as in \sref{sec:stabana} but now we always expand around the trivial solution $\bar\sigma=0$ and restrict ourselves to spatially constant fluctuations, i.e., the $\mathbf{q}=\mathbf{0}$ mode. 
Defining
\begin{align}
	\Gamma_0^{-1}(\mathbf{q}^2) 
	=
	\Gamma^{-1}(\mathbf{q}^2))\big|_{\bar\sigma = 0}
	=
	\Nf \bigg(\frac{1}{\lambda}-\ell_1\Big|_{\bar\sigma = 0} -\frac{1}{2}\mathbf{q}^2\ell_2(\mathbf{q}^2)\Big|_{\bar\sigma = 0}\bigg) ,
	\label{eq:Gamma0}
\end{align}
the phase boundary is then given by the condition
\begin{equation}
\label{eq:2ndorder} \Gamma_0^{-1}(0) = 0 \quad \Leftrightarrow \quad \frac{1}{\lambda}-\ell_1\Big|_{\bar\sigma = 0} = 0 ,
\end{equation}
provided the system is stable against finite fluctuations, related to a first-order phase transition. 
The latter can in general be identified as described above. An exception are the regions in the close vicinity of a TCP, where a first-order phase boundary turns into a second-order one, so that their distinction becomes difficult in practice. 
Therefore, in order to identify these points, we perform a Ginzburg-Landau analysis.
To this end we expand $\Seff^{(0)}$ around the restored solution $\bar\sigma=0$ in powers of $\bar\sigma$, 
\begin{equation}
	\label{EQN695} \Seff^{(0)}(\bar\sigma)
	= \Seff^{(0)}(0) + \frac{1}{2} \frac{d^2\Seff^{(0)}}{d\bar\sigma^2}\bigg|_{\bar\sigma = 0} \bar\sigma^2
	+   \frac{1}{4!} \frac{d^4\Seff^{(0)}}{d\bar\sigma^4}\bigg|_{\bar\sigma = 0} \bar\sigma^4 + \dots
\end{equation}
Odd powers vanish because of the symmetry (\ref{eq:sig_to_minus_sig}).
Assuming that the fourth- and all higher-order derivatives are positive, the phase transition is of second order and takes place 
at the points where the second derivative vanishes. On the other hand, if the fourth-order derivative is negative, there can be a 
first-order phase transition. TCPs are thus located at the points where both the second and the fourth derivative of $\Seff^{(0)}(0)$ vanish.\footnote{Here we still assume that higher-order derivative terms are positive or unimportant. In principle this may not be true, and there could be first-order phase transitions even if the fourth-order derivative is positive. 
	In practice, however, we never encountered such a situation in our numerical calculations,
	i.e., the TCPs resulting from the Ginzburg-Landau analysis always show up where we roughly expect them to be from the numerical minimization of $\Seff^{(0)}$.}
Explicit evaluation of the second derivative yields
\begin{equation}
	\frac{d^2\Seff^{(0)}}{d\bar\sigma^2}\bigg|_{\bar\sigma = 0} 
	= \Nf \beta V \left(\frac{1}{\lambda} -\ell_1\big|_{\bar\sigma = 0}  \right)
	= \beta V \Gamma_0^{-1}(0),
	\label{eq:dSdsig2}
\end{equation}
confirming our previous result (\ref{eq:2ndorder}) for the second-order phase boundary. 
For the fourth-order derivative we find
\begin{equation}
	\frac{d^4\Seff^{(0)}}{d\bar\sigma^4}\bigg|_{\bar\sigma = 0} 
	= -6\Nf \beta V \ell_2(0)\big|_{\bar\sigma = 0} ,
	\label{eq:dSdsig4}
\end{equation}
i.e., the additional condition for the TCP is $\ell_2(0)|_{\bar\sigma = 0} = 0$.

Finally, let us include again the possibility of inhomogeneous phases into these considerations.
Obviously, \eqref{eq:2ndorder} only describes the location of a second-order phase boundary between 
a homogeneous symmetry-broken and a restored phase if the corresponding region is stable against inhomogeneities. 
A necessary condition for this is that the momentum dependent term in Eq.~(\ref{eq:Gamma0}) does not turn
$\Gamma_0^{-1}$ negative at $\mathbf{q}^2 > 0$.
The LP, i.e., the point where the second-order phase boundary between the two homogeneous phases is smoothly connected to the second-order instability line with respect to inhomogeneous fluctuations is thus given by \eqref{eq:2ndorder},
together with the condition 
\begin{equation}
	\frac{d\Gamma_0^{-1}}{d\mathbf{q}^2}\bigg|_{\mathbf{q}^2 = 0} = 0
	\quad \Leftrightarrow \quad \ell_2(0) = 0 .
	\label{eq:LP}
\end{equation}
and, hence, the LP coincides with the TCP (cf.\ \eqref{eq:dSdsig4}).
This is a well-known result from the 1+1-dimensional GN model \cite{Schnetz:2004vr} 
and the 3+1-dimensional NJL model \cite{Nickel:2009ke}.
Indeed, the number of spatial dimensions did not enter in an essential way into the considerations above. 

It is known from the homogeneous analysis of the 2+1-dimensional GN model that the phase transition is second-order
at any non-zero temperature, while at $T=0$ the order parameter changes discontinuously at the phase boundary~\cite{Klimenko:1987gi,Rosenstein:1988dj,Inagaki:1995xw}. Since there is thus no TCP, we conclude that there is no LP either, and we may anticipate that we will not find an inhomogeneous phase in the model at non-zero temperature. We stress, however, that these arguments do not exclude inhomogeneous phases which are reached via first-order phase transitions or which have second-order boundaries without LP (e.g., inhomogeneous ``islands'' surrounded by a single homogeneous phase).


\subsection{Regularization and renormalization}
\label{sec:regren}

The integral $\ell_1$ defined in \eqref{eq:l1} diverges linearly and needs to be regularized.
It is well known, however, that the model is renormalizable in the large-$\Nf$ expansion \cite{Rosenstein:1988dj}. 
The renormalization is done by employing the gap equation (\ref{eq:gap}) to 
relate the coupling constant $\lambda$ to a cutoff $\Lambda$, where the scale is set by 
the ground-state value $\sigma_0$ of $\bar\sigma$ at $T=\mu=0$,
\begin{equation}
	\lambda(\Lambda) = \frac{1}{\ell_1(\Lambda)}\bigg|_{T=\mu=0,\, \bar\sigma = \sigma_0} .
	\label{eq:lambdaren}      
\end{equation} 
Since the model is renormalizable, all observables remain finite, even in the limit $\Lambda \rightarrow \infty$. In the following sections we refer to that limit as renormalized GN model. 

It is, however, instructive to study the phase diagram of the model not only in this renormalized limit but also for 
finite values of the cutoff parameter, in particular in view of our lattice field theory computations discussed in \srefs{SEC500} and \ref{sec:results}, which have to be carried out always at finite lattice spacing. Specifically, we use Pauli-Villars (PV) regularization with one regulator term, 
\begin{align}
	\int\frac{\D^2p}{(2\pi)^2}f(\bar\sigma^2)\longrightarrow\int\frac{\D^2p}{(2\pi)^2}\bigg(f(\bar\sigma^2)-f(\sqrt{\bar\sigma^2+\Lambda^2})\bigg),
	\label{eq:PV}
\end{align}
i.e., we subtract from the integrand a term with the same structure but with $\bar\sigma^2$ replaced by $\bar\sigma^2+\Lambda^2$.
We apply this prescription only to the vacuum parts, i.e., the parts which survive at $T=\mu=0$, but not to the medium contributions,
which are related to Fermi occupation numbers in the integrand and are finite without regularization.\footnote{This is a common but arbitrary choice which becomes irrelevant in the limit $\Lambda \rightarrow \infty$. An advantage of leaving the medium parts unregularized is that many integrals can be performed analytically.}
For the integral $\ell_1$ we then obtain from \eqref{eq:l1}
\begin{eqnarray}
\nonumber & & \hspace{-0.7cm}	\ell_1 \longrightarrow
	N_d\int \frac{\mathrm{d}^2p}{(2\pi)^2}
	\bigg(
	\frac{1}{2E_\mathbf{p}}\Big(1-n(E_\mathbf{p})-\bar{n}(E_\mathbf{p})\Big)	- \frac{1}{2 \sqrt{E^2_\mathbf{p} + \Lambda^2}}
	\bigg) = \\
 & & = \frac{N_d}{4\pi} \bigg(\sqrt{\bar\sigma^2+\Lambda^2} -|\bar\sigma|	-T\ln\Big(1+\e^{-(|\bar\sigma|-\mu)/T}\Big)-T\ln\Big(1+\e^{-(|\bar\sigma|+\mu)/T}\Big)\bigg)
	\label{eq:l1reg}
\end{eqnarray}
and thus, from \eqref{eq:lambdaren}, for the coupling constant
\begin{align}
	\lambda=\frac{4\pi}{N_d (\sqrt{\sigma_0^2+\Lambda^2} -|\sigma_0|)} .
	\label{eq:lambdareg}
\end{align}
For the integral $\ell_2$ we get
\begin{align}
	\ell_2(\mathbf{q}^2)
	=\frac{N_d}{4\pi q}\bigg(
	\arctan\bigg(\frac{2|\bar\sigma|}{q}\bigg)-\arctan\bigg(\frac{2\sqrt{\bar\sigma^2+\Lambda^2}}{q}\bigg)
	+ \int\limits_{|\bar\sigma|}^{\sqrt{\bar\sigma^2 + q^2/4}} dE \, \frac{n(E)+\bar n(E)}{\sqrt{q^2/4 + \bar\sigma^2-E^2}}
	\bigg),
\end{align}
where $q = |\mathbf{q}| = \sqrt{\mathbf{q}^2}$.
Note that $\ell_2$ is finite even for $\Lambda\rightarrow\infty$ but we regularize it nevertheless for consistency.
In particular one can show that the unregularized expressions satisfy the relation
\begin{equation}
	\ell_2(0) = \frac{d}{d\bar\sigma^2} \ell_1,
\end{equation}
which underlies the calculation leading to \eqref{eq:dSdsig4} and, related to this, the coincidence of a TCP and a LP.
This relation is only preserved if $\ell_1$ and $\ell_2$ are regularized in the same way.
One finds
\begin{equation}
	\ell_2(0) = 
	\frac{N_d}{8\pi}\bigg(\frac{1}{\sqrt{\bar\sigma^2+\Lambda^2}} -\frac{1}{|\bar\sigma|}
	\Big(1- n(|\bar\sigma|) - \bar n(|\bar\sigma|) \Big)\bigg) .
	\label{eq:l20reg}
\end{equation}
Note that this expression remains also finite in the restored phase, where $\bar\sigma = 0$,
\begin{equation}
	\ell_2(0)\Big|_{\bar\sigma=0} = 
	\frac{N_d}{8\pi}\bigg(\frac{1}{\Lambda} -\frac{1}{T(1+\cosh(\frac{\mu}{T}))}\bigg) .
\end{equation}

A more intuitive regularization scheme is to apply a sharp cutoff to the momentum integrals,  
\begin{align}
	\int\frac{\D^2p}{(2\pi)^2} f(\mathbf{p})\longrightarrow \int\frac{\D^2p}{(2\pi)^2} f(\mathbf{p}) \theta(\Lambda - |\mathbf{p}|) .
	\label{eq:cut}
\end{align}
Compared to PV regularization, this prescription has the disadvantage that it violates Lorentz invariance.
Related to this, the result for $\ell_2$  is in general not unique but changes under a shift of the loop momentum 
$\mathbf{p} \rightarrow \mathbf{p} + \Delta\mathbf{p}$ in \eqref{eq:ell2}. 
We therefore stay with the PV regularization scheme as described above. It turns out however, that for both $\ell_1$ and
$\ell_2(0)$, which are relevant for the homogeneous second-order phase boundary as well as for the LP (see
Eqs.~(\ref{eq:2ndorder}) and  (\ref{eq:LP})), the cutoff regularization gives identical results.\footnote{This is a special property of $2+1$ dimensions. It holds irrespective of whether or not the finite medium parts are regularized as well.}
In this sense the PV regularization parameter $\Lambda$ can be interpreted as a Lorentz invariant generalization of a
momentum cutoff.

Finally, we note that one regulator term is not sufficient to render the cubically divergent effective action finite. 
The physically more relevant difference $\Delta\Seff^{(0)}(\bar\sigma) = \Seff^{(0)}(\bar\sigma) - \Seff[0]$ is however finite in this regularization scheme.
In particular the determination of the homogeneous ground state as the basis of our stability analysis can always be done by
calculating $\Delta\Seff^{(0)}$  for the different solutions of the gap equation. 
One finds
\begin{eqnarray}
\nonumber   \frac{\Delta\Seff^{(0)}(\bar\sigma)}{\beta V\Nf} 
\hspace{-1mm}
&=&
\hspace{-1mm}
\frac{\bar\sigma^2}{2\lambda}
-\frac{N_d}{12\pi}\bigg(|\bar\sigma|^3-\Lambda^3+\left(\bar\sigma^2+\Lambda^2\right)^{\frac{3}{2}}
+\frac{3|\bar\sigma|}{\beta^2}\Big(\Li_2\big(-\e^{\beta(|\bar\sigma|-\mu)}\big)+\Li_2\big(-\e^{\beta(|\bar\sigma|+\mu)}\big)\Big) \\
 & & -\frac{3}{\beta^3}\Big(\Li_3\big(-\e^{\beta(|\bar\sigma|-\mu)}\big)+\Li_3\big(-\e^{\beta(|\bar\sigma|+\mu)}\big)-\Li_3\big(-\e^{-\beta\mu}\big)-\Li_3\big(-\e^{\beta\mu}\big)\Big)\bigg),
\end{eqnarray}
where the $\Li$ are polylogarithms.
Inserting \eqref{eq:lambdareg} for $\lambda$ and evaluating this expression at $T=\mu=0$,
we obtain for the vacuum solution $\bar\sigma = \sigma_0$
\begin{equation}
\frac{\Delta\Seff^{(0)}(\sigma_0)}{\beta V\Nf}
\ \overset{\mu,T\rightarrow 0}{=} \
-\frac{N_d}{24\pi} \bigg( \sigma_0^3
+(2\Lambda^2 - \sigma_0^2) \sqrt{\sigma_0^2 + \Lambda^2} - 2\Lambda^3
\bigg),
\end{equation}
which is always negative, confirming the stability of the vacuum solution $\sigma_0$ 
with respect to the restored phase $\sigma = 0$. 
For $\Lambda \rightarrow \infty$, the right-hand side of this equation stays finite
and becomes $-(N_d / 24\pi) \sigma_0^3$.


\subsection{Analytical results for the phase diagram}
\label{sec:anapd}

We are now in the position to evaluate the relations derived in \sref{sec:conthomo} more explicitly. 
In particular we can determine the second-order phase boundary 
between the restored and the homogeneous symmetry-broken phase
by inserting \eqref{eq:l1reg} with $\bar\sigma = 0$ and \eqref{eq:lambdareg} into \eqref{eq:2ndorder}.     
This leads to
\begin{align}
	T\ln\Big(1+\e^{\mu/T}\Big) + T\ln\Big(1+\e^{-\mu/T}\Big) = s,
	\label{eq:Tmupb}
\end{align}
where we have defined
\begin{align}
	s =\sigma_0  + \Lambda  -  \sqrt{\sigma_0^2+\Lambda^2} .
\end{align}
Here and in the following we assume the vacuum condensate $\sigma_0$ to be positive.

Solving \eqref{eq:Tmupb} for $\mu=0$ yields the critical temperature
\begin{align}
	T_c(\mu = 0) = \frac{s}{2\ln 2} ,
	\label{eq:Tc}
\end{align}
which can be expanded in powers of the inverse cutoff,
\begin{align}
	\frac{T_c}{\sigma_0}  
	= \frac{1}{2\ln 2} \bigg( 1 - \frac{\sigma_0}{2\Lambda} + \dots \bigg) .
	\label{eq:Tclin}
\end{align}
This is consistent with the known result $T_c/\sigma_0 = 1/ 2 \ln2$ in the limit $\Lambda \rightarrow\infty$ \cite{Klimenko:1987gi,Rosenstein:1988dj}.

For $T\leq T_c$ we can also solve \eqref{eq:Tmupb} for the critical chemical potential. One finds
\begin{align}
	\mu_c(T) = T \arcosh\bigg(\frac{1}{2} \e^{s/T} - 1\bigg) .
	\label{eq:muc}
\end{align}
Again this is consistent with the critical phase boundary found in Refs.\ \cite{Klimenko:1987gi,Rosenstein:1988dj} 
in the limit $\Lambda \rightarrow\infty$, corresponding to $s \rightarrow \sigma_0$.
In this case the line reaches the $\mu$ axis at $\mu_c(T=0) = \sigma_0$, which is replaced by $s$ in the case of
finite $\Lambda$.

We must keep in mind, however, that the above equations have been derived under the assumption that there is a 
second-order phase transition from the restored to the homogeneous symmetry-broken phase. 
In Ref.~\cite{Rosenstein:1988dj} it was shown for $\Lambda \rightarrow\infty$ that this is indeed the case for all $T>0$ (at least 
for homogeneous condensates), while at $T=0$ the condensate changes discontinuously at $\mu_c$, corresponding to a first-order phase transition. 
For arbitrary values of $\Lambda$ we can compute the corresponding TCP (which, as shown above,  is equal to the LP of
an inhomogeneous phase)  as the simultaneous solution of Eqs.~(\ref{eq:2ndorder}) and (\ref{eq:LP}), i.e., as the point where the lines defined by \eqref{eq:muc} and by the condition $\ell_2(0) = 0$ cross. Using \eqref{eq:l20reg} and taking the limit $\bar\sigma \rightarrow 0$ one finds that the temperature at the LP is given by the equation
\begin{equation}
	\frac{T_\mathrm{LP}}{s} e^{s/T_\mathrm{LP}} = \frac{2\Lambda}{s} ,	\label{eq:TLPeq}
\end{equation}
which has the solution
\begin{equation}
	T_\mathrm{LP}(\Lambda) = \frac{-s}{W_{-1}\big(-\frac{s}{2\Lambda}\big)} ,
	\label{eq:TLP}
\end{equation}
where $W_{-1}$ is the lower branch of the Lambert $W$ function.
Inserting \eqref{eq:TLPeq} into \eqref{eq:muc} one then obtains for the chemical potential
\begin{align}
	\mu_\mathrm{LP}(\Lambda)=T_\mathrm{LP}\arcosh\bigg(\frac{\Lambda}{T_\mathrm{LP}}-1\bigg) .
	\label{eq:muLP}
\end{align}

\begin{figure}[htb]
	\begin{center}
		\includegraphics[width=10.0cm]{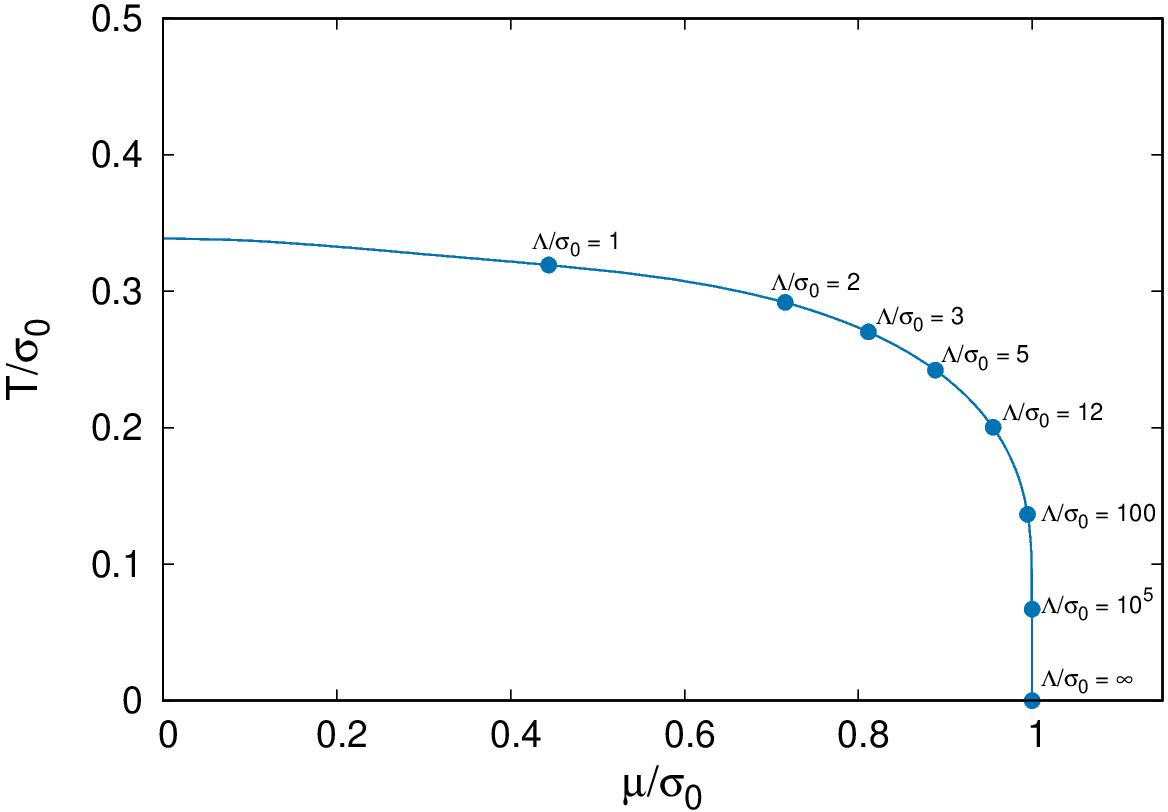}
	\end{center}	
	\caption{\label{fig:LP}Line of LPs (which coincide with the TCPs if the analysis is restricted to 
		homogeneous phases)  given by \eqref{eq:TLP} and \eqref{eq:muLP} 
		for continuously varied cutoff parameters $\Lambda$ (solid line). The dots indicate specific
		values of $\Lambda$.
	}
\end{figure}

The line of LPs in the $\mu$-$T$ plane for varying cutoff parameter $\Lambda$ is shown in \fref{fig:LP},
where the dots indicate specific examples of $\Lambda$.
For finite values of $\Lambda$ there is a LP at non-zero temperature, signaling the existence of an inhomogeneous phase.\footnote{At $\Lambda / \sigma_0 < 2 (1 - \ln 2) / (1 - (1-\ln2)^2) \approx 0.68$ the LP reaches the $T$ axis at $T = \Lambda/2$, and for even smaller cutoff values it disappears from the phase diagram. In this case the stability analysis predicts that the inhomogeneous phase reaches all the way down to $\mu = 0$.}
In the limit $\Lambda \rightarrow \infty$, on the other hand, $T_\mathrm{LP} \rightarrow 0$, and we expect a  second-order phase transition between homogeneous phases which are stable against inhomogeneous fluctuations for all $T>0$.  
The figure shows, however, that this limit is reached extremely slowly.



\section{\label{SEC500}Lattice field theory techniques}


\subsection{Lattice discretization} 

We consider a $2$-dimensional spatial volume $V$ of extent $L$, i.e., $V = L^2$, with periodic boundary conditions. This volume is discretized using lattice field theory, where the lattice spacing is denoted by $a$ and the number of lattice sites is $\Ns^2$, i.e., $\Ns$ lattice sites in each of the two directions and $L = \Ns a$. Since we are interested in studying spontaneous chiral symmetry breaking, it is essential to use a chirally symmetric fermion discretization. We decided to use the naive discretization (see, e.g., the textbook \cite{Rothe:1992nt}). Naive fermions imply fermion doubling, i.e., in our case of two spatial dimensions the number of fermion flavors $\Nf$ is restricted to multiples of $4$. For our work this is not a problem, because we are interested in the limit $\Nf \rightarrow \infty$.

The extent of the temporal direction corresponds to the inverse temperature $\beta = 1/T$ and boundary conditions are antiperiodic. In temporal direction we do not use lattice field theory, but regularize the fermion fields by a superposition of $2 N_0$ plane waves as discussed in detail below and in Refs.\ \cite{Wagner:2007he,Heinz:2015lua}. Since the chiral condensate does not depend on $x_0$, i.e., $\sigma = \sigma(\textbf{x})$ as discussed in previous sections, plane waves allow straightforward analytical simplifications of the fermion determinant. Moreover, the chemical potential can be introduced as in the continuum by adding $\gamma_0 \mu$ to the Dirac operator. In particular an exponential coupling as typically used in lattice field theory is not necessary. As a consequence we expect smaller discretization errors (see Ref.\ \cite{Lenz:2020cuv} for a detailed discussion).


\subsubsection{\label{SEC672}Free fermions}

We define the plane-wave expansion of a fermion field representing a single flavor as
\begin{eqnarray}
\label{eq:plane_wave_ansatz} \psi(x_0,\textbf{x}) = \sum_{n_0=-\Nt+1}^{\Nt} \frac{1}{\sqrt{2 \Nt}} \psi(n_0,\textbf{x}) \e^{\ii \omega_{n_0} x_0} ,
\end{eqnarray} 
where $2 \Nt$ represents the number of modes used in temporal direction. The frequencies \\ $\omega_{n_0} = 2 \pi (n_0 - 1/2) / \beta$ with $n_0 = -\Nt + 1,-\Nt + 2,\ldots,+\Nt-1,+\Nt$ imply antiperiodic boundary conditions in temporal direction. We use $1 / a \equiv 1$ as density of degrees of freedom \footnote{Throughout this section we express all dimensionful quantities in units of the lattice spacing, e.g., $L \equiv L/a$, $\mu \equiv \mu a$ or $\sigma \equiv \sigma a$.}, i.e., $2 \Nt / \beta = 1$. Consequently, the inverse temperature and the number of modes are related according to $\beta = 2 \Nt$. Inserting the plane-wave expansion into the free fermion action leads to
\begin{eqnarray}
\nonumber & & \hspace{-0.7cm} S_{\text{free}}[\bar{\psi},\psi] = \int d^3x \, \bar{\psi}(x_0,\mathbf{x}) \Big(\gamma_0 (\partial_0 + \mu) + \gamma_1 \partial_1 + \gamma_2 \partial_2\Big) \psi(x_0,\mathbf{x}) = \\
\label{EQN693} & & = \sum_{n_0=-\Nt+1}^{\Nt} \int d^2x \, \bar{\psi}(n_0,\mathbf{x}) \Big(\gamma_0 (\ii \omega_{n_0} + \mu) + \gamma_1 \partial_1 + \gamma_2 \partial_2\Big) \psi(n_0,\mathbf{x}) .
\end{eqnarray} 
For the two spatial directions we apply the naive lattice discretization,%
\begin{eqnarray}
\nonumber & & \hspace{-0.7cm} S_{\text{free}}[\bar{\chi},\chi] = \\
\nonumber & & = \sum_{n_0=-\Nt+1}^{\Nt} \sum_{\textbf{x}} \bar{\chi}(n_0,\textbf{x}) \bigg(\gamma_0 (\ii \omega_{n_0} + \mu) \chi(n_0,\textbf{x}) + \sum_{\nu=1,2} \gamma_\nu \frac{\chi(n_0,\textbf{x} + \textbf{e}_{\nu}) - \chi(n_0,{\textbf{x}-\textbf{e}_{\nu}})}{2}\bigg) . \\
\label{eq:S_free_hybrid_disc} & &
\end{eqnarray} 
Note that, due to fermion doubling, $\chi$ represents four fermion flavors and, thus, cannot be interpreted as a standard fermion field $\psi$ as, e.g., used in Eq.\ (\ref{EQN693}) or in \sref{SEC597}. The relation between the components of $\chi$ and of $\psi$ is non-trivial. $\Ns$ must be even to have periodic boundary conditions in the spatial directions for all flavors. We refer to Refs.\ \cite{Cohen:1983nr,Lenz:2020bxk}, where this is discussed in detail.


\subsubsection{The GN model\label{sec:lattice_GN}}

A possible lattice discretization of the effective action (\ref{eq:partboson_action}) of the GN model with $\Nf$ flavors (where $\Nf$ must be a multiple of $4$) with naive fermions is
\begin{eqnarray}
\nonumber & & \hspace{-0.7cm} S_\sigma[\bar{\chi}_f,\chi_f,\sigma] = \\%
\nonumber & & = \sum_{f=1}^{\Nf/4} \bigg(S_{\text{free}}[\bar{\chi}_f,\chi_f] + \sum_{n_0=-\Nt+1}^{\Nt} \sum_{\textbf{x}, \textbf{y}} \bar{\chi}_f(n_0, \textbf{x}) W_2(\textbf{x} - \textbf{y}) \sigma(\textbf{y}) \chi_f(n_0, \textbf{x})\bigg) + \frac{\Nf \Nt}{\lambda} \sum_{\textbf{x}} \sigma^2(\textbf{x}) . \\
\label{eq:bos_action_disc} & &
\end{eqnarray}
$W_2$ is the Fourier transform of a function $\tilde{W}_2$ with the following properties:
\begin{itemize}
\item $\tilde{W}_2 (\textbf{k}) \rightarrow 1$ for $(|k_1|,|k_2|) \approx (0,0)$,

\item $\tilde{W}_2 (\textbf{k}) \rightarrow 0$ for $(|k_1|,|k_2|) \approx (\pi,0)$, $(|k_1|,|k_2|) \approx (0,\pi)$ and $(|k_1|,|k_2|) \approx (\pi,\pi)$
\end{itemize} 
with $\textbf{k} = (k_1,k_2)$.

In the following we explore two suitable choices, which have the same correct continuum limit:
\begin{eqnarray}
\label{eq:corr_disc} W'_2(\textbf{x}) = \prod_{\nu=1,2} W'_1(x_\nu) \quad , \quad W'_1(x_\nu) = \frac{1}{4} \delta_{x_\nu,-1} + \frac{1}{2} \delta_{x_\nu,0} + \frac{1}{4} \delta_{x_\nu,+1} ,
\end{eqnarray}
where the corresponding Fourier transform is $\tilde{W}'_1(k) = (\cos(k) + 1) / 2$, and
\begin{eqnarray}
\label{eq:corr_disc_2} W''_2(\textbf{x}) = \prod_{\nu=1,2} W''_1(x_\nu) \quad , \quad W''_1(x_\nu) = \frac{1}{\Ns}\bigg(1 + \sum_{n=1}^{\Ns/4-1} 2 \cos\bigg(\frac{2 \pi n x_\nu}{L}\bigg) + \cos\bigg(\frac{\pi x_\nu}{2}\bigg)\bigg) ,
\end{eqnarray}
where $\Ns$ is a multiple of 4 and the corresponding Fourier transform is $\tilde{W}'_1(k) = \Theta(\pi/2 - |k|)$.

Because of fermion doubling $\Nf/4$ naive fermion fields $\chi_f$ are needed to represent $\Nf$ fermion flavors. We stress that a specific non-diagonal structure of $W_2(\textbf{x})$ is mandatory for a valid discretization of the GN model with naive fermions, i.e., a discretization with the correct continuum limit. The straightforward and probably more intuitive choice $W_2(\textbf{x}) = \delta_{x_1,0} \delta_{x_2,0}$, which we used at an early stage of this work \cite{Winstel:2019zfn} and which was also used in Ref.\ \cite{Narayanan:2020uqt}, introduces additional four fermion interactions which are not part of the GN model, e.g., couplings between different flavors like $\bar{\psi}_1 \psi_2 \sigma$, $\bar{\psi}_1 \psi_3 \sigma$, etc. For a more detailed discussion we refer to \sref{SEC650} of this work and to appendix~A of Ref.\ \cite{Lenz:2020bxk}.

After integrating over the fermion fields in the partition function, as discussed in \sref{SEC597}, one obtains the discretized effective action
\begin{eqnarray}
\label{eq:eff_action_discr} \frac{S_{\text{eff}}[\sigma]}{\Nf} = \frac{\Nt}{\lambda} \sum_{\textbf{x}} \sigma^2(\textbf{x}) - \frac{1}{4} \ln(\det(\Q)) .
\end{eqnarray}
The Dirac operator $\Q$ is a matrix of size $2 \Nt \Ns^2 N_d \times 2 \Nt \Ns^2 N_d$ with rows and columns representing spacetime and spin,
\begin{eqnarray}
\nonumber & & \hspace{-0.7cm} \Q(n_0,\textbf{x};n'_0,\textbf{x}') = \\
\nonumber & & = \delta_{n_0,n'_0} \bigg(\gamma_0 (\ii \omega_{n_0} + \mu) \delta_{\textbf{x}, \textbf{x}'} + \sum_{\nu = 1,2} \gamma_\nu \frac{\delta_{\textbf{x} + \textbf{e}_\nu, \textbf{x}'} - \delta_{\textbf{x}  - \textbf{e}_\nu, \textbf{x}'}}{2} + \sum_{\textbf{y}} W_2(\textbf{x} - \textbf{y}) \sigma(\textbf{y}) \delta_{\textbf{x}, \textbf{x}'}\bigg) . \\%
\label{eq:disc_dirac_op} & &
\end{eqnarray}
This matrix is diagonal with respect to the temporal indices $n_0$ and $n'_0$. Thus, one can factorize the fermion determinant in \eqref{eq:eff_action_discr} according to
\begin{eqnarray}
\label{eq:ln_det_Q} \ln(\det(\Q)) = \sum_{n_0=-\Nt+1}^{\Nt} \ln\Big(\det\Big(\Q(n_0,\textbf{x};n_0,\textbf{x}')\Big)\Big) ,
\end{eqnarray}
i.e., the problem is reduced to the computation of the determinants of $2 \Nt$ smaller matrices of size $\Ns^2 N_d\times \Ns^2 N_d $.

When restricting the dependence of $\sigma$ to only one of the two spatial coordinates, i.e., $\sigma = \sigma (x_1)$, the numerical costs of computing $S_{\text{eff}}$ can be further reduced. Similar to the plane-wave expansion in temporal direction one can diagonalize the Dirac operator (\ref{eq:disc_dirac_op}) with respect to $x_2$ by writing
\begin{equation}
\chi(x_0,\textbf{x}) = \sum_{n_0=-\Nt+1}^{\Nt} \sum_{n_2 = 0}^{\Ns-1} \frac{1}{\sqrt{2 \Nt \Ns}} \chi(n_0,x_1,n_2) \, \e^{\ii (\omega_{n_0} x_0 + k_{n_2} x_2)}
\end{equation} 
with $x_2$ still restricted to the sites of the spatial lattice\footnote{In principle the $x_2$ direction could be treated in the continuum, exactly in the same way as the temporal direction discussed in \sref{SEC672}. However, since we carry out computations with $\sigma = \sigma(x_1)$ and $\sigma = \sigma(x_1,x_2)$ in \sref{sec:results}, we prefer to use the same lattice regularization in both cases.
} and $k_{n_2} = 2 \pi n_2 / L$. The Dirac operator (\ref{eq:disc_dirac_op}) then becomes
\begin{eqnarray}
\nonumber & & \hspace{-0.7cm} \Q(n_0,x_1,n_2;n'_0,x'_1,n'_2) = \\
\nonumber & & = \delta_{n_0,n'_0} \delta_{n_2,n'_2} \bigg(\gamma_0 (\ii \omega_{n_0} + \mu) \delta_{x_1, x'_1} + \gamma_1 \frac{\delta_{x_1 + 1, x'_1} - \delta_{x_1 - 1, x'_1}}{2} + \gamma_2 \ii \sin(k_{n_2}) \delta_{x_1, x'_1} \\
\label{eq:disc_dirac_op2} & & \hspace{0.675cm} + \sum_{y_1} W_1(x_1 - y_1) \sigma(y_1) \delta_{x_1, x'_1}\bigg) ,
\end{eqnarray}
where $W_1 = W'_1$ or $W_1 = W''_1$. This is a diagonal matrix with respect to the temporal indices $n_0$ and $n'_0$ as well as the spatial indices $n_2$ and $n'_2$. The computation of the fermion determinant in \eqref{eq:eff_action_discr} is, thus, reduced to the computation of the determinants of $2 \Nt \Ns$ matrices of size $\Ns N_d \times \Ns N_d$,
\begin{eqnarray}
\label{eq:ln_det_Q_1d} \ln(\det(\Q)) = \sum_{n_0=-\Nt+1}^{\Nt} \sum_{n_2 = 0}^{\Ns-1} \ln\Big(\det \Big(\Q(n_0,x_1,n_2;n_0,x'_1,n_2)\Big)\Big) .
\end{eqnarray}


\subsection{\label{SEC588}Numerical evaluation of the effective action}

We perform all computations with the $2 \times 2$ representation of $\gamma$ matrices (\ref{eq:2comp_gamm1}), i.e., $N_d=2$. An important part of these computations is the numerical evaluation of the effective action (\ref{eq:eff_action_discr}) for a given field configuration $\sigma$. Typically, this has to be repeated many times, e.g., when minimizing $S_\textrm{eff}$ with respect to $\sigma$, or when checking the stability of a homogeneous condensate $\sigma = \bar{\sigma} = \textrm{const}$ with respect to inhomogeneous perturbations. Particularly time consuming is the computation of $\ln(\det(\Q))$. To maximize efficiency, we distinguish the following three cases:
\begin{itemize}
\item $\sigma = \sigma(\textbf{x})$: \\
The Dirac operator (\ref{eq:disc_dirac_op}) is a block-diagonal matrix with $2 \Nt$ blocks of size $2 \Ns^2 \times 2 \Ns^2$. The determinant of each block is computed via a standard LU-decomposition. We use the publicly available GSL library \cite{GSL}.
	
\item $\sigma = \sigma(x_1)$: \\
The Dirac operator (\ref{eq:disc_dirac_op2}) is a block-diagonal matrix with $2 \Nt \Ns$ blocks of size $2 \Ns \times 2 \Ns$. Again the determinant of each block is computed via a standard LU-decomposition.
	
\item $\sigma = \bar{\sigma} = \textrm{const}$: \\
For homogeneous condensates we obtain identical results for the discretizations $W'_2$ and $W''_2$, as can be seen from Eqs.\ (\ref{eq:bos_action_disc}) to (\ref{eq:corr_disc_2}). The eigenvalues of $\Q$ can be calculated analytically in a straightforward way. $\ln(\det(\Q))$ is then computed by summing over the eigenvalues, leading to
\begin{eqnarray}
\label{eq:Seff_hom_series} \frac{S_{\text{eff}}[\bar{\sigma}]}{\Nf} = \frac{\Nt \Ns^2}{\lambda} \bar{\sigma}^2 - \frac{1}{4} \sum_{n_0=1}^{\Nt} \sum_{n_1 = 0}^{\Ns-1} \sum_{n_2 = 0}^{\Ns-1} \ln\Big(A^2(\bar{\sigma}) + B^2\Big)
\end{eqnarray}
(note that in contrast to previous equations the sum over $n_0$ is restricted to positive integers) with
\begin{eqnarray}
A(\bar{\sigma}) = \bar{\sigma}^2 - \mu^2 + \omega_{n_0}^2 + \sin^2(k_{n_1}) + \sin^2(k_{n_2}) \quad , \quad B = 2 \mu \omega_{n_0} ,
\end{eqnarray}
$\omega_{n_0} = 2 \pi (n_0 - 1/2) / \beta$ and $k_{n_j} = 2 \pi n_j / L$.

In \sref{sec:scale_setting} we also need the second derivative of $S_{\text{eff}}$ with respect to $\bar{\sigma}$ at $\bar{\sigma} = 0$, which can be calculated as
\begin{eqnarray}
\label{eq:ddSeff_sigmasq} & & \hspace{-0.7cm} \frac{\partial^2}{\partial \bar{\sigma}^2} \frac{S_{\text{eff}}[\bar{\sigma}]}{\Nf}\bigg|_{\bar{\sigma} = 0} = \frac{2 \Nt \Ns^2}{\lambda} - \sum_{n_0=1}^{\Nt} \sum_{n_1 = 0}^{\Ns-1} \sum_{n_2 = 0}^{\Ns-1} \frac{A(0)}{A^2(0) + B^2} .
\end{eqnarray} 
Note that for $\sigma = \bar{\sigma} = \textrm{const}$ the effective action corresponds to the term $\Seff^{(0)}$ of the expansion (\ref{eq:Seffexpansion}). Hence, \eqref{eq:ddSeff_sigmasq} is the lattice field-theory version of \eqref{eq:dSdsig2}. Indeed, since $2 \Nt \Ns^2 = \beta V$ in lattice units, this is immediately obvious for the first term on the right-hand side, while the lattice version of the integral (\ref{eq:l1def}) is given by
\begin{eqnarray}
\label{eq:l1lat} \ell_1 \rightarrow \frac{1}{2 \Nt \Ns^2} \sum_{n_0=1}^{\Nt} \sum_{n_1 = 0}^{\Ns-1} \sum_{n_2 = 0}^{\Ns-1} \frac{A(\bar{\sigma})}{A^2(\bar{\sigma}) + B^2}
\end{eqnarray}
in consistency with the second term.
\end{itemize}


\subsection{Stability analysis on the lattice}

In \sref{sec:stabana} we have discussed in detail within the continuum approach, how to perform a stability analysis with respect to inhomogeneous perturbations of a homogeneous condensate minimizing the effective action. The same steps and calculations can also be carried out in the lattice field theory approach.

The important analytical result is an expression for the second-order contributions of the fluctuations to the effective action, when using the lattice Dirac operator (\ref{eq:disc_dirac_op}).
It reads
\begin{eqnarray}
\label{eq:Seff2q_lattice} \Seff^{(2)} = \frac{\beta L^2}{2} \sum_{\mathbf{q}_k} |\delta\tilde{\sigma}_{\mathbf{q}_k}|^2 \Gamma^{-1}(\mathbf{q}_k),
\end{eqnarray}
 where
\begin{eqnarray}
\label{EQN784} \Gamma^{-1}(\mathbf{q}_k) = \Nf \bigg(\frac{1}{\lambda} - \frac{\tilde{W}_2(\mathbf{q}_k)\tilde{W}_2(-\mathbf{q}_k)}{2 \beta L^2} \sum_{n_0=-\Nt+1}^{\Nt} \sum_{n_1=0}^{\Ns-1} \sum_{n_2=0}^{\Ns-1} \ell(n_0,n_1, n_2,\mathbf{q}_k)\bigg)
\end{eqnarray}
with $\mathbf{q}_k = \left(q_{k,1}, q_{k,2}\right)$ and $q_{k,j} = 2 \pi k_j/L$ with $k_j = 0,1,\ldots, \Ns-1$,
and
\begin{eqnarray}
\nonumber & & \hspace{-0.7cm} \ell(n_0,n_1, n_2,\mathbf{q}_k) = \frac{(\omega_{n_0} - \ii \mu)^2 + \sin(k_{n_1}) \sin(k_{n_1}-q_{k,1}) + \sin(k_{n_2}) \sin(k_{n_2}-q_{k,2}) - \bar{\sigma}^2}{((\omega_{n_0} - \ii \mu)^2 + \sin^2(k_{n_1}) + \sin^2(k_{n_2}) + \bar{\sigma}^2)} 
\; \times
\\
\label{eq:ell} & & \hspace{2.5cm} \times
\hspace{0.675cm} \frac{1}{((\omega_{n_0} - \ii \mu)^2 + \sin^2(k_{n_1}-q_{k,1}) + \sin^2(k_{n_2}-q_{k,2}) + \bar{\sigma}^2)}
\end{eqnarray}
with $\omega_{n_0} = 2\pi(n_0 - 1/2)/\beta$ and $k_{n_j} = 2\pi n_j/L$. These equations are the analogs of Eqs.\ (\ref{eq:Seff2q}) and (\ref{eq:Gamma}). As discussed in \sref{sec:stabana}, negative $\Gamma^{-1} (\mathbf{q}_k)$ with $\mathbf{q}_k \neq 0$ indicate inhomogeneous perturbations which decrease $S_{\text{eff}}$.


\subsection{\label{sec:scale_setting}Coupling constant, lattice spacing and scale setting}

As usual in a renormalizable lattice field theory, the lattice spacing $a$ can be set by tuning the coupling constant $\lambda$, i.e., $a = a(\lambda)$.\footnote{This equation is analogous to \eqref{eq:lambdaren}, which relates $\lambda$ and $\Lambda$.} Moreover, in our particular regularization the number of modes in temporal direction $\Nt$ is proportional to the inverse temperature $\beta$, i.e., $\Nt = \beta / 2 a$, as discussed in \sref{SEC672}. Thus, the temperature can be adjusted by either changing $\Nt$ or $\lambda$.

\begin{figure}[htb]
\begin{center}
\includegraphics[height=5.0cm]{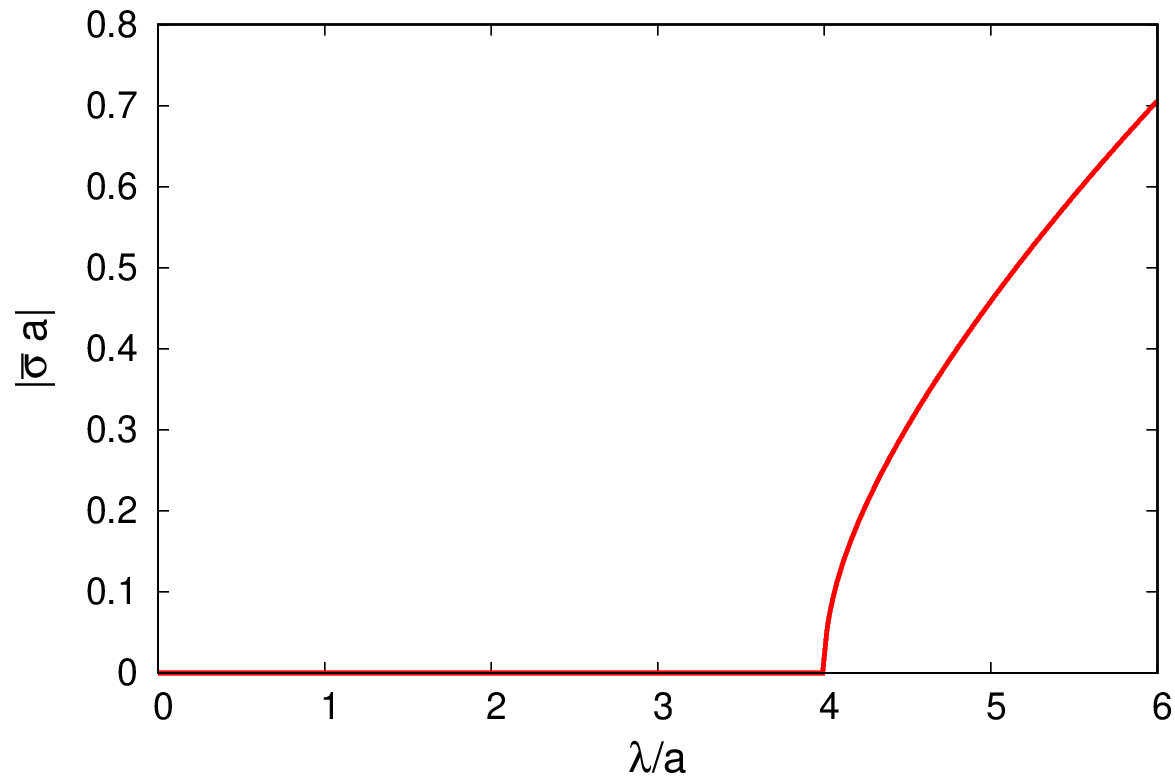}	
\end{center}
\caption{\label{sigma_lambda.png}$|\bar{\sigma}|$ as a function of $\lambda$ for $\Nt = 5$ and $\Ns = 100$. $N_{0,c} = 5$ corresponds to $\lambda = 3.998$, where $|\bar{\sigma}|$ starts to deviate from $0$.}
\end{figure}

For our computations we first fix the number of modes at the critical temperature, denoted by $N_{0,c}$, where we typically use a small number, $2 \leq N_{0,c} \leq 5$ (throughout this section $\Ns = L$ is chosen sufficiently large, such that finite volume corrections are essentially negligible). This in turn fixes the coupling constant $\lambda$ and the lattice spacing $a$, where the former has to be tuned in such a way that $2 N_{0,c} a(\lambda) = \beta_c$. An obvious possibility is to determine $\lambda$ such that $\bar{\sigma}(\lambda-\epsilon) = 0$ and $\bar{\sigma}(\lambda+\epsilon) \neq 0$ for infinitesimal $\epsilon$ (see \fref{sigma_lambda.png}, where $|\bar{\sigma}|$ is plotted as a function of $\lambda$).\footnote{This is based on Refs.\ \cite{Klimenko:1987gi,Rosenstein:1988dj,Urlichs:2007zz}, where it was found that at $\mu = 0$ there is a homogeneous symmetry-broken phase with $\sigma = \bar{\sigma} \neq 0$ for $T < T_c$ and a symmetric phase with $\sigma = \bar{\sigma} = 0$ for $T > T_c$ connected by a phase transition of second order.} Mathematically equivalent, but more practical from a numerical point of view is to consider $(\partial^2/\partial \bar{\sigma}^2) S_\textrm{eff}[\bar{\sigma}] / \Nf|_{\bar{\sigma} = 0}$ (see \eqref{eq:ddSeff_sigmasq}) as a function of $\lambda$ and to determine its root, which leads to
\begin{eqnarray}
\label{eq:lambdarenlat} \lambda = \bigg(\frac{1}{2 N_{0,c} \Ns^2} \sum_{n_0=1}^{N_{0,c}} \sum_{n_1 = 0}^{\Ns-1} \sum_{n_2 = 0}^{\Ns-1} \frac{A(0)}{A^2(0) + B^2}\bigg)^{-1}. 
\end{eqnarray} 
Even though $\lambda$ is now fixed, the temperature can still be changed in discrete steps by increasing or decreasing the number of modes, since $T = 1 / 2 \Nt$.

\eqref{eq:lambdarenlat} is similar to the renormalization condition (\ref{eq:lambdaren}) used in our continuum approach, as can be seen by inserting the correspondence given in \eqref{eq:l1lat}. In \eqref{eq:lambdaren} the integral $\ell_1$ is evaluated in the vacuum at $T = \mu = 0$ for $\bar{\sigma} = \sigma_0$, while in \eqref{eq:lambdarenlat} it is evaluated at $T = T_c$ and $\mu = 0$ for $\bar{\sigma} = 0$. However, both conditions correspond to the non-trivial solution of the gap equation (\ref{eq:gap}), which is equal to $\sigma_0$ in vacuum and goes to zero at $T_c$.

\begin{figure}[htb]
\includegraphics[height=5.0cm]{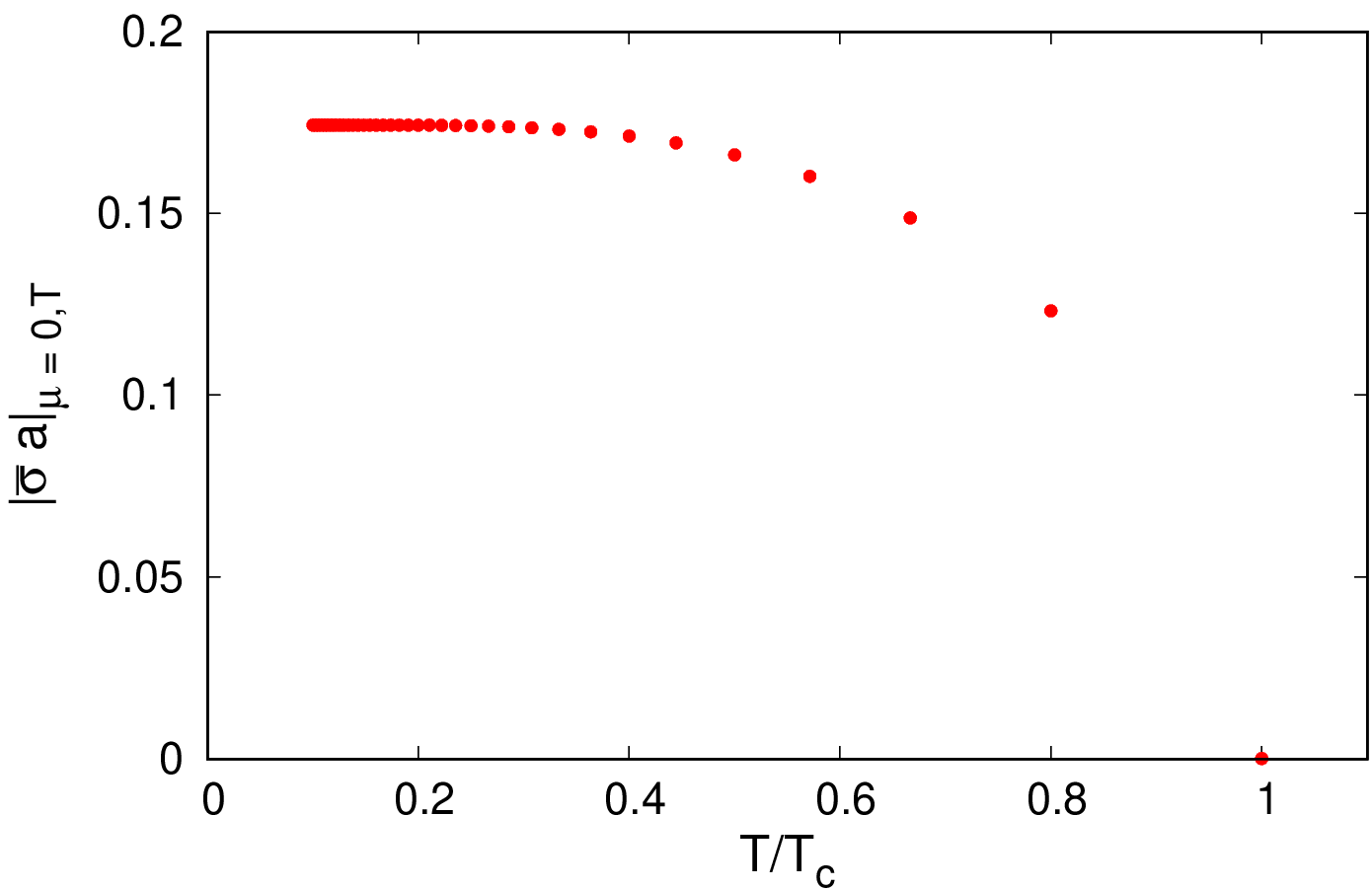}
\includegraphics[height=5.0cm]{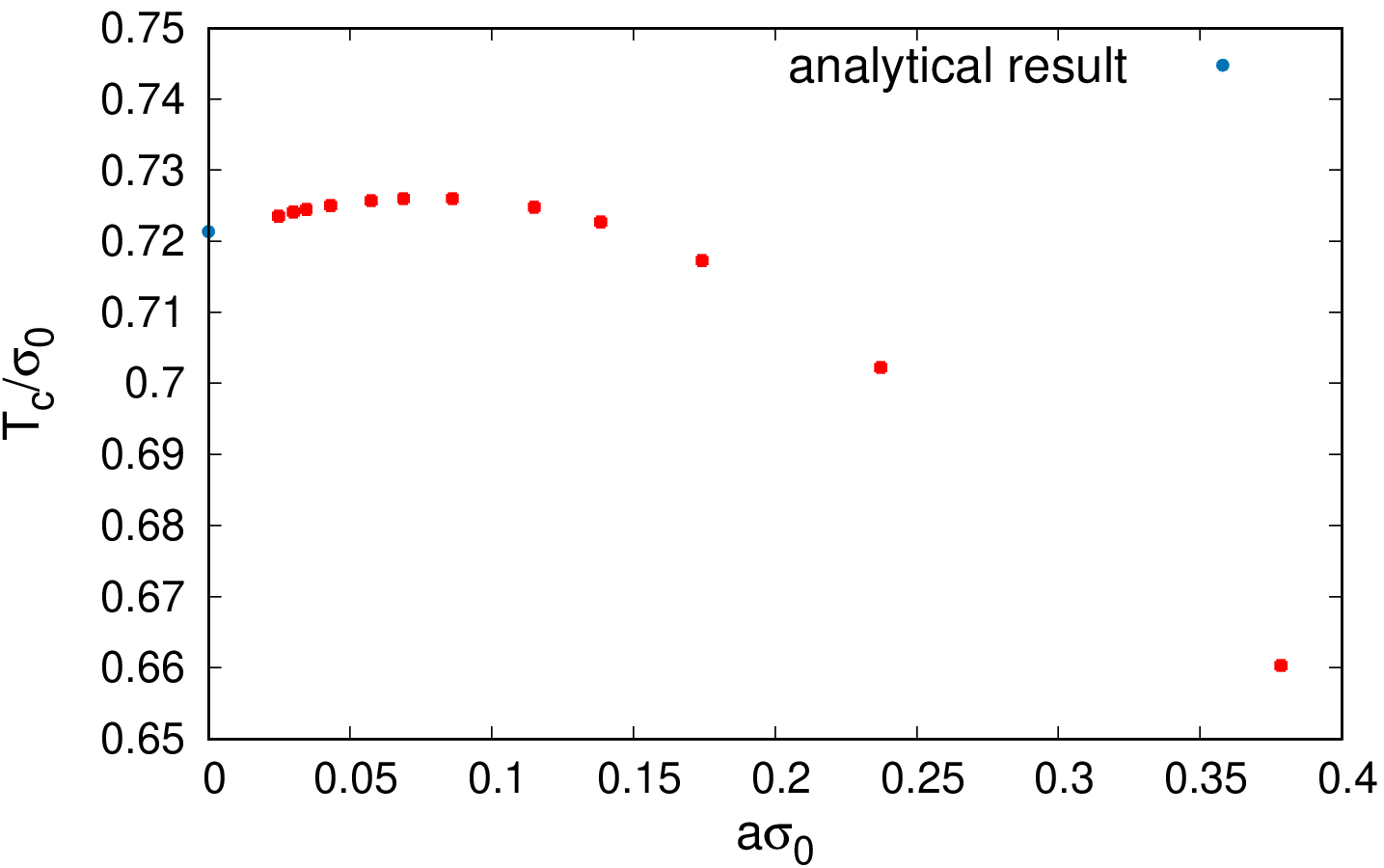}	
\caption{\label{sigma_1_2Nt.png}\textbf{(left)}~$|\bar{\sigma}|_{\mu=0,T}$ as a function of $T / T_c$ for $N_{0,c} = 4$ and $\Ns = 80$. \textbf{(right)}~$T_c / \sigma_0$ as a function of the lattice spacing $a$ (the data points correspond to $2 \leq N_{0,c} \leq 28$). For small values of $a$, $T_c / \sigma_0$ approaches the analytical result $1 / 2 \ln(2) \approx 0.721\ldots$  \cite{Klimenko:1987gi}.}
\end{figure}

In principle, the scale could now be set via the critical temperature $T_c = 1 / 2 N_{0,c}$, i.e., we could express all dimensionful quantities in units of $T_c$. However, we prefer to set the scale via $\sigma_0$, which is common in the existing literature. To determine $\sigma_0$, we compute $|\bar{\sigma}|_{\mu=0,T}$ for several small $T$ by minimizing $S_\text{eff}$ with respect to $\bar{\sigma}$. Using \eqref{eq:Seff_hom_series} this is numerically rather simple and can by done by a standard golden section search. $|\bar{\sigma}|_{\mu=0,T}$ quickly approaches a plateau, when decreasing $T$, i.e., $|\bar{\sigma}|_{\mu=0,T}$ is almost constant for $T \ltapprox T_c/4$, and the plateau value is identical to $\sigma_0$ (see \fref{sigma_1_2Nt.png}, left plot).

In the following we express all dimensionful quantities in units of $\sigma_0$. For example for $T_c / \sigma_0$ we obtain values rather close to the analytically known result $1 / 2 \ln(2)$ \cite{Klimenko:1987gi} also at finite lattice spacing. When increasing $N_{0,c}$, which amounts to decreasing $\lambda$ as well as decreasing $a$, and approaching the continuum limit, our results for $T_c / \sigma_0$ approach $1 / 2 \ln(2) \approx 0.721\ldots$, as can be seen in the right plot of \fref{sigma_1_2Nt.png}. Note, however, that in contrast to our continuum approach, where this limit is reached from below when the cutoff $\Lambda$ is sent to infinity (see \eqref{eq:Tclin}), here the data points first overshoot the limiting value and then approach the latter from above.



\section{Results\label{sec:results}}

After having introduced our different approaches,
we finally turn to the discussion of our results for the phase diagram.
From Refs.\ \cite{Klimenko:1987gi,Rosenstein:1988dj,Urlichs:2007zz,Winstel:2019zfn,Narayanan:2020uqt} 
as well as from our analytical studies in \sref{sec:anapd}
we expect up to three phases, each characterized by a different behavior of the field $\sigma$:
\begin{itemize}
	\item A symmetric phase with $\sigma = 0$ at large $\mu$ and/or large $T$.
	
	\item A homogeneous symmetry-broken phase with a spatially constant (but $\mu$ and $T$ dependent) field
	$\sigma = \bar{\sigma}  \neq 0$ at small $\mu$ and small $T$.
	
	\item Possibly an inhomogeneous phase, where $\sigma(\textbf{x})$ is a varying function of the spatial coordinates, at intermediate $\mu$ and small $T$ \cite{Winstel:2019zfn}. This phase might only be present at a finite value of the regulator (e.g., Pauli-Villars cutoff $\Lambda$ or lattice spacing $a$), as indicated by recent lattice field theory results reported in Ref.\ \cite{Narayanan:2020uqt} and the behavior of the LP found in \sref{sec:anapd}.
\end{itemize}

In the following we present and compare results, which we computed using our three regularizations, the continuum approach with Pauli-Villars cutoff $\Lambda$ discussed in \sref{SEC499} and the two lattice discretizations with lattice spacing $a$ discussed in \sref{SEC500}. 
The advantage of the lattice approach compared to the continuum approach is that one can 
perform minimizations of $S_\text{eff}$ with respect to $\sigma(\textbf{x})$,
allowing for arbitrary modulations of the condensate.
There are, however, also drawbacks, namely computations are typically expensive and have to be carried out at finite spatial volume $V$ and finite lattice spacing $a$, while in the continuum approach both $V$ and the Pauli-Villars cutoff $\Lambda$ can be sent to infinity. Thus, both approaches complement each other.

Our main goal is to explore and to understand the dependence of a possibly existent inhomogeneous phase in the $2+1$-dimensional GN model on the regulators $\Lambda$ and $a$. 
In particular we expect to get insights concerning the $a$ dependence of 
the lattice field theory results by analogy to the $\Lambda$ dependence of the continuum 
results, for which we have derived analytical expressions in 
\sref{sec:anapd}.
For a crude quantitative comparison of continuum and lattice results we relate the corresponding regulators via
\begin{eqnarray}
	\label{EQN698} \Lambda = \frac{\sqrt{\pi}}{a} .
\end{eqnarray}
This equation can be obtained by equating the considered regions of spatial momenta, which are $\pi \Lambda^2$ (a circle in momentum space in the continuum approach; see \sref{sec:regren}) and $(\pi/a)^2$ (a square in momentum space in the lattice approach; see appendix~A in Ref.\ \cite{Lenz:2020bxk}).

\begin{table}[htb]
    \centering
    \renewcommand{\arraystretch}{0.8}
\begin{tabular}{rc|rc}
        \toprule
\multicolumn{2}{c|}{lattice field theory} & \multicolumn{2}{c}{continuum approach} \\
        \midrule
        $a\sigma_0$ & $\lambda \sigma_0$ &
$\Lambda/\sigma_0$ & $\lambda \sigma_0$ \\
        \midrule
        $0.379$ & $1.78$ & 4.68 & 1.66\\
        $0.237$ & $1.01$ & 7.48 & 0.96\\
        $0.115$ & $0.45$ & 15.41 & 0.43\\
        $0.086$ & $0.33$ & 20.61 & 0.32\\
        $0.049$ & $0.19$ & 36.17 & 0.18\\
        \bottomrule
    \end{tabular}
    \caption{Lattice spacings $a$ used in the numerical calculations, and     
    the corresponding values of the Pauli-Villars cutoff $\Lambda$ according to \eqref{EQN698}.
    We also list the corresponding coupling constants $\lambda$ for both the lattice field theory and the continuum approach (see \eqref{eq:lambdarenlat} and \eqref{eq:lambdareg}).
    All quantities are made dimensionless with the help of the vacuum value $\sigma_0$ of the scalar field $\sigma$.
    \label{tab:a_lambda}
    }
\end{table}

The lattice spacings used in the numerical calculations presented below,
together with the corresponding values of the Pauli-Villars cutoffs, are listed in \tref{tab:a_lambda}.
We also list the  corresponding coupling constants $\lambda$ for both cases. 
Despite the rather different approaches, the values of $\lambda\sigma_0$ turn out to be quite similar,
which may be considered as an additional justification of \eqref{EQN698}.


\subsection{\label{sec:hom_boundary}Phase diagram for homogeneous condensate {\boldmath $\sigma(\textbf{x}) = \bar\sigma = \textrm{const}$}}

\begin{figure}[htb]
	\includegraphics[width=7.8cm]{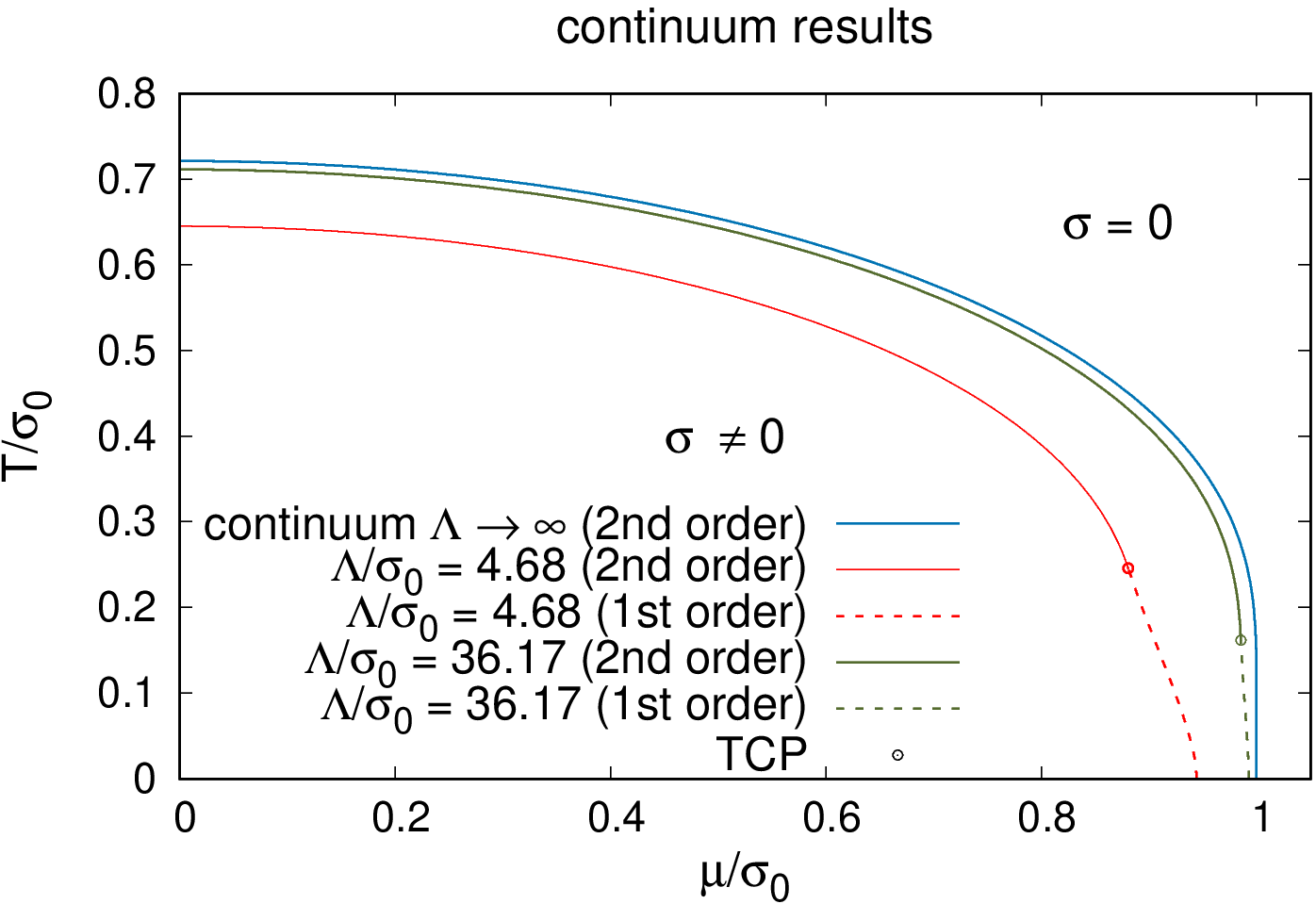}	
	\hfill
	\includegraphics[width=7.8cm]{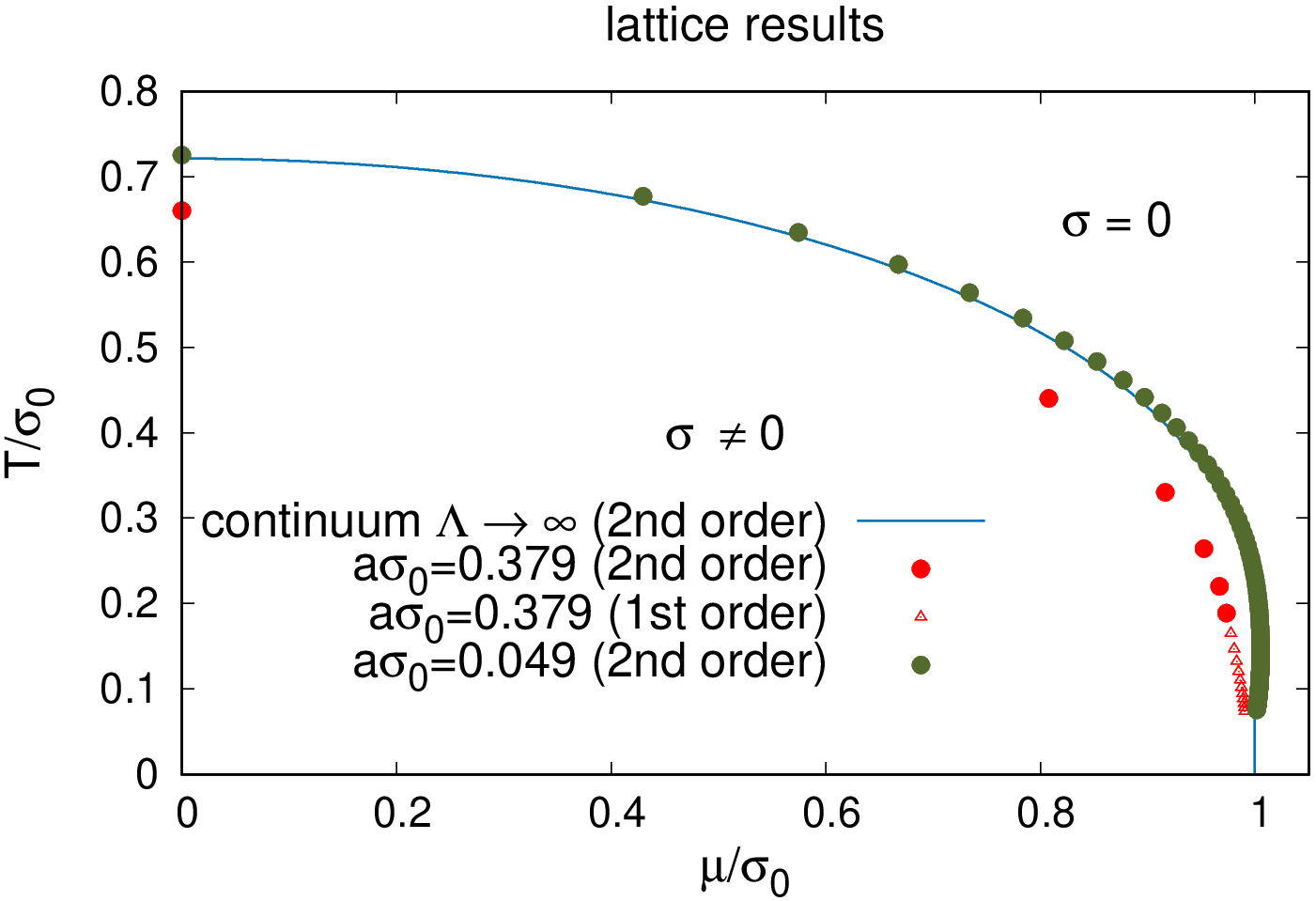}	
	\caption{\label{fig:hom_pd}Phase diagram of the $2+1$-dimensional GN model in the $\mu$-$T$ plane for $\sigma(\textbf{x}) = \bar\sigma = \textrm{const}$. The blue lines shown in both plots represent the phase boundary from Ref.\ \cite{Klimenko:1987gi}, which we also obtain in our continuum approach for $\Lambda \rightarrow \infty$.
		\textbf{(left)} Continuum results at finite Pauli-Villars cutoffs, $\Lambda/\sigma_0 = 4.68$ (red) and $\Lambda/\sigma_0 = 36.17$ (green). Second-order (first-order) phase boundaries are indicated by solid (dashed) lines. The open circles mark the 
		locations of the TCPs.
		\textbf{(right)} Lattice results at finite lattice spacings, $a \sigma_0 = 0.379$ (red) and $a \sigma_0 = 0.049$ (green), and finite spatial volume, $(L T_c)^2 = 10^2$. Full circles (open triangles) indicate second-order (first-order) phase transitions. 
	}
\end{figure}

In this subsection we only allow spatially constant condensates, i.e., $\sigma(\textbf{x}) = \bar\sigma = \textrm{const}$. 
The resulting homogeneous phase diagrams are presented in \fref{fig:hom_pd}.
For $\Lambda \rightarrow \infty$ the phase boundary obtained in the continuum approach is of second order (with the exception of the endpoint at $T = 0$, where it is of first order) and given by \eqref{eq:muc} with $s \rightarrow \sigma_0$. 
It is identical to the known result of Ref.\ \cite{Klimenko:1987gi} and shown as solid blue line in both plots of \fref{fig:hom_pd}.

The two lattice discretizations with $W_2'$ and $W_2''$ become identical for homogeneous condensates, 
as discussed in \sref{SEC588}. In the right plot of \fref{fig:hom_pd} we show results for two different lattice spacings, $a \sigma_0 = 0.379$ and $a \sigma_0 = 0.049$, but identical spatial volume (in units of the critical temperature), $(L T_c)^2 = (N_s /2 N_{0,c})^2 = 10^2$. The phase boundary obtained with the finer lattice spacing is much closer to the result from Ref.\ \cite{Klimenko:1987gi} and a combined continuum and infinite-volume extrapolation indicates consistency with that result. It is interesting to note that at the coarse lattice spacing $a \sigma_0 = 0.379$ the phase transition is of second order for $T/\sigma_0 \geq 0.189$, while it is of first order for $T/\sigma_0 \leq 0.165$. At the fine lattice spacing $a \sigma_0 = 0.049$ we only observe second-order phase transitions, but we expect that somewhere below $T/\sigma_0 = 0.075$ (the smallest temperature we have investigated) these transitions change to first order.

Our continuum results, which are displayed in the left panel of \fref{fig:hom_pd},
exhibit a qualitatively similar behavior. 
We use the finite Pauli-Villars cutoffs $\Lambda/\sigma_0 = 4.68$ and 
$\Lambda/\sigma_0 = 36.17$,
which, according to the matching  formula (\ref{EQN698}), correspond to the two lattice spacings used in the right panel of the figure (see also \tref{tab:a_lambda}).
As discussed in  \sref{sec:anapd} and illustrated in \fref{fig:LP}, at finite $\Lambda$ there always exists a TCP, separating
first- and second-order phase boundaries.  In \fref{fig:hom_pd} these points are marked by open circles. 
For  $\Lambda/\sigma_0 = 4.68$ the TCP is located at $\mu/\sigma_0 = 0.881$ and $T/\sigma_0= 0.245$, 
which is somewhat lower in chemical potential and higher in temperature than observed 
in the lattice approach for $a \sigma_0 = 0.379$.
A similar tendency is seen at $\Lambda/\sigma_0 = 36.17$, where the TCP is found at 
$\mu/\sigma_0 = 0.985$ and $T/\sigma_0= 0.166$ in the continuum approach,
while no first-order phase transition at temperatures above  $T/\sigma_0 = 0.075$ was found in the corresponding lattice calculation.
We emphasize, however, that  \eqref{EQN698} is only a crude prescription to compare two very different approaches and there is no reason to expect perfect agreement. 
In general we find that the lattice points at low $T$ converge faster to the $\Lambda \rightarrow \infty$ line than the 
corresponding  continuum results, while the second-order phase boundaries at low $\mu$ agree fairly well, even quantitatively. 


\subsection{\label{sec:inhom_boundary}Instabilities with respect to spatially inhomogeneous perturbations}

We now relax the restriction to a homogeneous field $\sigma$ and investigate the possible appearance of inhomogeneous phases
in the phase diagram.
To do this in a rigorous way, one has to consider arbitrary modulations of the condensate and minimize the effective action $\Seff$ with respect to $\sigma(\textbf{x})$. 
In the lattice approach this is possible, but numerically a very hard problem. It amounts to minimizing $\Seff$ in \eqref{eq:eff_action_discr}, which is a function in $\Ns^2$ variables $\sigma(\mathbf{x})$. Finding the global minimum of a function in a large number of variables (in our case $\Ns^2 = \mathcal{O}(10^2) \ldots \mathcal{O}(100^2)$) is an extremely challenging task.
Therefore, in a first step, we check, whether the constant $\sigma = \bar{\sigma}$ determined in \sref{sec:hom_boundary} is stable with respect to spatially inhomogeneous perturbations $\delta \sigma$.
We do this both in the continuum approach introduced in \sref{SEC499} as well as in the lattice approach of 
\sref{SEC500}. 
Steps towards a  rigorous minimization of the effective action with respect to arbitrarily varying fields will be discussed afterwards,
in \sref{sec:fullmini}.

The starting point for the 
stability analysis is $\Seff^{(2)}$ at $\sigma = \bar{\sigma}$, which is given by \eqref{eq:Seff2q} in the continuum approach and by \eqref{eq:Seff2q_lattice} in the lattice approach. In both cases the perturbation $\delta \sigma$ is expressed as a sum of plane waves. This leads to a particularly simple form for $\Seff^{(2)}$, where an instability with respect to inhomogeneous perturbations is indicated by a negative value of $\Gamma^{-1}$ for at least one momentum $\mathbf{q} \neq 0$. 
(For $\mathbf{q} = 0$ we have $\Gamma^{-1} \geq 0$ because $\sigma = \bar{\sigma}$ minimizes $\Seff$ when restricting $\sigma$ to a constant.) 
Thus, in order to determine numerically whether there is an instability at given $\mu$ and $T$, one just has to find the minimum of $\Gamma^{-1}$. This is rather straightforward, because $\Gamma^{-1}$ is a function with only a single argument $\mathbf{q}^2$ in the continuum approach and a finite set of numbers, each corresponding to one of the discrete lattice momenta $\mathbf{q}_k$, in the lattice field theory approach. Instability lines separating regions of stability from regions of instability can then be computed by a simple bisection in $\mu$ direction.

\begin{figure}[p]
	\centering   
	\includegraphics[width=0.49\textwidth]{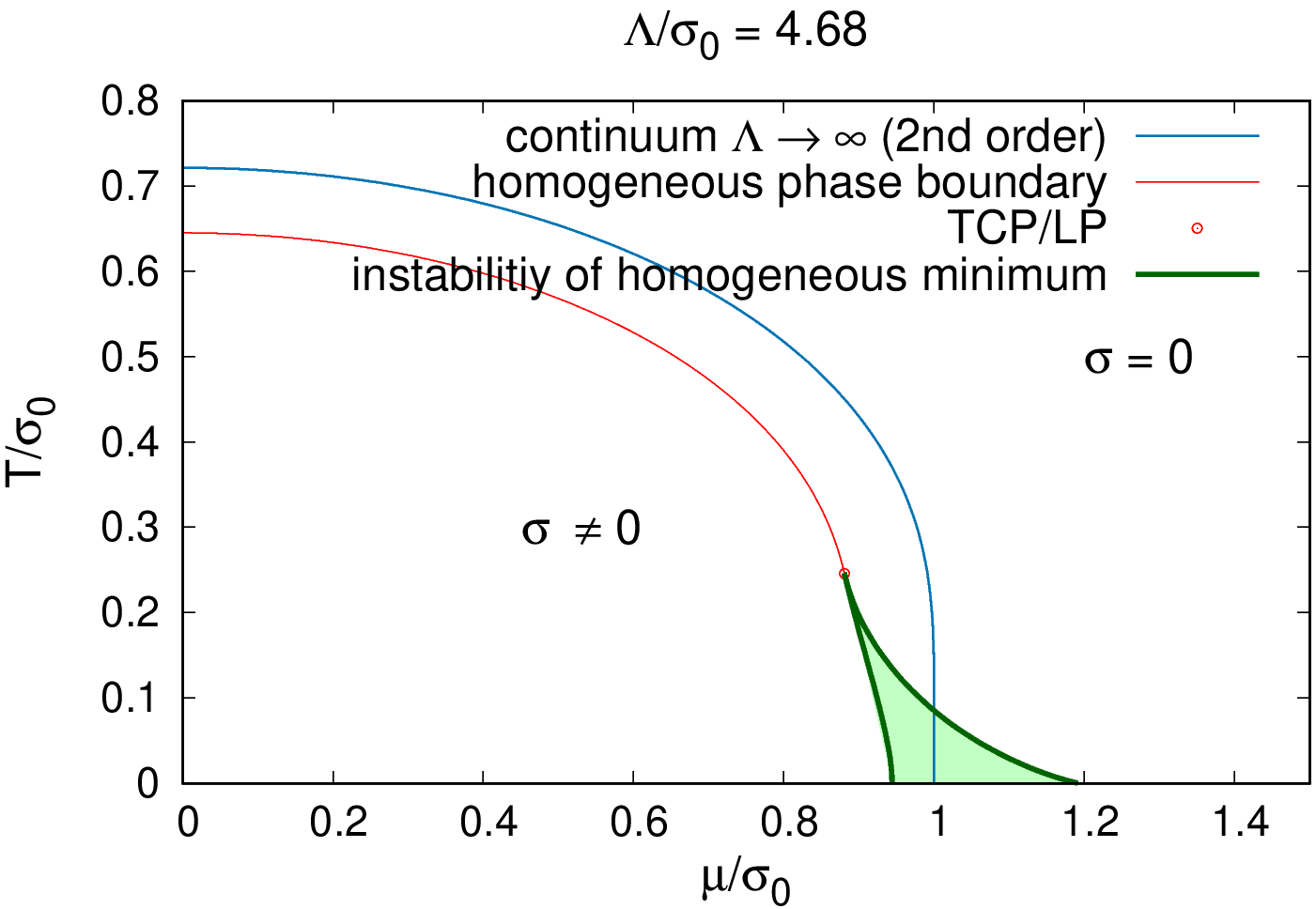}
	\hfil
	\includegraphics[width=0.49\textwidth]{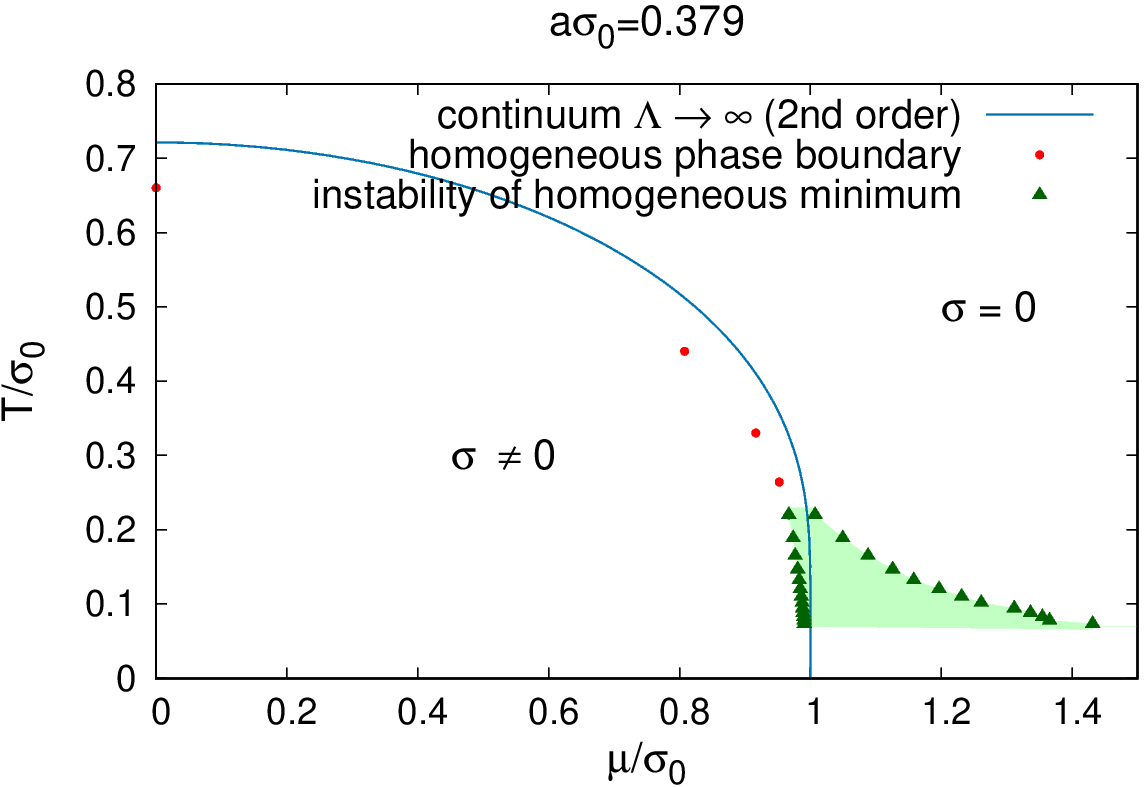} \\
	\includegraphics[width=0.49\textwidth]{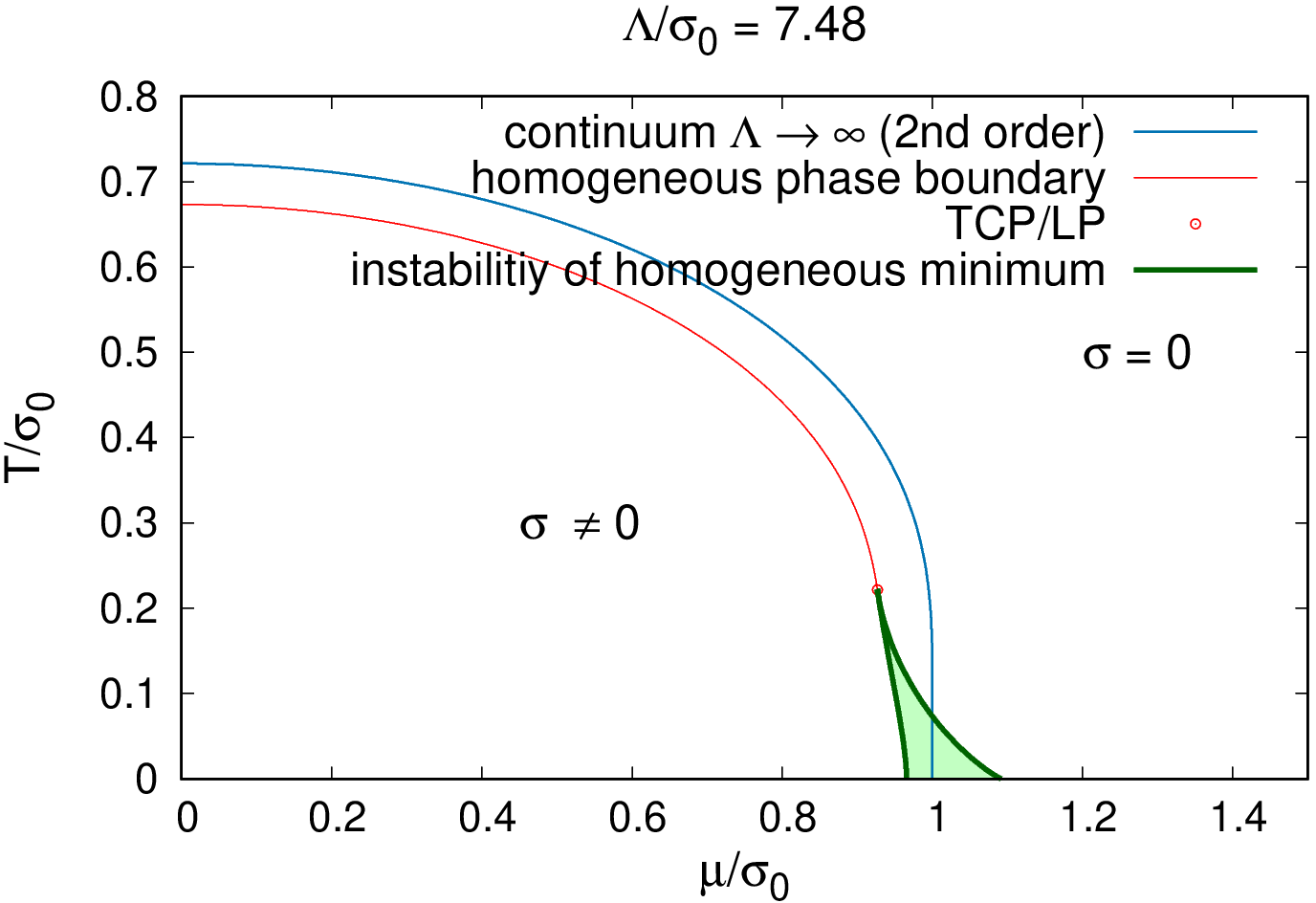}
	\hfil
	\includegraphics[width=0.49\textwidth]{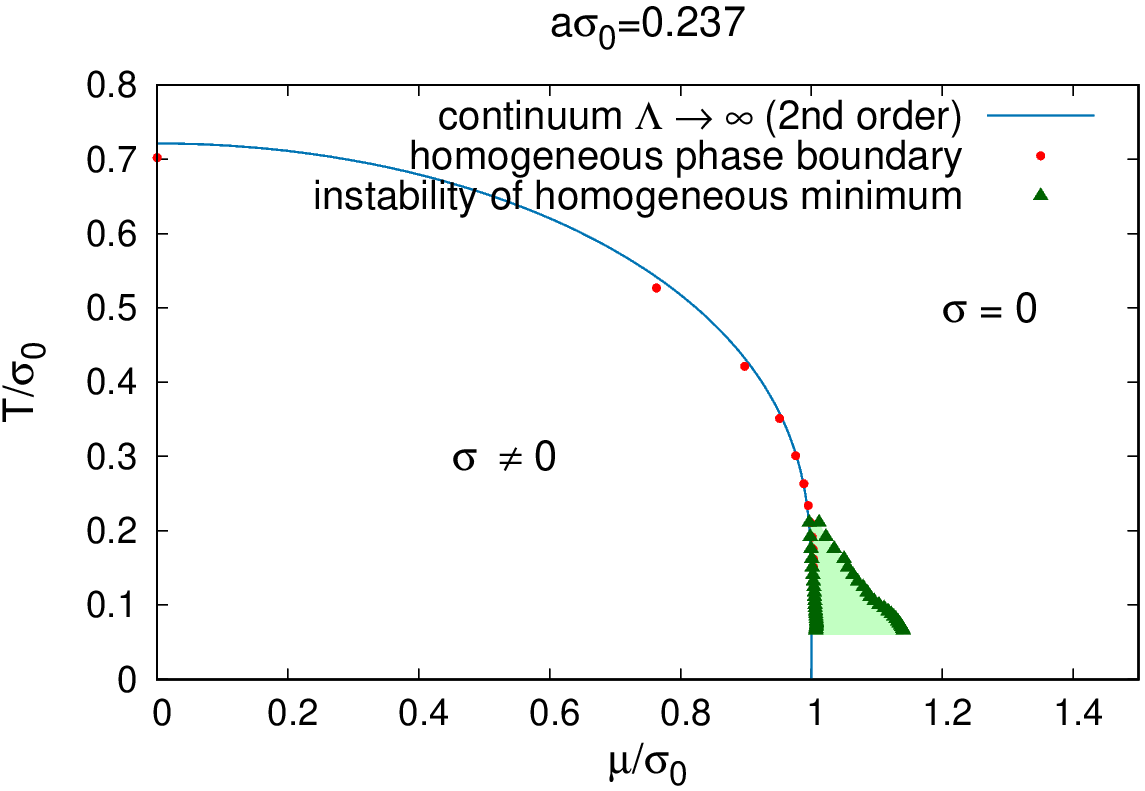} \\
	\includegraphics[width=0.49\textwidth]{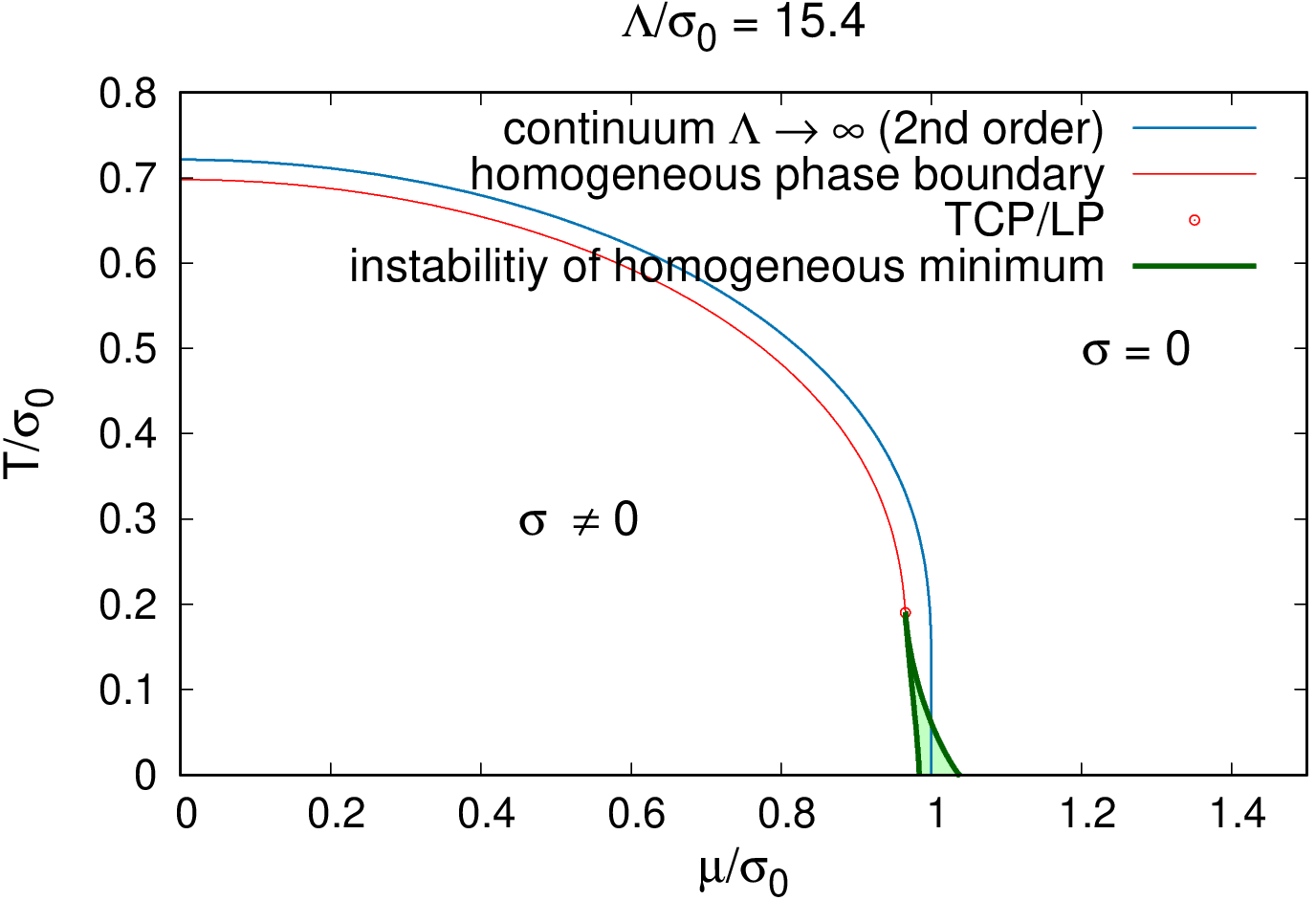}
	\hfil
	\includegraphics[width=0.49\textwidth]{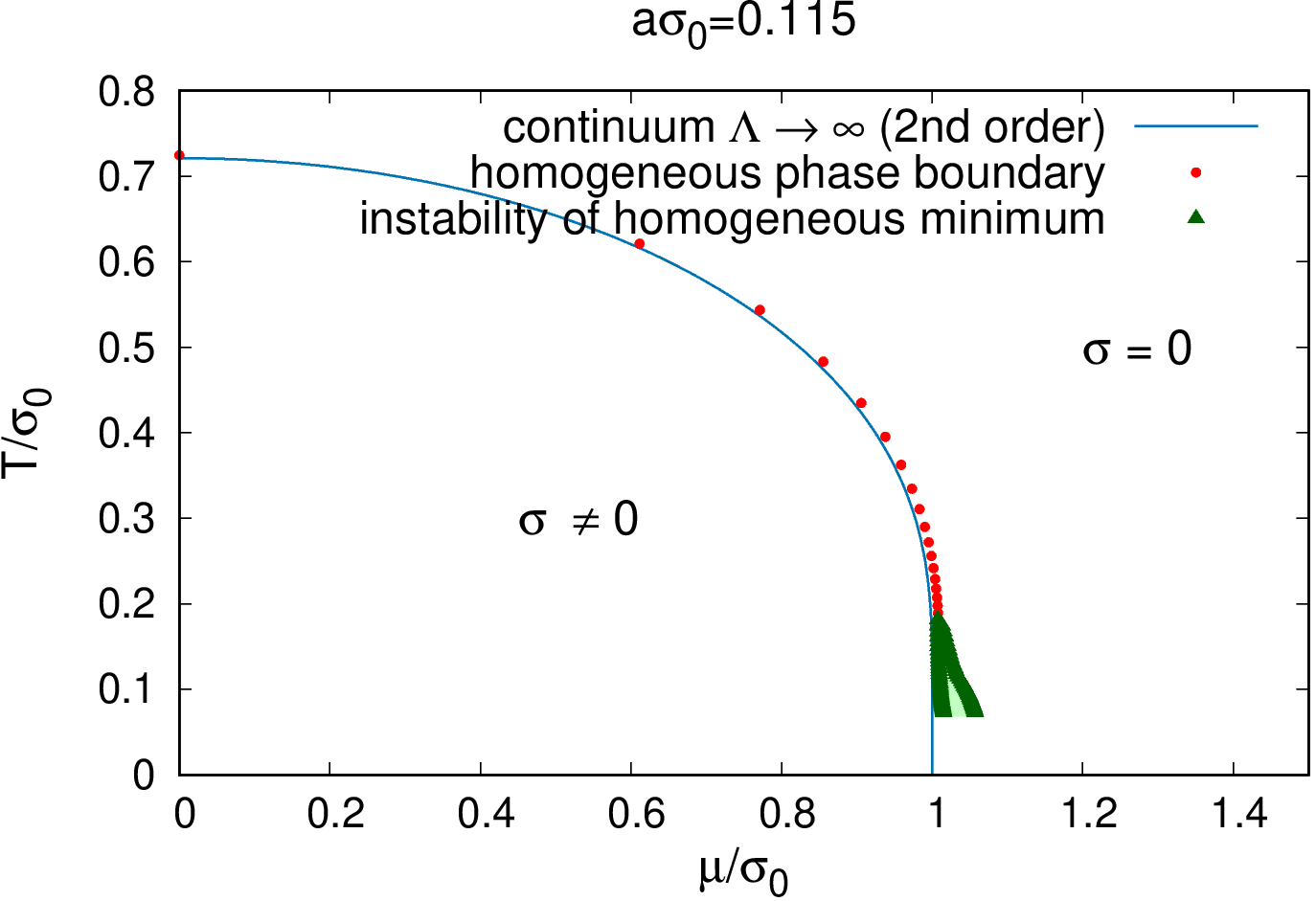}
	\caption{\label{fig:stabanaly_for_large_a_zoom}``Phase diagram'' of the $2+1$-dimensional GN model in the $\mu$-$T$ plane       obtained by stability analyses with respect to spatially inhomogeneous perturbations.  
	Lattice field theory results obtained with $W''_2$ for three different lattice spacings $a$  are shown in the right column, continuum results for the corresponding Pauli-Villars cutoff  $\Lambda = \sqrt{\pi} / a$ in the left column. 
	The blue line appearing in all six plots represents the known phase boundary from Ref.\ \cite{Klimenko:1987gi}, which we also obtain in our continuum approach for $\Lambda \rightarrow \infty$. Regions of instability, which are part of and may coincide with inhomogeneous phases, are shaded in green.}
\end{figure}

In \fref{fig:stabanaly_for_large_a_zoom} we show results of such analyses obtained in 
lattice field theory with the discretization  $W_2= W''_2$ (see \eqref{eq:corr_disc_2}) for three different lattice spacings 
as well as in the continuum approach with the corresponding values of the Pauli-Villars cutoff $\Lambda$.
Regions where the homogeneous ground state is unstable against inhomogeneous fluctuations and which are therefore parts of an inhomogeneous phase are shaded in green.  

In the continuum approach (left column of \fref{fig:stabanaly_for_large_a_zoom})
these instability regions are found in a temperature regime from $T=0$ up to the LP,
which, as shown in \sref{sec:conthomo},
coincides with the TCP of the corresponding homogeneous phase diagram.
Accordingly, as already expected from the behavior of the LP shown in \fref{fig:LP}, the instability region shrinks when $\Lambda$
is increased, and disappears completely for $\Lambda \rightarrow \infty$. 
At finite $\Lambda$ we find that instabilities only occur in the symmetric phase of the homogeneous phase diagrams,
while the homogeneous symmetry-broken phase remains stable against small inhomogeneous fluctuations. In particular the
``left'' phase boundaries of the instability region coincide with the first-order phase boundaries of the homogeneous
phase diagrams. It is quite likely, however, that the true inhomogeneous phase (at finite $\Lambda$)
reaches to somewhat lower values of $\mu$, as known, e.g., from the GN model in $1+1$ dimensions~\cite{Thies:2003kk,Schnetz:2004vr,Thies:2006ti} or the NJL model in $3+1$ dimensions~\cite{Nakano:2004cd,Nickel:2009wj}.

\begin{figure}[htb]
	\includegraphics[width=0.49\textwidth]{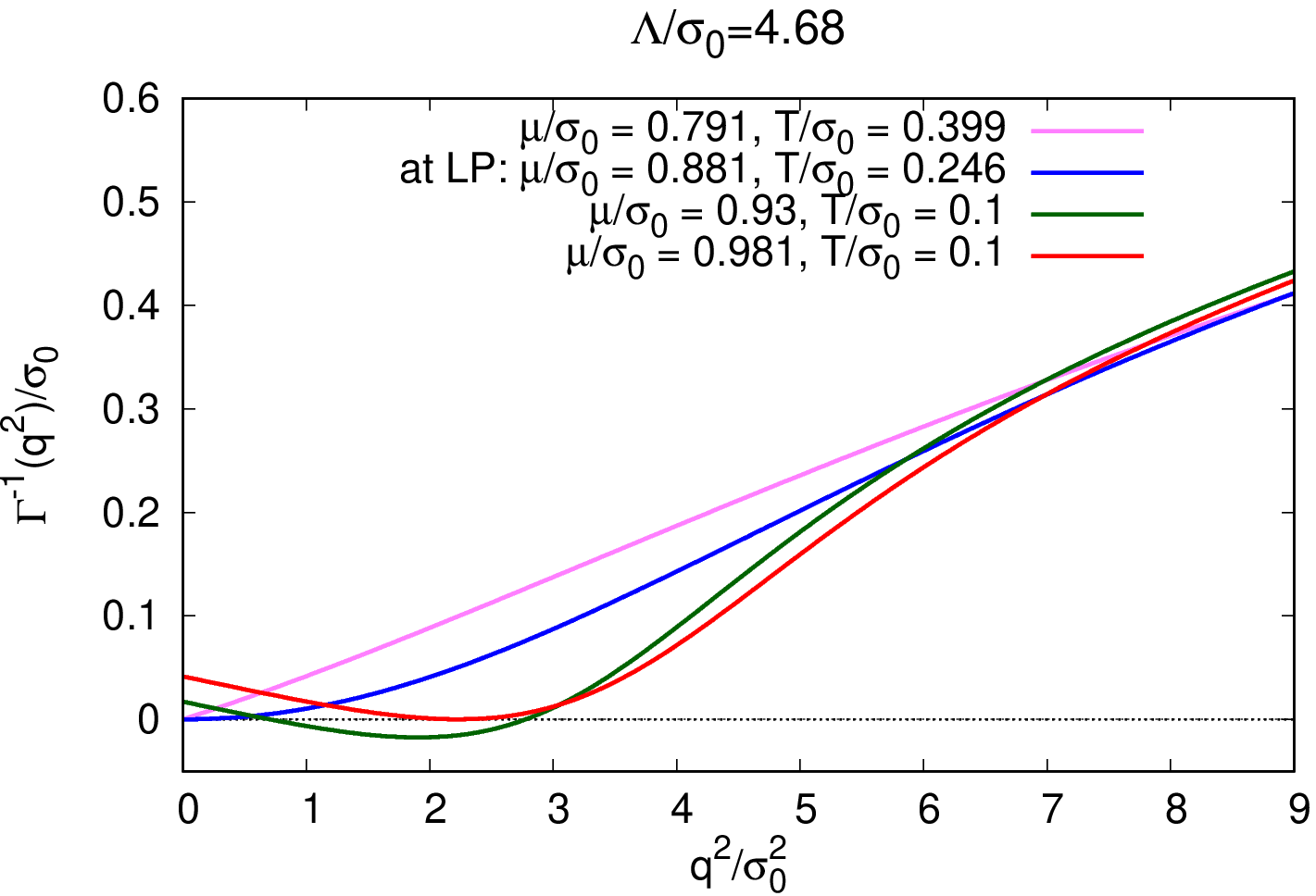}	
	\hfill
	\includegraphics[width=0.49\textwidth]{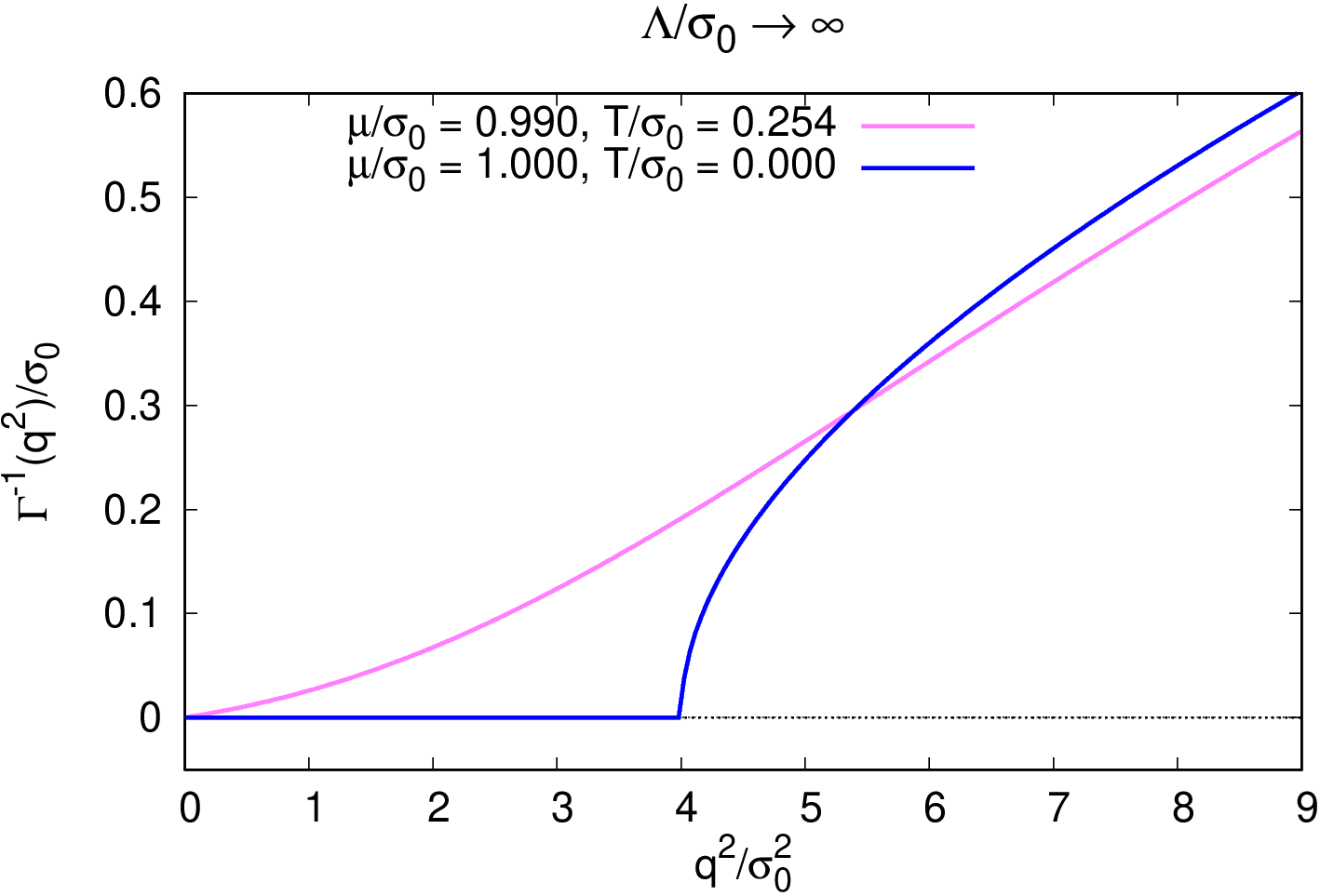}	
	\caption{\label{fig:Gamma} $\Gamma^{-1}$ as a function of the squared spatial momentum in the continuum approach
	for selected values of $\mu$ and $T$ (see also the upper left panel of \fref{fig:stabanaly_for_large_a_zoom}). 
	\textbf{(left)} $\Lambda/\sigma_0 = 4.68$. 
	\textbf{(right)} $\Lambda \rightarrow \infty$. 
	}
\end{figure}

For further illustration of the continuum-approach results of \fref{fig:stabanaly_for_large_a_zoom}
we show in \fref{fig:Gamma} selected examples of the function $\Gamma^{-1}$
for $\Lambda/\sigma_0 = 4.68$ (left plot) and $\Lambda \rightarrow \infty$ (right plot).
The green curve in the left plot corresponds to a point inside the instability region, as obvious from the fact that
$\Gamma^{-1}$ is negative in some momentum interval.
The red curve, on the other hand, just touches the $\Gamma^{-1} = 0$ axis at some non-zero $\mathbf{q}^2$
and therefore corresponds to a  point on the boundary between the stable and the unstable region with respect
to inhomogeneous fluctuations.
The blue curve has a similar minimum as the red curve, however at $\mathbf{q}^2 = 0$, thus indicating a LP (cf.\ \eqref{eq:LP}).
Finally, the magenta curve corresponds to a point on the second-order boundary between the homogeneous symmetry-broken 
and the symmetric phase, where $\Gamma^{-1}(0)=0$ but with a non-vanishing derivative with respect to $\mathbf{q}^2$.

The magenta curve in the right plot of \fref{fig:Gamma} is qualitatively similar and corresponds to a point on the 
second-order boundary between the homogeneous symmetry-broken and the symmetric phase in the renormalized model, i.e., for
$\Lambda \rightarrow \infty$. 
As pointed out above, for $\Lambda \rightarrow \infty$ there is no instability with respect to inhomogeneous fluctuations
at any $T>0$, and therefore the blue curve is characteristic for the entire phase boundary. 
An exceptional case is the situation at $T=0$ and $\mu = \sigma_0$, indicated by the blue curve. 
Here $\Gamma^{-1}(\mathbf{q}^2)=0$ in the whole interval $0 \leq \mathbf{q}^2 \leq 4\sigma_0^2$.
This point may thus be interpreted as a point on the instability boundary, 
but instead of having a single unstable mode, the instability is driven by any momenta $0 \leq |\mathbf{q}| \leq 2\mu$
or superposition of them.
This behavior, which clearly deserves further investigation, could be related to an observation 
reported in Ref.\ \cite{Urlichs:2007zz},
where it was found that at $T=0$ a specific inhomogeneous modulation based on Jacobi elliptic functions 
is degenerate with the favored homogeneous solutions.

Going back to  \fref{fig:stabanaly_for_large_a_zoom}, we now turn to the diagrams shown in the right column, 
which have been obtained within the lattice field theory approach, using the discretization with $W_2 = W''_2$.
As one can see by comparison with the left panels, the results are qualitatively similar to those from the continuum approach.
Again, the regions of instability are only found in the symmetric phase of the homogeneous phase diagram, and they shrink when the lattice spacing is decreased. 
We therefore expect that in the continuum limit $a \rightarrow 0$ the instability regions vanish (at least for $T>0$), 
as they do for $\Lambda \rightarrow \infty$ in the continuum approach.

When carefully inspecting the instability line towards the symmetric phase, one can identify small deviations from a smooth behavior. This is a finite-volume effect, similar to that observed in numerical studies of the $1+1$-dimensional GN model \cite{deForcrand:2006zz,Wagner:2007he}. The reason is that close to the phase boundary only non-vanishing momenta inside a small interval would lead to negative $\Gamma^{-1}$, but none of the numerically considered momenta, which are quantized by the finite spatial volume, is inside that interval. We also note that the LP has a higher temperature than the TCP, i.e., they do not coincide as it is the case in the continuum approach. 
We attribute this to the lattice discretization, which invalidates some of the steps we have performed in \sref{sec:conthomo} to
show the coincidence of the two points in the continuum approach.  
In particular the momentum derivative in \eqref{eq:LP}
is not well defined on the lattice, where the allowed momenta are discrete. 
Interestingly, the positions of the LP for the different lattice spacings agree fairly well with the corresponding ones in the continuum approach, when we apply our matching prescription (\ref{EQN698}), while the agreement of the positions of the TCP is not as good (see \fref{fig:hom_pd}).

\begin{figure}[p]
	\centering   
	\includegraphics[width=7.7cm]{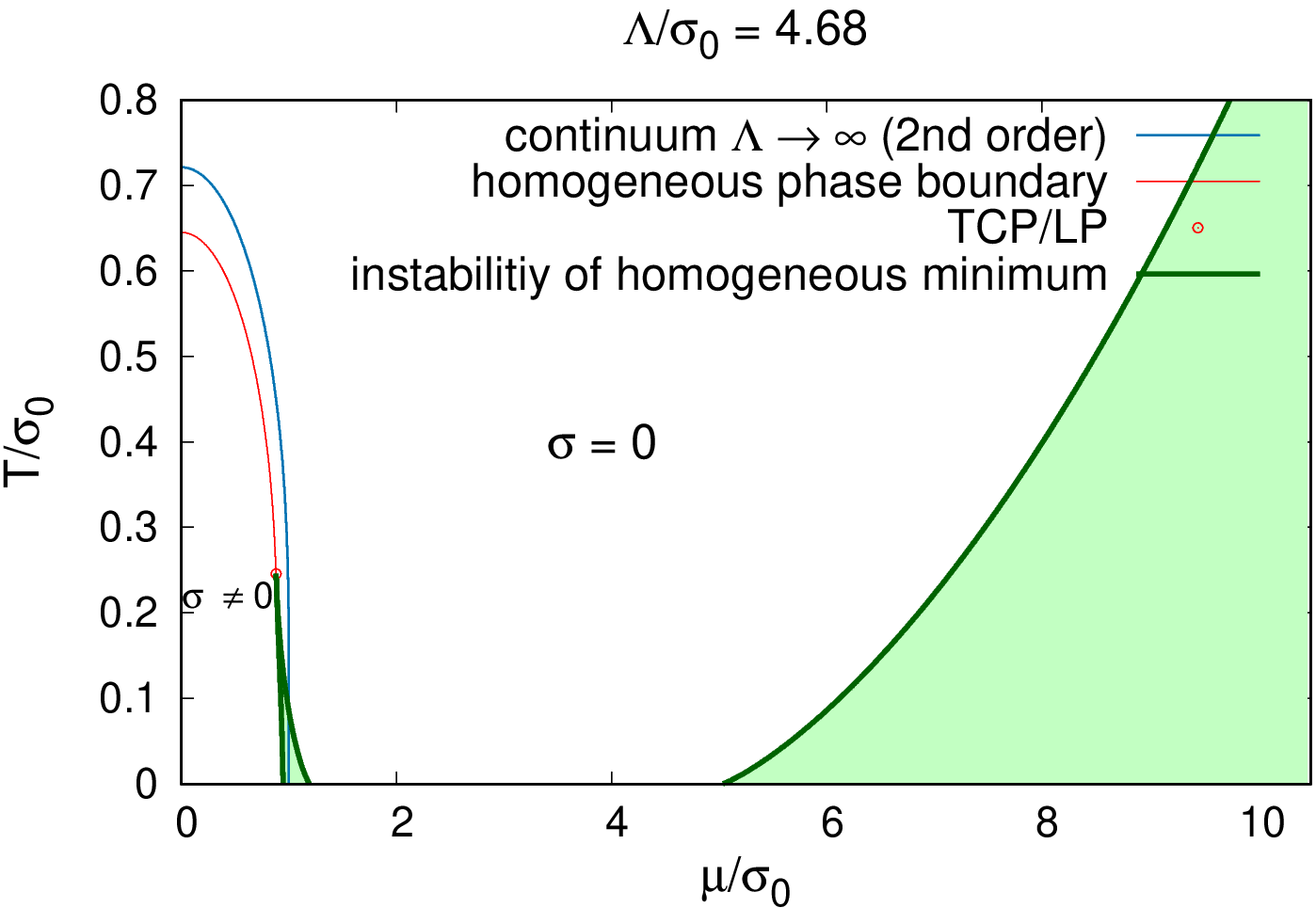}
	\hfill
	\includegraphics[width=7.7cm]{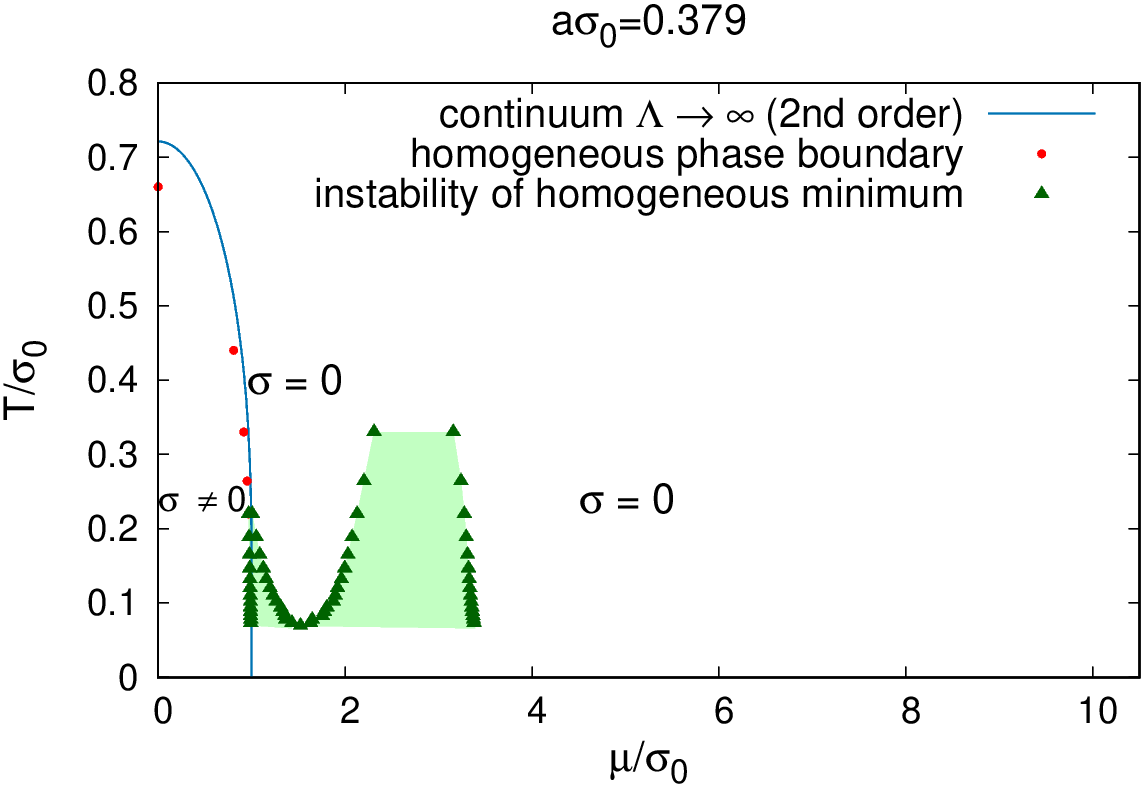} \\
	\includegraphics[width=7.7cm]{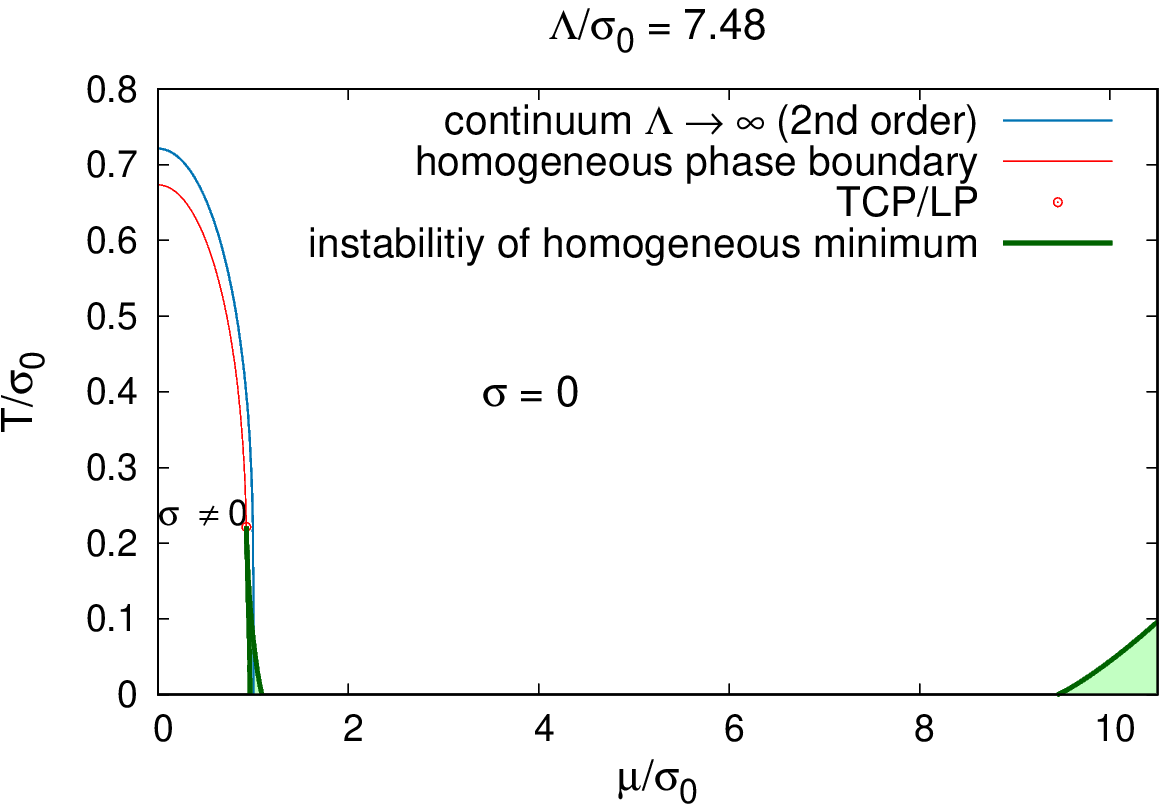}
	\hfill
	\includegraphics[width=7.7cm]{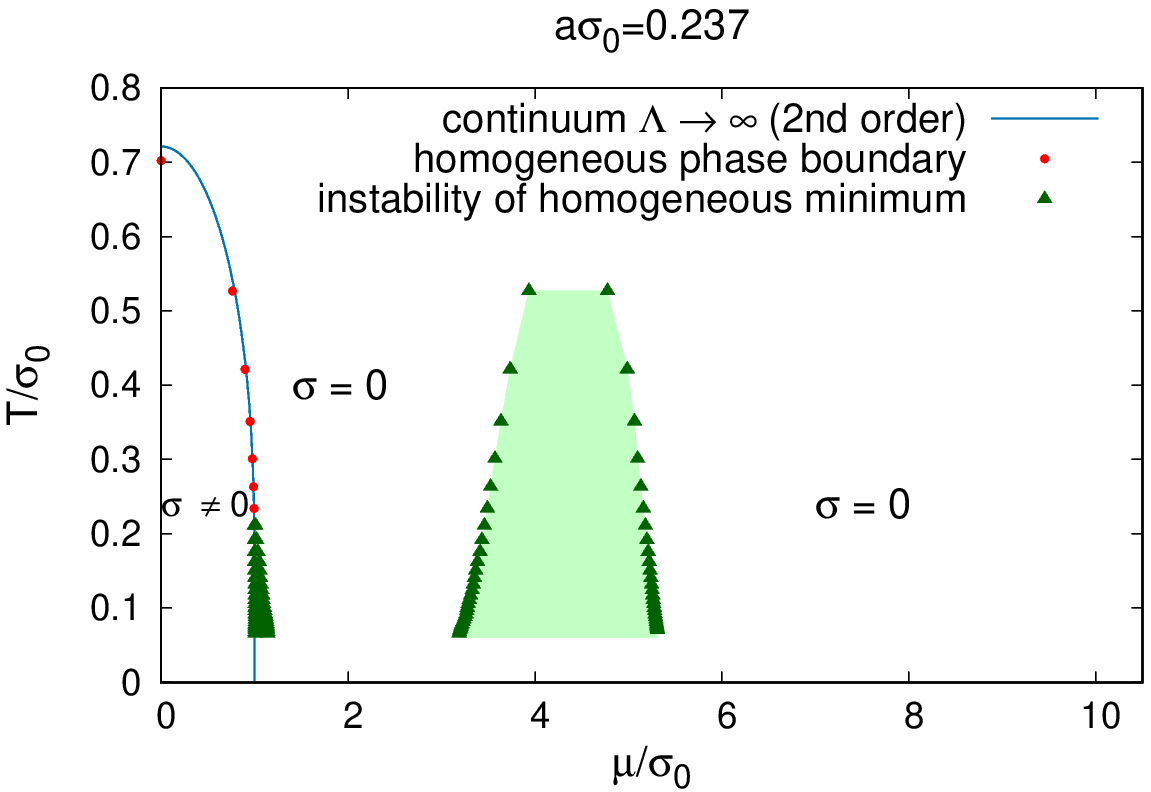} \\
	\includegraphics[width=7.7cm]{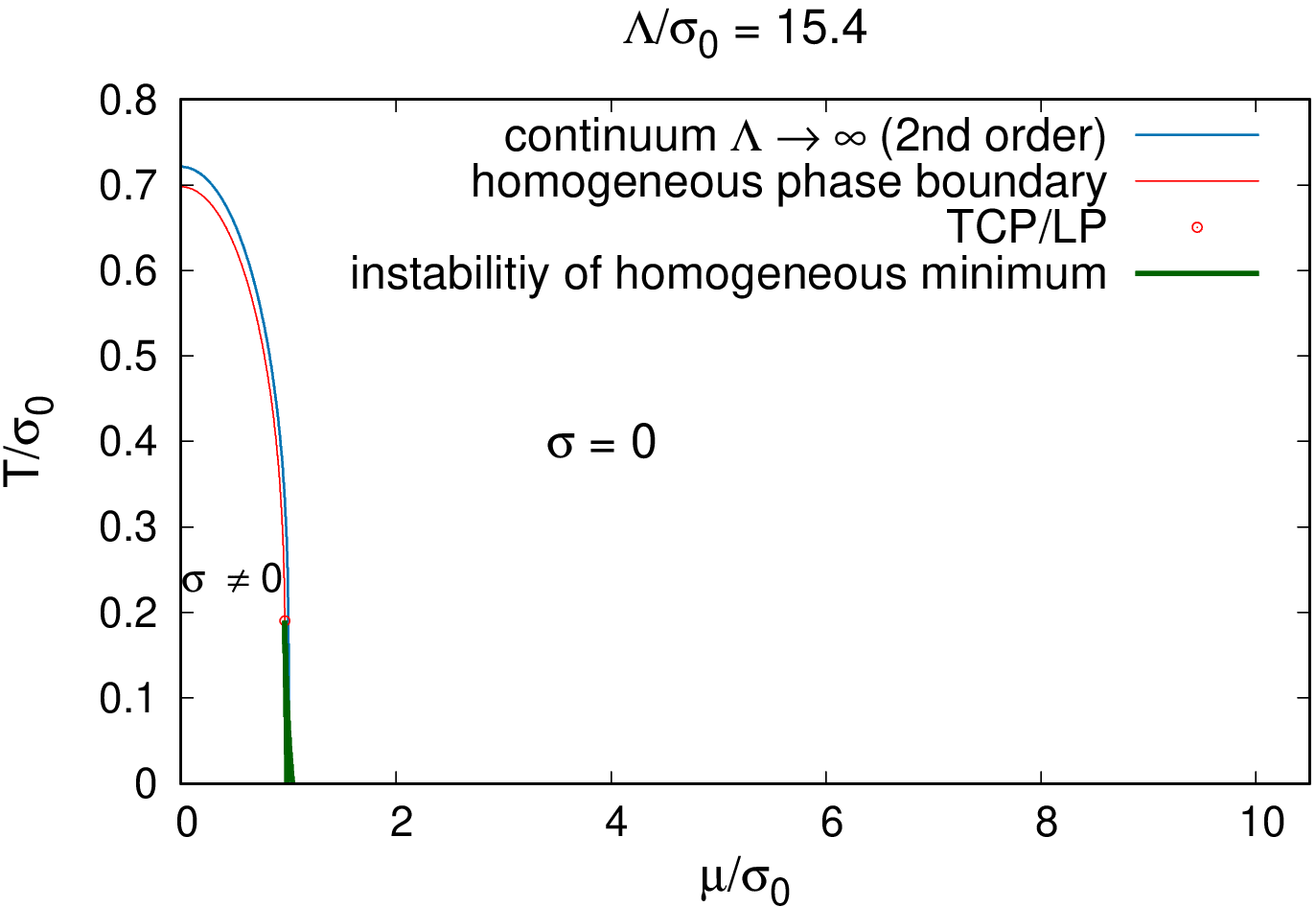}
	\hfill
	\includegraphics[width=7.7cm]{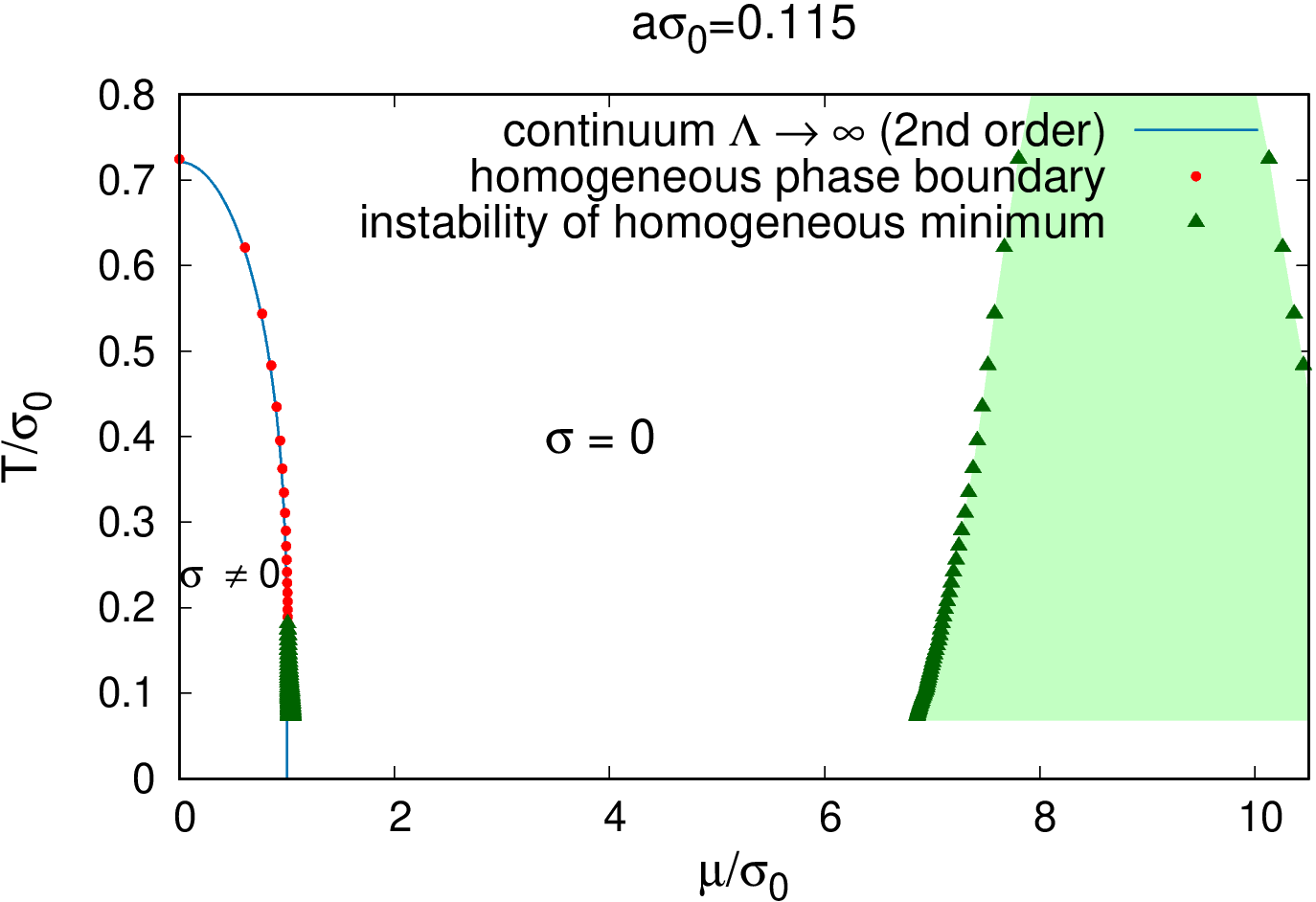}
	\caption{\label{fig:stabanaly_for_large_a} The same as \fref{fig:stabanaly_for_large_a_zoom} but showing a larger region
	of chemical potentials.
	}
\end{figure}

For $a\sigma_0 = 0.379$, despite the similar positions of the LP,
the instability region from the lattice approach extends to considerably higher
chemical potentials than in the corresponding continuum calculation. In fact, extending the stability analysis to higher values
of $\mu$ it turns out that the temperature of the boundary to the stable region rises again, leading to a much larger
instability region, before it drops towards $T=0$ at around $\mu/\sigma_0 = 3.5$. 
This can be seen in \fref{fig:stabanaly_for_large_a} (upper right plot), where the results of the stability analysis are shown
for the same parameters as in \fref{fig:stabanaly_for_large_a_zoom} but for a much larger $\mu$ range.
In the corresponding continuum plot (upper left) we see that around $\mu/\sigma_0 = 5$ a second instability region
appears, which grows to very high temperatures and does not seem to be bounded at all. 
Such a type of inhomogeneous region is also known from the $3+1$-dimensional NJL model, where it has been termed
``inhomogeneous continent''. 
Lowering the cutoff below $\Lambda / \sigma_0 \approx 3$ (not shown in the figure), the continent merges with the finite instability region at smaller $\mu$, which is connected to the homogeneous symmetry-broken phase (sometimes called ``inhomogeneous island''), 
leading to a single large instability region, similar as in 
the lattice result at $a\sigma_0 = 0.379$ but without a decreasing boundary at large $\mu$. 
On the other hand, if we increase the cutoff, the onset of the continent is pushed to higher values of $\mu$, and it disappears
completely for $\Lambda \rightarrow \infty$, so that no instability region is left in the renormalized model.
The behavior on the lattice is qualitatively similar, with the exception that the size of the second inhomogeneous region is finite.
This difference is most likely due to the fact that the momenta of the unstable modes increase with $\mu$, and that, in contrast to the 
continuum approach, the available momentum modes on the lattice are restricted.

So far we have only discussed lattice results with the discretization $W_2 = W''_2$. 
Using the lattice discretization with $W_2 = W'_2$ (see \eqref{eq:corr_disc}) we did not find instabilities anywhere in the $\mu$-$T$ plane, neither for $a \sigma_0 = 0.379$ nor for $a \sigma_0 = 0.115$.
When using naive lattice fermions, momenta $|k_1| \gtapprox \pi/2$ or $|k_2| \gtapprox \pi/2$ in the Fourier expansion of $\sigma$ lead to unphysical interaction terms not present in the GN model, as discussed in \sref{sec:lattice_GN} and in more detail in Ref.\ \cite{Lenz:2020bxk}. Thus, these large momenta need to be suppressed, to obtain lattice actions with the correct continuum limit. This is done in different ways for the two lattice discretizations we are using. $W'_2$ is a smooth cosine function in momentum space, while $W''_2$ is a step function. Both discretizations yield identical results in the continuum limit, but $W'_2$ generates a slight suppression $\propto a^2$ also for small momenta, which are relevant for the formation of inhomogeneous instabilities. In contrast to that $W''_2$ does not generate such a suppression. This could be the reason why the discretization with $W_2 = W''_2$ leads to instability regions at finite $a$, while the discretization with $W_2 = W'_2$ does not.
We will come back to this issue in \sref{sec:fullmini}.


\subsection{\label{sec:fullmini}Minimization of the effective action allowing inhomogeneous modulations}

As pointed out before,  the true inhomogeneous phases could be larger than the instability regions found 
in \sref{sec:inhom_boundary}. 
The latter are  are fully included in the former, but the former might extend further if somewhere 
$\bar{\sigma}$ is just a local minimum of $\Seff$ with the global minimum given by an inhomogeneous $\sigma(\mathbf{x})$.
In particular the non-existence of instability regions in the renormalized model
(at least for $T \neq 0$) does not generally exclude
the existence of an inhomogeneous phase in this limit. 
In the following, in order to explore this possibility, 
we perform numerical minimizations of the lattice discretized effective action (\ref{eq:eff_action_discr}), allowing inhomogeneous modulations of the condensate. To this end, we use $W'_2$ as well as $W''_2$.


\subsubsection{\label{SEC650}Minimization using a cosine ansatz for the condensate}

We start by restricting the condensate to
\begin{eqnarray}
\label{EQN421} \sigma(x_1) = \alpha \cos\bigg(\frac{2 \pi n (x_1/a)}{\Ns}\bigg)
\end{eqnarray}
with fixed integer $n$ and minimize $S_\text{eff}$ with respect to $\alpha$. For small $\alpha$ this ansatz corresponds to one of the terms in the sum of \eqref{eq:Seff2q_lattice}, i.e., it represents a particular perturbation investigated in the stability analyses of \sref{sec:inhom_boundary}. The motivation for this subsection is to provide further support that the two lattice discretizations corresponding to $W_2 = W'_2$ and $W_2 = W''_2$ coincide in the continuum limit. Moreover, this particular cosine ansatz allows to make contact to and perform a cross check with existing lattice results from Ref.\ \cite{Narayanan:2020uqt}.

\begin{figure}[htb]
\begin{center}
\includegraphics[width=10.0cm]{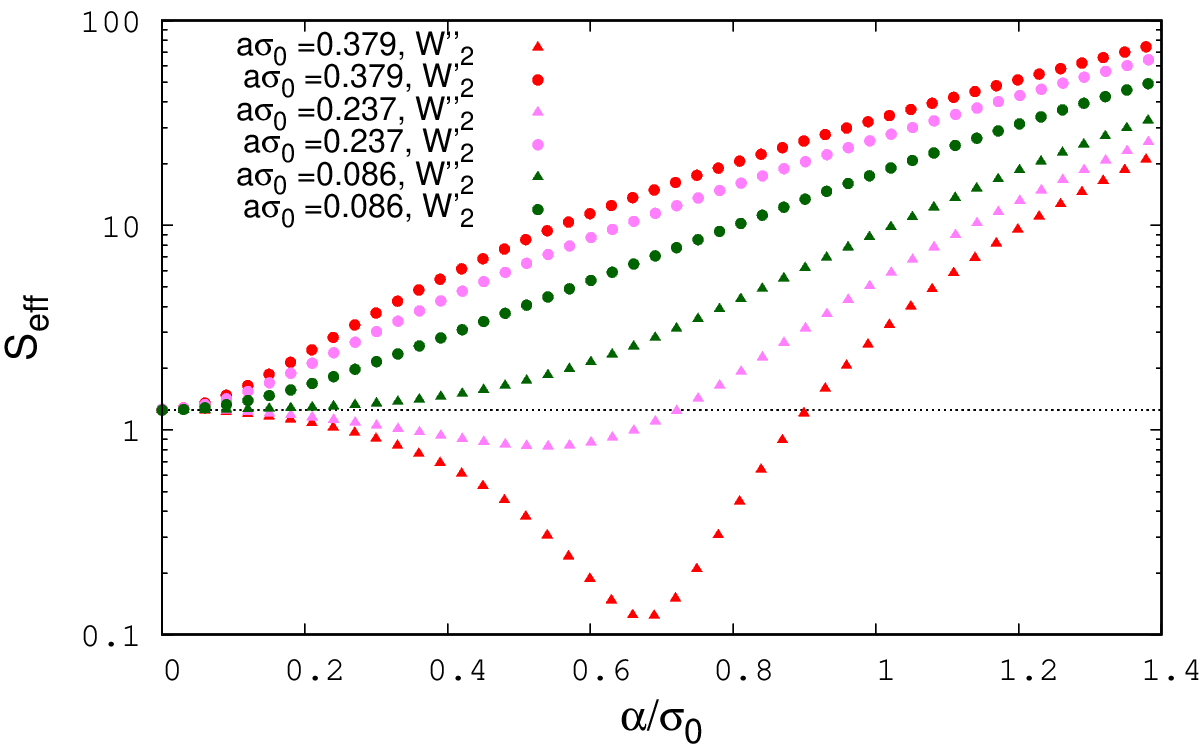}
\caption{\label{fig:S_eff_comp}$S_{\text{eff}} - S_0$ for $(\mu/\sigma_0 , T/\sigma_0) = (1.035 , 0.110)$ as a function of $\alpha$, where the condensate is restricted according to $\sigma(x_1) = \alpha \cos(6 \pi (x_1/a) /\Ns)$. We show results for both lattice discretizations $W_2 = W'_2$ and $W_2 = W''_2$ and three lattice spacings $a$. $S_0$ is a physically irrelevant $a$-dependent shift chosen such that $S_{\text{eff}}|_{\alpha = 0}$ is identical for all three lattice spacings.} 
\end{center}
\end{figure}

As an example we focus on $n = 3$ at $(\mu/\sigma_0 , T/\sigma_0) = (1.035 , 0.110)$. For the lattice discretization with $W_2 = W''_2$ and the two coarser lattice spacings $a \sigma_0 \in \{ 0.379 , 0.237 \}$ the associated $\Gamma^{-1}$ (see \eqref{EQN784}) is negative, i.e., in these cases the ansatz leads to an inhomogeneous phase. For $W_2 = W''_2$ and $a \sigma_0 = 0.086$ as well as for $W_2 = W'_2$ the associated $\Gamma^{-1}$ is positive. In \fref{fig:S_eff_comp} we plot the effective action as a function of $\alpha$. The plot confirms our findings from \sref{sec:inhom_boundary}: $\alpha = 0$ is the location of a maximum of $S_\text{eff}$ for $W_2 = W''_2$ and $a \sigma_0 \in \{ 0.379 , 0.237 \}$ and of a minimum in the other cases. The qualitatively different behavior of $S_\text{eff}$ for $W_2 = W'_2$ and $W_2 = W''_2$, in particular for larger $a$, is the reason, why the discretization with $W_2 = W''_2$ leads to an inhomogeneous phase at finite $a$, while $W_2 = W'_2$ does not. From \fref{fig:S_eff_comp} one can also see that the effective actions for $W_2 = W'_2$ and $W_2 = W''_2$ are approaching each other for decreasing $a$. This indicates that in the limit $a \rightarrow 0$ both actions converge to the same action, the action of the renormalized $2+1$-dimensional GN model, as theoretically expected (see the discussion in \sref{sec:lattice_GN} and Ref.\ \cite{Lenz:2020bxk}). 

The ansatz (\ref{EQN421}) together with a numerical minimization was already used in the 
previous lattice field theory study of the $2+1$-dimensional GN model 
in Ref.\ \cite{Narayanan:2020uqt}. 
There, however, a different discretization was used, equivalent to \eqref{eq:bos_action_disc} with $W_2(\mathbf{x}-\mathbf{y}) = \delta_{\mathbf{x},\mathbf{y}}$. While this discretization does not correspond to the GN model for an arbitrary spatially varying condensate $\sigma(\mathbf{x})$ (see \sref{sec:lattice_GN} and the detailed discussion in Ref.\ \cite{Lenz:2020bxk}), it becomes identical to $W_2 = W''_2$ if $\sigma$ is restricted to a cosine-shaped modulation as in \eqref{EQN421} with $|n| < \Ns / 4$. Of particular interest is a comparison of Fig.\ 8 of Ref.\ \cite{Narayanan:2020uqt} and our results from \fref{fig:S_eff_comp} for $W_2 = W''_2$.
Since we do not use the same lattice spacings and spatial volumes as used in Ref.\ \cite{Narayanan:2020uqt}, a precise quantitative comparison is not possible. However, the two figures are qualitatively identical and indicate consistency of the results presented in Ref.\ \cite{Narayanan:2020uqt} and our results.


\subsubsection{\label{SEC599}Minimization allowing arbitrary 1-dimensional modulations of the condensate}

Finding the global minimum of the effective action for arbitrary $\sigma = \sigma(x_1,x_2)$ is time consuming, because it is quite expensive to evaluate $S_{\text{eff}}$. Thus, we restrict $\sigma$ to arbitrary 1-dimensional modulations, i.e., consider $\sigma = \sigma(x_1)$. $S_{\text{eff}}$ can then be evaluated more efficiently, because the determinant of the Dirac operator factorizes into determinants of smaller matrices of size $\Ns N_d \times \Ns N_d$ (see \eqref{eq:ln_det_Q_1d}). Moreover, the number of variables for the minimization is reduced from $\Ns^2$ to $\Ns$.
To search for the global minimum of $S_{\text{eff}}$ for given $(\mu,T)$, a Fletcher-Reeves conjugate gradient algorithm is used, as implemented in the GNU Scientific Library \cite{GSL}. This algorithm is suited to compute local minima of a given function. We try to find the global minimum of $S_{\text{eff}}$ by carrying out several local minimizations with different initial field configurations $\sigma(x_1)$. Some of these initial configurations are proportional to a cosine as in \eqref{EQN421}, but also randomly generated initial configurations are used. For values $(\mu,T)$, where we find local minima, the corresponding condensates are always periodic and oscillating. We note that local minimizations with randomly generated initial configurations do not lead to additional local minima, but to the same minima already found with initial configurations proportional to a cosine. We interpret this as indication that the found $\sigma(x_1)$ with the smallest corresponding value for $S_{\text{eff}}$ represents the global minimum.


\subsubsection*{The shape of the condensate inside the instability region for $W_2 = W''_2$}

\begin{figure}[htb]
	\includegraphics[width=7.7cm]{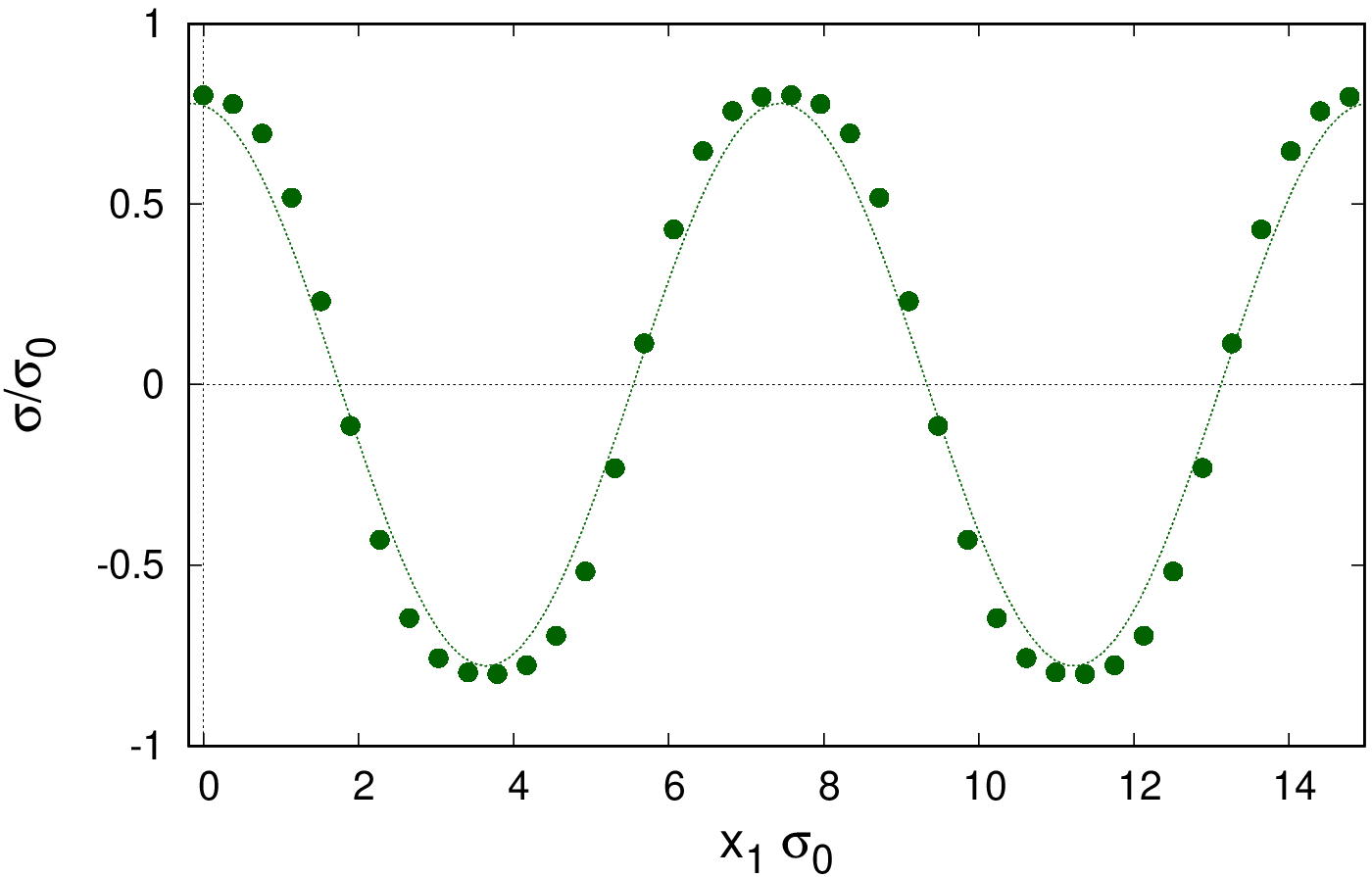}
	\hfill
	\includegraphics[width=7.7cm]{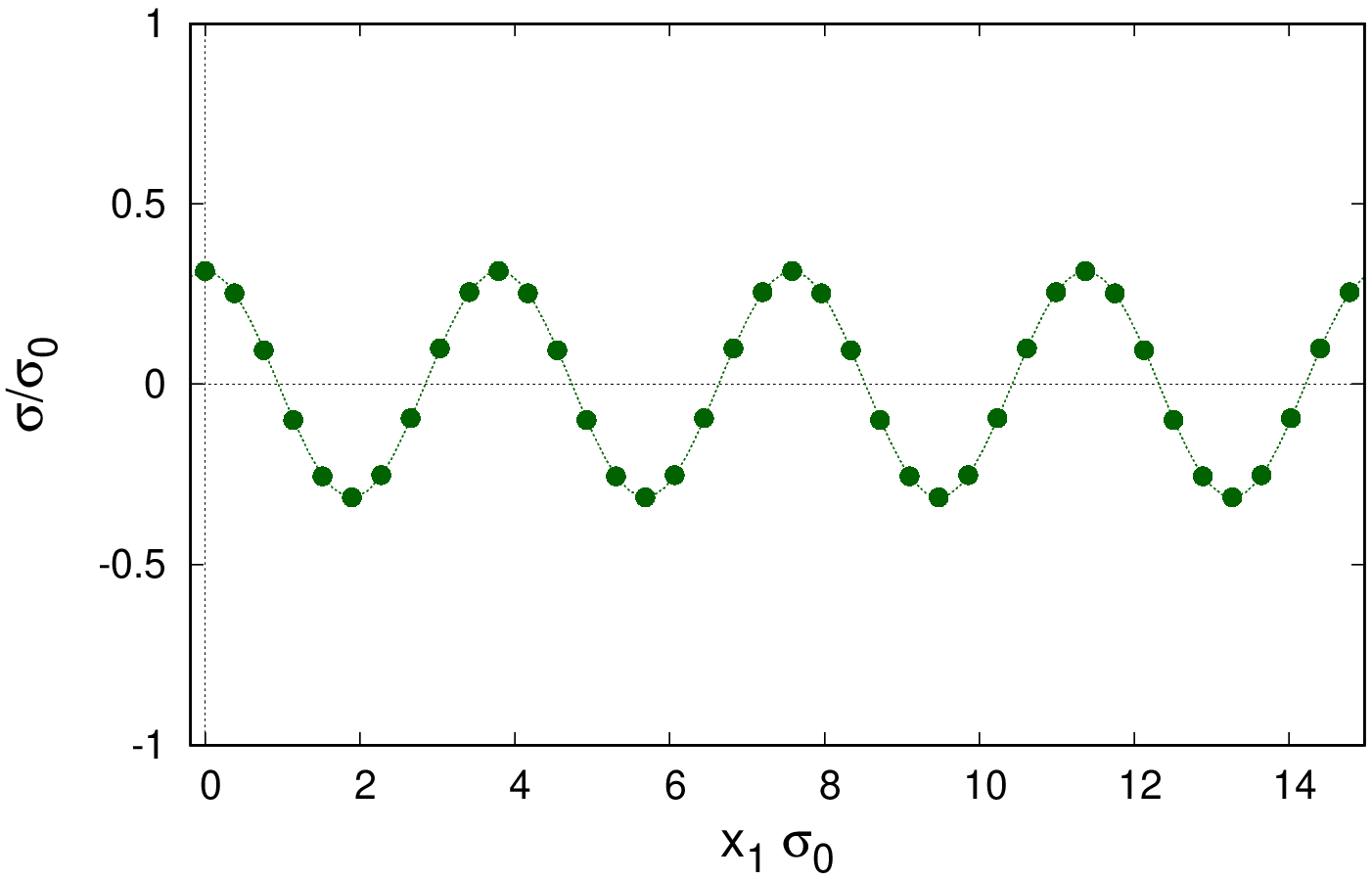} 
	\caption{\label{fig:global_min}The condensate $\sigma$ as a function of $x_1$ for the lattice discretization with $W_2 = W''_2$, $a \sigma_0 = 0.379$, $T/\sigma_0 = 0.132$ and $\mu/\sigma_0 = 0.97$ (left plot) and $\mu/\sigma_0 = 1.11$ (right plot). The dotted lines represent cosine functions with the same wavelengths and amplitudes.}
\end{figure}

As discussed in \sref{sec:inhom_boundary}, the instability regions found for the lattice discretization with $W_2 = W''_2$ and finite lattice spacing (see \fref{fig:stabanaly_for_large_a_zoom} and \fref{fig:stabanaly_for_large_a}) are regions where the condensate exhibits inhomogeneous modulations. By minimizing $S_{\text{eff}}$ we determine the shape of these modulations. In \fref{fig:global_min} we show the condensate $\sigma$ as a function of $x_1$ for $a \sigma_0 = 0.379$, $T/\sigma_0 = 0.132$ and two values of the chemical potential, $\mu/\sigma_0 = 0.97$ (left plot) and $\mu/\sigma_0 = 1.11$ (right plot). For the smaller value of $\mu$ the number of oscillations is smaller and the amplitude is larger than for the larger value of $\mu$. Moreover, at $\mu/\sigma_0 = 0.97$ the condensate deviates from a cosine,
while at $\mu/\sigma_0 = 1.11$ the modulation is essentially a cosine. For even larger $\mu$, when approaching the boundary of the instability region at $\mu/\sigma_0 = 1.134$, the amplitude tends to zero, whereas the wavelength is still finite and consistent with the unstable momentum mode in the stability analysis. Note that the same behavior was observed for the $1+1$-dimensional GN model inside the inhomogeneous phase \cite{Thies:2003kk,Schnetz:2004vr}.

\subsubsection*{Local minima of $S_{\text{eff}}$ in the homogeneous symmetry-broken phase}

\begin{figure}[htb]
	\centering
	\includegraphics[width=7.7cm]{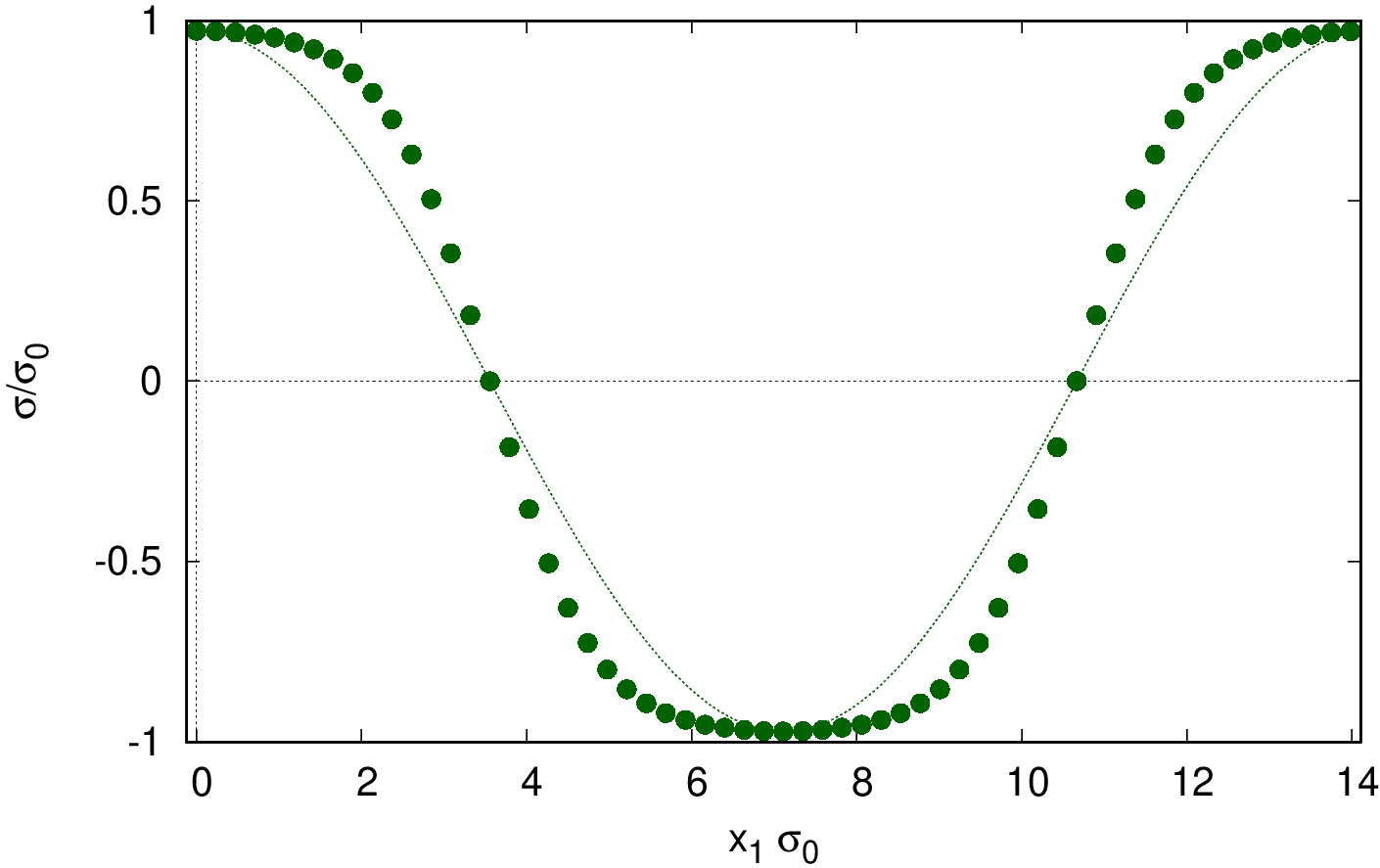}
	\caption{\label{fig:local_min}The condensate $\sigma(x_1)$ corresponding to one of the local minima at $(\mu/\sigma_0,T/\sigma_0) = (0.60,0.176)$ and lattice spacing $a\sigma_0 = 0.237$. There is essentially no difference between the two lattice discretizations $W_2 = W'_2$ and $W_2 = W''_2$. The dotted line represents a cosine function with the same wavelength and amplitude.}
\end{figure}

Inside the $\sigma \neq 0$ regions shown in the right plot of \fref{fig:hom_pd} we find in addition to 
the favored homogeneous configuration
$\sigma = \bar{\sigma}$ several local minima corresponding to inhomogeneous modulations. None of these minima leads to a smaller value of $S_{\text{eff}}$ than the constant condensate $\sigma = \bar{\sigma}$, which seems to represent the global minimum. Only rather close to the boundary of the homogeneous symmetry-broken phase the existence and properties of the inhomogeneous minima depend on the lattice discretization, i.e., whether we use $W_2 = W'_2$ or $W_2 = W''_2$, and on the lattice spacing $a$. Farther inside the homogeneous symmetry-broken phase they are almost independent of $W_2$ and of $a$. This indicates that these local inhomogeneous minima are also present in the renormalized $2+1$-dimensional GN model. As an example we show in \fref{fig:local_min} the condensate $\sigma(x_1)$ corresponding to one such local minimum at $(\mu/\sigma_0,T/\sigma_0) = (0.60,0.176)$ and lattice spacing $a\sigma_0 = 0.237$. As one might expect from the minimizations in the instability region 
discussed above, the wavelengths of the condensates corresponding to the local inhomogeneous minima are quite large, their amplitudes are close to $\sigma_0$ and the shape is somewhere between a cosine and a kink-antikink structure.

We also studied the differences between $S_{\text{eff}}(\bar{\sigma})$ and $S_{\text{eff}}$ evaluated at the local minima. These differences are positive in the homogeneous symmetry-broken phase, since $\sigma = \bar{\sigma}$ represents the global minimum of $S_{\text{eff}}$. For fixed $T$ and increasing $\mu$ the differences become smaller. For the discretization with $W_2 = W''_2$ and $T < T_\mathrm{LP}$ the difference to the local minimum corresponding to a kink-antikink with wavelength $L$ approaches $0$ extremely close to or exactly at the boundary to the instability region. For the discretization with $W_2 = W'_2$ inhomogeneous minima cease to exist near the boundary to the instability region. For fixed $\mu$ and decreasing $T$ the differences between $S_{\text{eff}}(\bar{\sigma})$ and $S_{\text{eff}}$ evaluated at the local minima also decrease. This is consistent with Ref.\ \cite{Urlichs:2007zz}, where the $2+1$-dimensional GN was studied at $T = 0$ using a specific ansatz for $\sigma(x_1)$ based on Jacobi elliptic functions. There it was found that such inhomogeneous modulations represent minima of $S_{\text{eff}}$, which are degenerate with the minimum at $\sigma = \bar{\sigma}$.


\subsubsection*{Phase diagram}

Within the $\sigma = 0$ regions shown in the right plot of \fref{fig:hom_pd} and (for $W_2 = W''_2$) outside the instability region we do not find any local minima corresponding to inhomogeneous modulations. Thus, the phase diagram of the $2+1$-dimensional GN model, when allowing arbitrary 1-dimensional modulations, is identical to that already found by stability analyses (see \fref{fig:hom_pd}, right plot for $W_2 = W'_2$ and \fref{fig:stabanaly_for_large_a_zoom} and \fref{fig:stabanaly_for_large_a}, right columns for $W_2 = W''_2$). In particular we confirm and consolidate the findings from Ref.\ \cite{Narayanan:2020uqt}, where some evidence was presented that there is no inhomogeneous phase in the continuum limit.

It is, however, important to note, that we did not yet carry out minimizations for 2-dimensional modulations.  Such 2-dimensional modulations of the condensate might lead to lower values of $S_{\text{eff}}$ and, thus, could generate larger inhomogeneous phases at finite $a$ or inhomogeneous phases, which survive the continuum limit. We plan to study this possibility in the near future.



\section{\label{SEC608}Conclusions}

In this work we explored in detail the phase diagram of the $2+1$-dimensional GN model in the limit of infinitely many flavors. We implemented three different regularizations, a continuum regularization with a Pauli-Villars cutoff $\Lambda$ and two lattice field theory regularizations with lattice spacing $a$, which are based on naive fermions. Particular focus was put on studying the possible existence of inhomogeneous phases, their location in the $\mu$-$T$ plane and their dependence on the regularization and the corresponding regulator, $\Lambda$ or $a$, respectively.

Our main results are the following:
\begin{itemize}
\item For finite values of the regulator, inhomogeneous phases may exist, depending on the details of the regularization: For the continuum regularization and one of the two lattice field theory regularizations we found an inhomogeneous phase, while for the other lattice field theory regularization there is no inhomogeneous phase.

\item Even if there is an inhomogeneous phase at finite values of the regulator, it seems to disappear, when the regulator is removed, i.e., in the limit $\Lambda \rightarrow \infty$ or $a \rightarrow 0$.
\end{itemize}
These results confirm existing results, e.g., continuum results for the homogeneous phase diagram \cite{Klimenko:1987gi,Rosenstein:1988dj} or lattice field theory results at a single lattice spacing \cite{Winstel:2019zfn} and at several lattice spacings, but with a specific cosine ansatz for the condensate \cite{Narayanan:2020uqt}. Our work also substantially extends these existing results, in particular by an analytical stability analysis of homogeneous phases with respect to arbitrary inhomogeneous perturbations and by a full numerical minimization of the effective action, where arbitrary 1-dimensional modulations of the condensate are allowed, i.e., without restriction to a specific ansatz like plane waves or Jacobi elliptic functions as done in existing work \cite{Urlichs:2007zz,Narayanan:2020uqt}.

It is important to note that our numerical minimization is currently limited to arbitrary 1-dimensional modulations. Thus, an important next step is to allow arbitrary 2-dimensional modulations. This could not only lead to larger or additional inhomogeneous phases at finite values of the regulator, but also to the existence of inhomogeneous phases in the renormalized $2+1$-dimensional GN model, i.e., in the limit $\Lambda \rightarrow \infty$ or $a \rightarrow 0$.

Our results call for a critical revision of the role of the regularization in the physically more relevant case of $3+1$ space-time dimensions. For instance, inhomogeneous phases have also been found in the $3+1$-dimensional NJL model.
However, unlike in $2+1$ dimensions, this model is non-renormalizable and therefore the studies have been performed 
using finite fixed regulators. 
Given that inhomogeneous phases exist in the renormalized $1+1$-dimensional GN and NJL models, 
while in the $2+1$-dimensional GN model they are only present at finite regulator values, one might suspect that the 
observed inhomogeneous phases at $3+1$ dimensions could be regularization artifacts.
The QM model, on the other hand, is renormalizable in $3+1$ dimensions, and inhomogeneous phases have been
reported to exist in that model even in the renormalized limit. Unfortunately, the model suffers from other instabilities at large cutoff values, at least in mean-field approximation. 
In the near future we therefore plan to extend our detailed investigations to $3+1$-dimensional models in order to shed light on these issues.



\section*{Acknowledgements}

We acknowledge useful discussions with J.\ Braun, P.\ de Forcrand, A.\ K\"onigstein, C.\ Niehof, L.\ Pannullo, M.\ Thies, and A.\ Wipf.

We acknowledge support by the Deutsche Forschungsgemeinschaft (DFG, German Research Foundation) through the CRC-TR 211 ``Strong interaction matter under extreme conditions'' -- project number 315477589-TRR 211. Marc Wagner acknowledges support by the Heisenberg Programme of the Deutsche Forschungsgemeinschaft (DFG, German Research Foundation) -- project number 399217702. Lennart Kurth and Marc Winstel are supported  by  the  GSI  F\&E.

Calculations on the GOETHE-HLR and on the on the FUCHS-CSC high-performance computers of the Frankfurt University were conducted for this research. We would like to thank HPC-Hessen, funded by the State Ministry of Higher Education, Research and the Arts, for programming advice.



\appendix

\section{Proof of $\det(\Q[-\sigma]) = \det(\Q[+\sigma])$ and $\det(\Q) \in \R$}

The calculations in this appendix are valid for the $2 \times 2$ fermion representations (\ref{eq:2comp_gamm1}) and (\ref{eq:2:comp_gamm2}) and for the $4 \times 4$ fermion representation (\ref{eq:eq:4comp_gamma}). Note that we restrict the dependence of $\sigma$ to the spatial coordinates, i.e., $\sigma = \sigma(x_1,x_2)$, as specified in \sref{SEC456}.


\subsection{\label{APP001}$\det(\Q[-\sigma]) = \det(\Q[+\sigma])$}

We start with the eigenvalue equation for $\Q^T$,
\begin{align}
\Q^T[+\sigma] f_j = \Big(+\gamma_0 \partial_0 - \gamma_0 \mu - \gamma_1 \partial_1 - \gamma_2 \partial_2 + \sigma(x_1,x_2)\Big) f_j(x_0,x_1,x_2) = \alpha_j f_j(x_0,x_1,x_2) ,
\end{align}
where we have used $\partial_\mu^T = -\partial_\mu$. The coordinate transformation $u = -x_0$ leads to
\begin{align}
\underbrace{\Big(\gamma_0 \partial_0 + \gamma_0 \mu + \gamma_1 \partial_1 + \gamma_2 \partial_2 - \sigma(x_1,x_2)\Big)}_{= \Q[-\sigma]} f_j(-u,x_1,x_2) = -\alpha_j f_j(-u,x_1,x_2) .
\end{align}
Thus, if $\alpha_j$ is an eigenvalue of $\Q^T[+\sigma]$, $-\alpha_j$ is an eigenvalue of $\Q[-\sigma]$. Consequently,
\begin{equation}
\label{EQN679} \det(\Q[-\sigma]) = \prod_j (-\alpha_j) = \prod_j \alpha_j = \det(\Q^T[+\sigma]) = \det(\Q[+\sigma]) ,
\end{equation}
where we have used that the number of eigenvalues is even. Note that \eqref{EQN679} implies
\begin{equation}
S_{\text{eff}}[-\sigma] = S_{\text{eff}}[+\sigma] .
\end{equation}


\subsection{\label{APP002}$\det(\Q) \in \R$}

We start with the eigenvalue equation for $\Q$,
\begin{align}
\Q[+\sigma] f_j = \Big(+\gamma_0 \partial_0 + \gamma_0 \mu + \gamma_1 \partial_1 + \gamma_2 \partial_2 + \sigma(x_1,x_2)\Big) f_j(x_0,x_1,x_2) = \alpha_j f_j(x_0,x_1,x_2) .
\end{align}
Complex conjugation leads to
\begin{align}
\Big(-\gamma_0 \partial_0 - \gamma_0 \mu + \gamma_1 \partial_1 + \gamma_2 \partial_2 + \sigma(x_1,x_2)\Big) f_j^\ast(x_0,x_1,x_2) = \alpha_j^\ast f_j^\ast(x_0,x_1,x_2)
\end{align}
and multiplication of this equation with $-\gamma_0$ to
\begin{align}
\underbrace{\Big(+\gamma_0 \partial_0 + \gamma_0 \mu + \gamma_1 \partial_1 + \gamma_2 \partial_2 - \sigma(x_1,x_2)\Big)}_{= Q[-\sigma]} \gamma_0 f_j^\ast(x_0,x_1,x_2) = -\alpha_j^\ast \gamma_0 f_j^\ast(x_0,x_1,x_2) .
\end{align}
Thus, if $\alpha_j$ is an eigenvalue of $\Q[+\sigma]$, $-\alpha_j^\ast$ is an eigenvalue of $\Q[-\sigma]$. Consequently,
\begin{equation}
\label{EQN680} \Big(\det(\Q[-\sigma])\Big)^\ast = \bigg(\prod_j (-\alpha_j^\ast)\bigg)^\ast = \prod_j \alpha_j = \det(\Q[+\sigma]) ,
\end{equation}
where we have again used that the number of eigenvalues is even. Combining \eqref{EQN679} and \eqref{EQN680} leads to
\begin{equation}
\Big(\det(\Q[+\sigma])\Big)^\ast = \det(\Q[+\sigma]) ,
\end{equation}
i.e., $\det(\Q) \in \R$.





\begin{thebibliography}{99}
	
	\bibitem{Kumar:2013cqa}%
	L.~Kumar, 
	``Review of recent results from the RHIC beam energy scan'',
	Mod.\ Phys.\ Lett.\ A \textbf{28}, 1330033 (2013) 
	[arXiv:1311.3426 [nucl-ex]].

	\bibitem{Friman:2011zz}
	B.~Friman et al. (ed.),
	``The CBM physics book: compressed baryonic matter in laboratory experiments'',
	Lect.\ Notes Phys.\ \textbf{814}, 1 (2011).

	\bibitem{Borsanyi:2010bp}
	S.~Borsanyi et al.,
	``Is there still any T\_c mystery in lattice QCD? Results with physical masses in the continuum limit III'',
	JHEP \textbf{09}, 073 (2010)
	[arXiv:1005.3508 [hep-lat]].
	
	\bibitem{Bazavov:2011nk}
	A.~Bazavov et al.,
	``The chiral and deconfinement aspects of the QCD transition'',
	Phys.\ Rev.\ D \textbf{85}, 054503 (2012)
	[arXiv:1111.1710 [hep-lat]].

	\bibitem{Bellwied:2015rza}
	R.~Bellwied, S.~Borsanyi, Z.~Fodor, J.~G\"unther, S.~D.~Katz, C.~Ratti and K.~K.~Szabo,
	``The QCD phase diagram from analytic continuation'',
	Phys.\ Lett.\ B \textbf{751}, 559 (2015)
	[arXiv:1507.07510 [hep-lat]].

	\bibitem{Bazavov:2018mes}
	A.~Bazavov et al.,
	``Chiral crossover in QCD at zero and non-zero chemical potentials'',
	Phys.\ Lett.\ B \textbf{795}, 15 (2019)
	[arXiv:1812.08235 [hep-lat]].
	
	\bibitem{Fischer:2018sdj}
	C.~S.~Fischer,
	``QCD at finite temperature and chemical potential from Dyson-Schwinger equations'',
	Prog.\ Part.\ Nucl.\ Phys.\ \textbf{105}, 1 (2019)
	[arXiv:1810.12938 [hep-ph]].

	\bibitem{Fu:2019hdw}
	W.~J.~Fu, J.~M.~Pawlowski and F.~Rennecke,
	``The QCD phase structure at finite temperature and density'',
	Phys.\ Rev.\ D \textbf{101}, 054032 (2020)
	[arXiv:1909.02991 [hep-ph]].
	
	\bibitem{Asakawa:1989bq}
	M.~Asakawa and K.~Yazaki,
	``Chiral Restoration at Finite Density and Temperature'',
	Nucl.\ Phys.\ A \textbf{504}, 668 (1989).

	\bibitem{Scavenius:2000qd}
	O.~Scavenius, A.~Mocsy, I.~N.~Mishustin and D.~H.~Rischke,
	``Chiral phase transition within effective models with constituent quarks'',
	Phys.\ Rev.\ C \textbf{64}, 045202 (2001)
	[arXiv:nucl-th/0007030].

	\bibitem{Schaefer:2006ds}
	B.~J.~Schaefer and J.~Wambach,
	``Susceptibilities near the QCD (tri)critical point'',
	Phys.\ Rev.\ D \textbf{75}, 085015 (2007)
	[arXiv:hep-ph/0603256].

	\bibitem{Buballa:2014tba}
	M.~Buballa and S.~Carignano,
	``Inhomogeneous chiral condensates'',
	Prog.\ Part.\ Nucl.\ Phys.\ \textbf{81}, 39 (2015)
	[arXiv:1406.1367 [hep-ph]].
	
	\bibitem{Thies:2003kk}
	M.~Thies and K.~Urlichs,
	``Revised phase diagram of the Gross-Neveu model'',
	Phys.\ Rev.\ D {\bf 67}, 125015 (2003)
	[arXiv:hep-th/0302092].
	
	\bibitem{Schnetz:2004vr}
	O.~Schnetz, M.~Thies and K.~Urlichs,
	``Phase diagram of the Gross-Neveu model: exact results and condensed matter precursors'',
	Annals Phys.\ {\bf 314}, 425 (2004)
	[arXiv:hep-th/0402014].

	\bibitem{Thies:2006ti}
	M.~Thies,
	``From relativistic quantum fields to condensed matter and back again: updating the Gross-Neveu phase diagram'',
	J.\ Phys.\ A \textbf{39}, 12707 (2006)
	[arXiv:hep-th/0601049].
	
	\bibitem{Nakano:2004cd}
	E.~Nakano and T.~Tatsumi,
	``Chiral symmetry and density wave in quark matter'',
	Phys.\ Rev.\ D \textbf{71}, 114006 (2005)
	[arXiv:hep-ph/0411350].

	\bibitem{Nickel:2009wj}
	D.~Nickel,
	``Inhomogeneous phases in the Nambu-Jona-Lasino and quark-meson model'',
	Phys.\ Rev.\ D \textbf{80}, 074025 (2009)
	[arXiv:0906.5295 [hep-ph]].

	\bibitem{Carignano:2014jla}
	S.~Carignano, M.~Buballa and B.~J.~Schaefer,
	``Inhomogeneous phases in the quark-meson model with vacuum fluctuations'',
	Phys.\ Rev.\ D \textbf{90}, 014033 (2014)
	[arXiv:1404.0057 [hep-ph]].

	\bibitem{Heinz:2015lua} 
	A.~Heinz, F.~Giacosa, M.~Wagner and D.~H.~Rischke,
	``Inhomogeneous condensation in effective models for QCD using the finite-mode approach'',
	Phys.\ Rev.\ D {\bf 93}, 014007 (2016)
	[arXiv:1508.06057 [hep-ph]].

	\bibitem{Muller:2013tya}
	D.~M\"uller, M.~Buballa and J.~Wambach,
	``Dyson-Schwinger study of chiral density waves in QCD'',
	Phys.\ Lett.\ B \textbf{727}, 240 (2013)
	[arXiv:1308.4303 [hep-ph]].
	
	\bibitem{Nickel:2009ke}
	D.~Nickel,
	``How many phases meet at the chiral critical point?'',
	Phys.\ Rev.\ Lett.\ \textbf{103}, 072301 (2009)
	[arXiv:0902.1778 [hep-ph]].

	\bibitem{Basar:2009fg}
	G.~Basar, G.~V.~Dunne and M.~Thies,
	``Inhomogeneous condensates in the thermodynamics of the chiral NJL$_2$ model'',
	Phys.\ Rev.\ D \textbf{79}, 105012 (2009)
	[arXiv:0903.1868 [hep-th]].

	\bibitem{Abuki:2011pf}
	H.~Abuki, D.~Ishibashi and K.~Suzuki,
	``Crystalline chiral condensates off the tricritical point in a generalized Ginzburg-Landau approach'',
	Phys.\ Rev.\ D \textbf{85}, 074002 (2012)
	[arXiv:1109.1615 [hep-ph]].
	
	\bibitem{Buballa:2018hux}
	M.~Buballa and S.~Carignano,
	``Inhomogeneous chiral phases away from the chiral limit'',
	Phys.\ Lett.\ B \textbf{791}, 361 (2019)
	[arXiv:1809.10066 [hep-ph]].
	
	\bibitem{Carignano:2019ivp}
	S.~Carignano and M.~Buballa,
	``Inhomogeneous chiral condensates in three-flavor quark matter'',
	Phys.\ Rev.\ D \textbf{101}, 014026 (2020)
	[arXiv:1910.03604 [hep-ph]].

	\bibitem{Buballa:2020xaa}
	M.~Buballa, S.~Carignano and L.~Kurth,
	``Inhomogeneous phases in the quark-meson model with explicit chiral-symmetry breaking'',
	[arXiv:2006.02133 [hep-ph]].
	
	\bibitem{deForcrand:2006zz} 
	P.~de Forcrand and U.~Wenger,
	``Baryon matter in the lattice Gross-Neveu model'',
	PoS LATTICE2007, \textbf{152} (2006)
	[arXiv:hep-lat/0610117].
	
	\bibitem{Wagner:2007he} 
	M.~Wagner,
	``Fermions in the pseudoparticle approach'',
	Phys.\ Rev.\ D {\bf 76}, 076002 (2007)
	[arXiv:0704.3023 [hep-lat]].
	
	\bibitem{Lenz:2020bxk}
	J.~Lenz, L.~Pannullo, M.~Wagner, B.~Wellegehausen and A.~Wipf,
	``Inhomogeneous phases in the Gross-Neveu model in 1+1 dimensions at finite number of flavors'',
	Phys.\ Rev.\ D \textbf{101}, 094512 (2020)
	[arXiv:2004.00295 [hep-lat]].
	
	\bibitem{Lenz:2020cuv}
	J.~J.~Lenz, L.~Pannullo, M.~Wagner, B.~Wellegehausen and A.~Wipf,
	``Baryons in the Gross-Neveu model in 1+1 dimensions at finite number of flavors'',
	Phys.\ Rev.\ D \textbf{102}, 114501 (2020)
	[arXiv:2007.08382 [hep-lat]].

	\bibitem{Klimenko:1987gi}
	K.~G.~Klimenko,
	``Phase structure of generalized {Gross-Neveu} models'',
	Z.\ Phys.\ C \textbf{37}, 457 (1988).
	
	\bibitem{Rosenstein:1988dj} 
	B.~Rosenstein, B.~J.~Warr and S.~H.~Park,
	``Thermodynamics of (2+1)-dimensional four Fermi models'',
	Phys.\ Rev.\ D {\bf 39}, 3088 (1989).
	
	\bibitem{Winstel:2019zfn} 
	M.~Winstel, J.~Stoll and M.~Wagner,
	``Lattice investigation of an inhomogeneous phase of the 2+1-dimensional Gross-Neveu model in the limit of infinitely many flavors'',
	J.\ Phys.: Conf.\ Ser.\ \textbf{1667}, 012044 (2020)
	[arXiv:1909.00064 [hep-lat]].
	
	\bibitem{Narayanan:2020uqt}
	R.~Narayanan,
	``Phase diagram of the large $N$ Gross-Neveu model in a finite periodic box'',
	Phys.\ Rev.\ D \textbf{101}, 096001 (2020)
	[arXiv:2001.09200 [hep-th]].
	
	\bibitem{Gross:1974jv} 
	D.~J.~Gross and A.~Neveu,
	``Dynamical symmetry breaking in asymptotically free field theories'',
	Phys.\ Rev.\ D {\bf 10}, 3235 (1974).
	
	\bibitem{Hubbard:1959ub} 
	J.~Hubbard,
	``Calculation of partition functions'',
	Phys.\ Rev.\ Lett.\  {\bf 3}, 77 (1959).
	
	\bibitem{Appelquist:1986fd} 
	T.~W.~Appelquist, M.~J.~Bowick, D.~Karabali and L.~C.~R.~Wijewardhana,
	``Spontaneous chiral symmetry breaking in three-dimensional QED'',
	Phys.\ Rev.\ D {\bf 33}, 3704 (1986).
	
	\bibitem{Gies:2010st}
	H.~Gies and L.~Janssen,
	``UV fixed-point structure of the three-dimensional Thirring model'',
	Phys.\ Rev.\ D \textbf{82}, 085018 (2010)
	[arXiv:1006.3747 [hep-th]].
	
	\bibitem{Scherer:2012nn} 
	D.~D.~Scherer and H.~Gies,
	``Renormalization group study of magnetic catalysis in the 3d Gross-Neveu model'',
	Phys.\ Rev.\ B {\bf 85}, 195417 (2012)
	[arXiv:1201.3746 [cond-mat.str-el]].
	
	\bibitem{Schmidt:2017} 
	D.~Schmidt,
	``Three-dimensional four-fermion theories with exact chiral symmetry on the lattice'', doctoral thesis at the Friedrich-Schiller-Universit\"at Jena (2017).
	
	\bibitem{Inagaki:1995xw}
	T.~Inagaki,
	``Phase structure of a four-fermion theory at finite temperature and chemical potential'',
	[arXiv:hep-ph/9511201].
	
	\bibitem{Rothe:1992nt} 
	H.~J.~Rothe,
	``Lattice gauge theories: An Introduction'',
	World Sci.\ Lect.\ Notes Phys.\ {\bf 82}, 1 (2012).
	
	\bibitem{Cohen:1983nr}
	Y.~Cohen, S.~Elitzur and E.~Rabinovici,
	``A Monte Carlo study of the Gross-Neveu model'',
	Nucl.\ Phys.\ B \textbf{220}, 102-118 (1983).
	
	\bibitem{GSL}
	M. Galassi et al.,
	``GNU Scientific Library Reference Manual (3rd Ed.)'',
	ISBN 0954612078.
	
	\bibitem{Urlichs:2007zz} 
	K.~Urlichs,
	``Baryons and baryonic matter in four-fermion interaction models'', doctoral thesis at the University of Erlangen-N\"urnberg (2007).

\end{thebibliography}
\end{document}